\documentclass[fleqn,usenatbib]{mnras}
\usepackage{newtxtext,newtxmath}

\usepackage[T1]{fontenc}
\usepackage{ae,aecompl}

\usepackage{graphicx}	
\usepackage{amsmath}	
\usepackage{amssymb}	
\usepackage{subfig}

\title[Paper I - size distributions]{Galaxy sizes and the galaxy-halo connection - I: the remarkable tightness of the size distributions}

\author[L. Zanisi ]{
Lorenzo Zanisi$^{1,2}$\thanks{E-mail: L.Zanisi@soton.ac.uk},
Francesco Shankar$^{1}$,
Andrea Lapi$^{3,4}$,
Nicola Menci$^{5}$,  \newauthor
Mariangela Bernardi$^{6}$,
Christopher Duckworth$^{7,8}$,
Marc Huertas-Company$^{9,10}$, \newauthor
Philip Grylls$^{1}$, Paolo Salucci$^{3}$
\\
$^{1}$ Department of Physics and Astronomy, University of Southampton, Highfield, SO17 1BJ, UK\\
$^{2}$ DISCnet Center for Doctoral Training, University of Southampton, Highfield, SO17 1BJ, UK\\
$^{3}$ SISSA, Via Bonomea 265, I-34136 Trieste, Italy\\
$^{4}$ IFPU - Institute for fundamental physics of the Universe, Via Beirut 2, 34014 Trieste, Italy\\
$^{5}$ INAF $-$ Osservatorio Astronomico di Roma, via Frascati 33, $I-00078$ Monteporzio, Italy\\
$^{6}$ Department of Physics and Astronomy, University of Pennsylvania, Philadelphia, PA 19104, USA\\
$^{7}$ School of Physics and Astronomy, University of St Andrews, North Haugh, St Andrews, KY16 9SS, UK\\
$^{8}$ Center for Computational Astrophysics, Flatiron Institute, 162 Fifth Avenue, New York, NY 10010, USA\\
$^{9}$ Instituto de Astrof\'isica de Canarias (IAC); Departamento de Astrof\'isica, Universidad de La Laguna (ULL), E-38200, La Laguna, Spain\\
$^{10}$ LERMA, Observatoire de Paris, CNRS, PSL, Universit\'e Paris Diderot, France
}

\date{Accepted 2019 December 4. Received 2019 November 26; in original form 2019 May 20}

\pubyear{2019}
\begin{document}
\label{firstpage}
\pagerange{\pageref{firstpage}--\pageref{lastpage}}
\maketitle

\begin{abstract}
The mass and structural assembly of galaxies is a matter of intense debate. Current theoretical models predict the existence of a linear relationship between galaxy size ($R_e$) and the host dark matter halo virial radius ($R_h$).\\
By making use of semi-empirical models compared to the size distributions of central galaxies from the Sloan Digital Sky Survey, we provide robust constraints on the normalization and scatter of the $R_e-R_h$ relation. We explore the parameter space of models in which the $R_e-R_h$ relation is mediated by either the spin parameter or the concentration of the host halo, or a simple constant the nature of which is in principle unknown.\\
We find that the data require extremely tight relations  for both early-type and late-type galaxies (ETGs,LTGs), especially for more massive galaxies. These constraints challenge models based solely on angular momentum conservation, which predict significantly wider distributions of galaxy sizes and no trend with stellar mass, if taken at face value. We discuss physically-motivated alterations to the original models that bring the predictions into better agreement with the data. We argue that the measured tight size distributions of SDSS disk galaxies can be reproduced by semi-empirical models in which the $R_e-R_h$ connection is mediated by the \emph{stellar} specific angular momenta $j_{star}.$ We find that current cosmological models of galaxy formation broadly agree with our constraints for LTGs, and justify the strong link between $R_e$ and $j_{star}$ that we propose, however the tightness of the $R_e-R_h$ relation found in such ab-initio theoretical models for ETGs is in tension with our semi-empirical findings.
\end{abstract}




\section{Introduction}

\label{Sect:introduction}
In the local Universe it is observed that the physical properties of galaxies belonging to different morphological types (i.e. the Hubble sequence, \citet{Hubble1926}) define different scaling relations. This evidence is widely interpreted as the signature of the different physical processes that have shaped galaxies from their formation to our epoch. Of particular interest are the scaling relations that link
galaxy structure and dynamics with galaxy stellar mass, the $R_e-M_{star}$ relation (\citealt{Shen+03}, \citealt{Bernardi+14}, \citealt{Lange+15}) and the $j_{star}-M_{star}$ relation (\citealt{Romanowsky&Fall2012}, \citealt{Obreschkow&Glazebrook2014}),  where $R_e$ is the radius that encloses half of the light of the galaxy and $j_{star}$ is the galaxy stellar angular momentum.
These relations are believed to bear a significant trace of how galaxies stellar mass is assembled through cosmic time (e.g., \citealt{Cappellari2016_review}, \citealt{Somerville&Dave2015}). \\
The standard scenario of the formation of galactic disks (\citet{MoMaoWhite98}, MMW hereafter) envisages that LTGs form out of gas that cools and falls towards the centre of a host dark matter halo. If the gas shares the specific angular momentum of dark matter and retains a fraction $f_{j}$ of it during the collapse, the model predicts that the scale lenght $R_d$ of the newly formed disk should be
\begin{equation}
\label{eq:MMW_tot}
R_d \simeq \frac{\lambda}{\sqrt{2}} f_c f_R f_{j} R_h.    \end{equation}
Here $f_c$ and $f_R$ are factors of order unity that provide minor corrections that account for adiabatic contraction (\citealt{Blumenthal+86}, \citealt{Shankar+17_bulgehalo}) and disk self-gravity, and $R_h$ is the dark matter halo radius, 
\begin{equation}
R_{h}=\Bigl( \frac{3M_{h}}{4\pi \cdot \Delta\rho_{\Delta}} \Bigr)^{\frac{1}{3}}
\label{eq:Rhalo}
\end{equation}
 where $\Delta$ is the virial overdensity with respect to the cosmological background density \citep{Bryan&Norman98}. The parameter $\lambda$ in eq. \ref{eq:MMW_tot} is the spin parameter of dark matter defined by \citet{Peebles1969} as
\begin{equation}
\label{eq:spinparamP}
\lambda_P=\frac{J \mid{E} \mid ^{\frac{1}{2}}}{G M_h^{\frac{5}{2}}},
\end{equation}
($J$,$E$ and $M_h$ are respectively the angular momentum, energy and mass of the halo)
or by \citet{Bullock+01} as
\begin{equation}
\label{eq:spinparamB}
\lambda=\frac{J}{\sqrt{2}M_{h}R_{h}}.
\end{equation}
The analysis of the Bolshoi-Planck and Multidark-Planck N-body numerical simulations (\citealt{Prada+12}, \citealt{Kyplin+11_bolshoi}, \citealt{Kyplin+16_multidark}) carried out in \citet{Rodriguez-Puebla+16} have revealed that when adopting the latter definition the distribution of the spin parameter is well fit by a log-normal distribution, while the spin parameter in eq. \ref{eq:spinparamP} follows a skewed distribution, closer to a ``Schechter-like" distribution (see \citealt{Rodriguez-Puebla+16} for details). The dispersion in both cases has been found to be $\sigma_{log\lambda}\sim 0.25$. It is crucial to note that in eq. \ref{eq:MMW_tot}
$\sigma_{log\lambda}$ is the dominant source of scatter in disc scale length at fixed halo radius. In the following, for completeness, we will adopt both of the above definitions of the spin parameter in our semi-empirical models.\\

 The success of the angular momentum conservation model put forward by MMW in predicting the structural properties of disc galaxies  has been suggested in several works (\citealt{Somerville+08},  \citealt{Lapi+18_disks}, \citealt{Kravtsov2013}, \citealt{Huang+17}, \citealt{Somerville+18}, \citealt{Straatman+17_tullyFisher}). For instance, \citealt{Kravtsov2013} (K13 hereafter) found that a linear relationship between the sizes of galaxies and their haloes, as predicted by eq. \ref{eq:MMW_tot}, indeed seems to exist. 
 \citet{Somerville+18} used  CANDELS (\citealt{Koekemoer+11}, \citealt{Gnedin+11}) and GAMA (\citealt{Driver+11_GAMA}, \citealt{Liske+15_GAMA}) observations to extend the results of K13 to $z \sim 3$, and claimed that the  MMW model is able to explain the size distributions of all galaxies, irrespective of morphology.   \citet{Lapi+18_disks} found that the normalization of the $R_e-R_{halo}$ relation of local disks is in good agreement with the predictions of the MMW model, although it is significantly offset with respect to that of the ETGs-dominated sample of K13 (see also \citet{Huang+17}). This is indeed reminiscent of the separation in angular momentum at fixed stellar mass reported by \citet{Romanowsky&Fall2012}. 
 On the other hand, some studies question the validity of the MMW model based on the fact that the scatter that it would predict (i.e. at least 0.25 dex) overestimates the one found in observations (\citealt{Lapi+18_disks}, \citealt{Desmond+15}).
 For example, \citet{Jiang+17} have used two suites of hydrodynamical cosmological zoom-in simulations (VELA, \citealt{Ceverino+14}, \citealt{Zolotov+15}  and NIHAO, \citealt{Wang+15_NIHAO} ) to study the connection between galaxy and halo size and found a weak link between galaxy and halo angular momenta, which would undermine the MMW model. 
 Fits to their simulations suggest instead a correlation between $R_e$ and the halo concentration $c$  \citep{NavarroFrenkWhite95} with $R_e \propto c^{-0.7}R_h$ and no correlation with the halo spin parameter. Nevertheless, the physical motivation behind this empirical finding is yet to be found. 
 
 \vspace{1 em}
 
  In this work we further investigate the role of dark matter in setting galaxy sizes and angular momenta from a semi-empirical point of view. In particular, we aim to directly test the dispersion predicted by the MMW model against a large photo-morphological catalogue \citep{Dominguez-Sanchez+18}. Our model builds on the assumption that a $R_e-R_h$ connection exists, to which we add an intrinsic scatter tuned to match observations of the size distributions. We then use the models proposed by \citet{MoMaoWhite98} and \citet{Jiang+17} to give empirical constraints on the galaxy-halo connection. We will show that in general observations require very small intrinsic scatters, a challenge for some of the assumed models. We will then discuss and interpret our results in the broader context of disc formation and angular momentum conservation  The formation of ETGs in cosmological models will also be briefly discussed in relation to our semi-empirical constraints.\\
  
  The paper is structured as follows. We first introduce the data set that we use to constrain the $R_e-R_h$ relation in Section \ref{Sect:data}. The details of the implementation of the models are given in Section \ref{Sect:model}. In Sections \ref{Sect:results} and \ref{Sect:discussion:size_distrib} respectively the results are presented and discussed. We discuss our results in the light of large scale cosmological simulations in Section \ref{Sect:discussion_stateoftheart}, and compare to other studies in Section \ref{sec:comparisons}.  We will give our final remarks in Section \ref{Sect:conclusions}.  Caveats and additional discussion on the models are given in the Appendices. \\
  In this work we adopt a standard flat $\Lambda CDM$ cosmology with $\Omega_m=0.3$, $\Omega_{\Lambda}=0.7$ and $h=0.7$ and $\sigma_8=0.82$. We note that our results are largely independent on the exact choice of cosmology within the current constraints (\citealt{Komatsu+11_cosmoparWmap}, \citealt{Planck_cosmopar}).

\section{Data}
\label{Sect:data}
\begin{figure}
\centering
\includegraphics[width=0.5\textwidth]{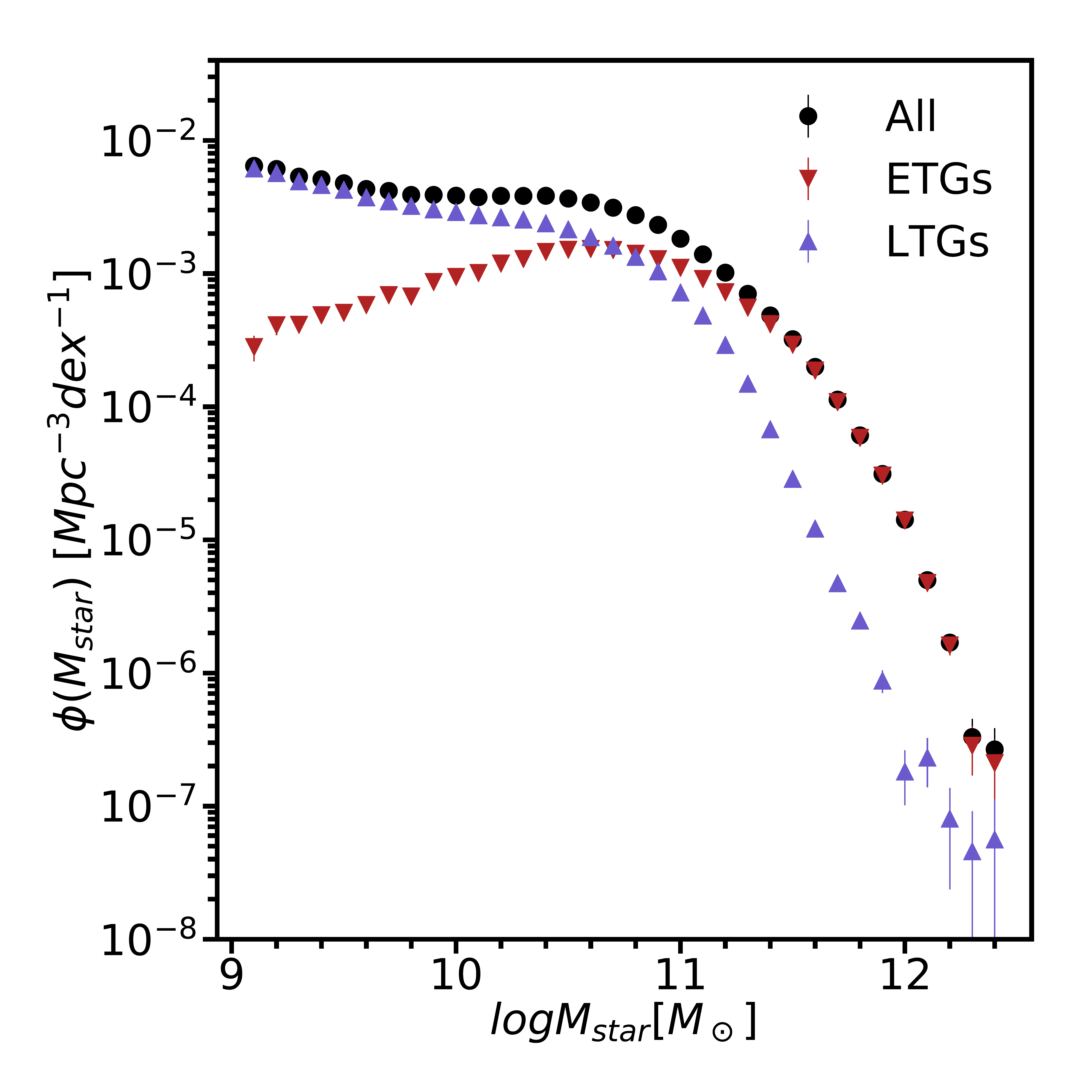}
\caption{The $V_{max}$ weighted morphological SMF from the M15/16 SDSS catalogues combined with the DS18 morphological catalogues.}
\label{fig:SMF_SDSS}
\end{figure}

In the following we will use the SDSS DR7 \citep{Abazajian+09}  spectroscopic sample as presented in \citet{Meert+15}, \citet{Meert+16} (hereafter M15/16). The Meert et al. catalogues consists of 670722 objects the photometry of which benefits of substantial improvement both in  background subtraction and  fits to the light profiles. In the M15/M16 catalogues galaxies are fit with a \texttt{S\'ersic+Exponential} as well as a \texttt{S\'ersic} profile.
In our work we only adopt the $r$-band best fit. The galaxy stellar masses  are computed adopting such light profiles and the mass-to-light ratio $M_{star}$/L by \citet{Mendel+14}, and the effective radius $R_e$ is the  truncated semi-major axis  half-light radius of the full fit (e.g., \citealt{Fischer+17_truncation}). \\
Recently, \citet{Dominguez-Sanchez+18} (hereafter DS18) have classified the morphology\footnote{more specifically, their TTypes \citep{Nair+10}} of the objects in the Meert et al. catalogues by means of Convolutional Neural Networks. 
Previous works (\citealt{Desmond+15}, \citealt{Lapi+18_disks}) rely on catalogues that are orders of magnitude smaller than that used in this work and therefore quantities such as the morphological stellar mass function $\phi(M_{star}$) (SMF) or  the morphological size function  $\phi(R_e)$ are not available. One of the main aims of our work is exploring whether current models of disk formation are able to explain the size distribution of disk galaxies and the catalogues of morphology by Dominguez-Sanchez et al. offers a unique testbed of such models. \\
In this work we define LTGs and ETGs as having $TType>0$ and $TType \leq 0$ respectively. Note that we include S0 galaxies as part of the ETGs populationWe further exclude from our selection Elliptical galaxies for which the \texttt{S\'ersic+Exponential} fits provide a bulge-to-total ratio lower than 0.5. Indeed, visual inspection of a sample of these objects reveal crowded fields, close companions or classification errors.  \\
We match the Meert et al. catalogues with the \citet{Yang+07} group catalogues. For each group we identify the central galaxy as the most luminous, while the remaining objects in that group are considered to be satellites. From the matched catalogues we compute the $V_{max}$-weighted stellar mass functions (SMF) of \emph{central} galaxies for the full catalogues and for both ETGs and LTGs. Error bars are computed via jackknife resampling\footnote{We adopt the publicly available library \textsc{astropy} \url{http://www.astropy.org/}}. The inferred SMFs are reported in Figure \ref{fig:SMF_SDSS} -  they  agree with the results of \citet{Bernardi+17_highmassend} and they compare well to the morphological SMF of \citet{Bernardi+13}. We then infer the fraction of late type galaxies in each stellar mass bin, $f_{L} (M_{star}) = \phi(M_{star})^{LTGs}/ \phi(M_{star})^{tot}$. \\

\begin{figure}
    \centering
    \includegraphics[width=0.5 \textwidth]{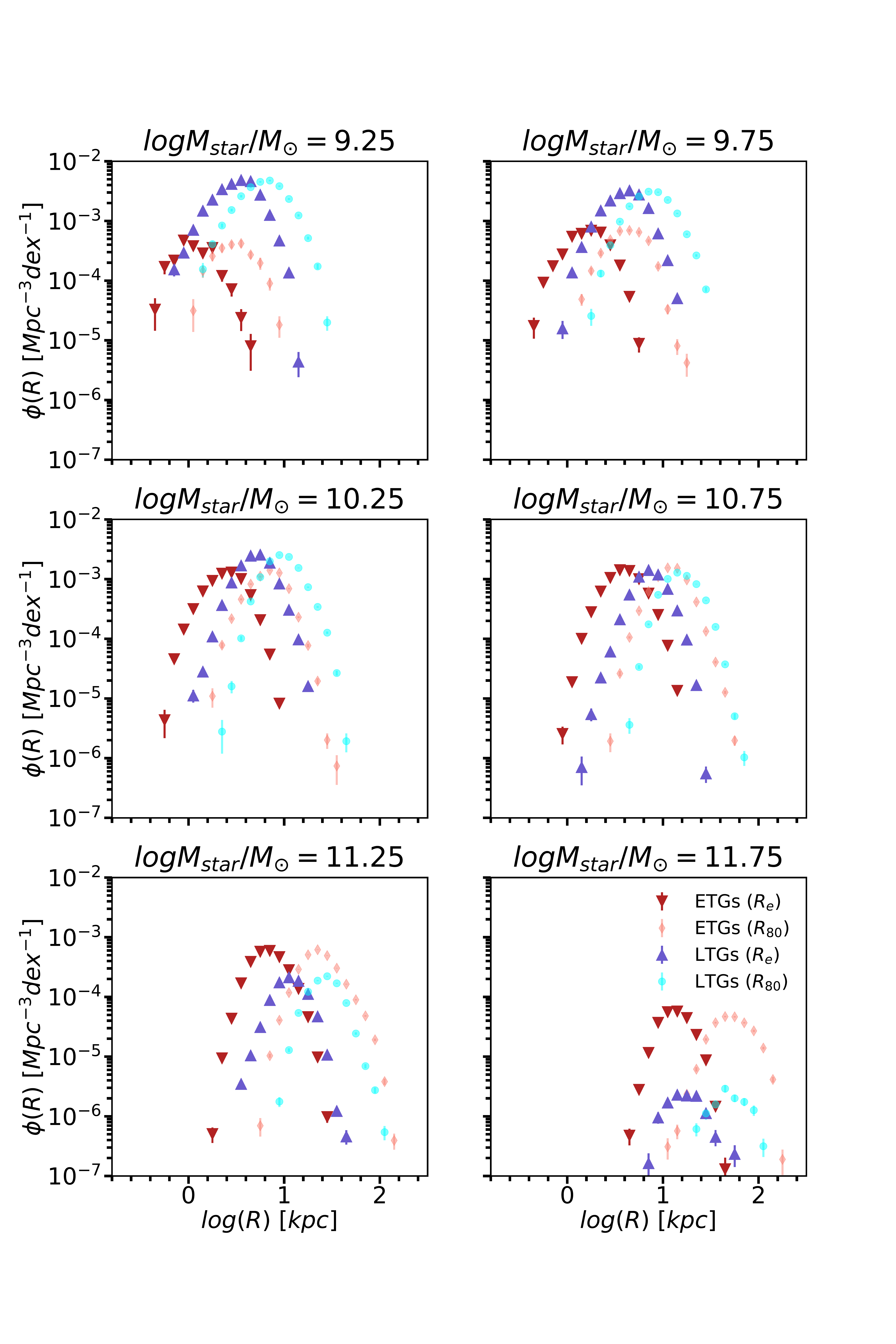}
    \caption{Size functions of ETGs and LTGs from the M15/16 SDSS catalogues combined with the DS18 morphological catalogues. Red downward triangles and blue upward trinagles are for the $R_e$ of ETGs and LTGs respectively, while light pink diamonds and light cyan circles show the results for $R_{80}$ for ETGs and LTGs. }
    \label{fig:sizefunctions_SDSS}
\end{figure}

As for the sizes, we compute the $V_{max}$-weighted size functions $\phi(R_e)$ similarly to the SMF.  Figure \ref{fig:sizefunctions_SDSS} shows that $\phi(R_e)$ is only weakly bimodal (red downward triangles and blue upward triangles). At low and high masses the distributions are dominated by LTGs and ETGs respectively, while the bimodality is most pronounced for $10^{10} \leq M_{star}/M_\odot \leq10^{11}$. However, most strikingly, we see that the width of the size functions of ETGs and LTGs are comparable at all masses. It is also worth noticing that the total size function has a larger scatter than those of LTGs and ETGs taken singularly, at least for $M_{star} \leq 10^{11} M_\odot$.

 We also notice that the mass dependence of the peak of the size function of ETGs is quite strong. 
Moreover, it is interesting to see that the size functions are somewhat skewed. While this feature was reported for LTGs also by \citet{vanderwel+2014} for galaxies in CANDELS  \citep{Koekemoer+11, Gnedin+11}, it is the first time that this is reported for ETGs. We have checked that using circularized sizes\footnote{Defined as $R_{e,circ}=R_{e,maj}\sqrt{b/a}$ where $b$ and $a$ are the semiminor and semimajor axis respectively.} leads to a reduced skeweness in the size functions of ETGs (not so for LTGs). However we choose to use semimajor-axis sizes to enable a more direct comparison with LTGs, for which circularized sizes would be difficult to interpret physically as LTGs are intrinsically two-dimensional structures.

The aim of our theoretical work is to explore the mass dependence of the scatter and normalization of the input $R_e$-$R_h$ relation to reproduce the measured SDSS size functions in different stellar mass bins..

To conclude, we also retrieve the size functions for the radius that encloses 80\% of the light, $R_{80}$, which is also shown in Figure \ref{fig:sizefunctions_SDSS}, and we comment on it in Section \ref{sec:R80}.

\section{The model}
\label{Sect:model}
 The aim of this paper is to predict the fine details of the size functions of low redshift galaxies. This  issue has been explored in fully cosmological models only in a few instances. For example, \citet{Shankar+10_sizefunct} showed that such level of accuracy was not yet achievable in semi-analytic models. To the best of our knowledge, the size functions have never been explored in hydrodynamical simulations of galaxy evolution.\\
 We here adopt the transparent and flexible approach offered by semi-empirical models. The latter are based on input observables, thus removing several degrees of freedom that characterize more standard approaches. Only a few parameters are required to match observables which are independent of the input.   From the values of the model parameters it is then possible, we will show, to infer constraints on the processes that shape galaxy formation and evolution.
  
 \vspace{1 em}
 
In brief, our semi-empirical model is structured as follows\footnote{To build our model we extensively rely on the open-source Python package \textsc{colossus} \citep{DiemerColossus}.}:
\begin{enumerate}
\item We extract catalogues of dark matter haloes from the \citet{Tinker+08} halo mass function.
\item We model the link between galaxies and dark matter via Abundance Matching (Sect. \ref{Sect:AM}), and produce large mock catalogues of galaxies with moderate-to-high stellar masses. 
\item We assign a half light radius $R_e$ to each galaxy according to diverse models of galaxy structure that exploit the galaxy-halo connection (Sect.  \ref{Sect:model_sizes}).
\end{enumerate}
We ultimately build a catalogue of dark matter haloes with mass $M_h$ and size $R_h$ and central galaxies with given stellar mass $M_{star}$ and effective radius $R_e$. An accurate comparison to data will be able to set valuable constraints on the assumptions and related parameters in input to each of our adopted models. 

For the remainder of the paper, we will consider dark matter haloes to follow a \citet{NavarroFrenkWhite95} density profile with scale radius $R_s$,
\begin{equation}
\rho(r) \propto \frac{1}{\frac{r}{R_s}\bigl[ 1+\frac{r}{R_s} \bigr]^2},
\end{equation}
and with $R_s=cR_h$ defining the \emph{concentration} parameter $c$.

\subsection{Abundance matching}
\label{Sect:AM}
The link between galaxies and their haloes is modelled following the popular \emph{Abundance Matching}
technique (AM) (e.g. \citealt{Vale&Ostriker2006}, \citealt{Shankar+06}), which is essentially based on assigning a galaxy of stellar mass $M_{star}$ to a host dark matter halo of mass $M_{h}$ via rank ordering of the cumulative relative number densities,
\begin{equation}
n(>M_{star}) = n(>M_{h})
\label{eq:AM}
\end{equation}
to infer a (mean) stellar-mass-to-halo-mass relation (SMHM, fig. \ref{fig:SMHM}). It has been shown that the simple ansatz of eq. \ref{eq:AM} is in excellent agreement with direct measurements of the SMHM with various techniques such as group finding algorithms  (e.g., \citealt{Yang+07}), satellite kinematics (\citealt{More+11}), X-ray measurements of galaxy clusters  (\citealt{Kravtsov+18}, \citealt{Erfanianfar+19}) and clustering analyses \citealt{Shankar+06}, \citealt{Shankar+14_SMHM}) as well as simulations (e.g., \citealt{Guo+11}, \citealt{Matthee+17}). 
The SMHM has been exploited by several authors for the most diverse purposes (only to mention a few, \citealt{Shi+17}, \citealt{Desmond&Wechsler2017}, \citealt{Diemer+13_pseudo}, \citealt{Shankar+17_bulgehalo}, \citealt{Posti+18_angmom}).  The SMHM is modelled as a lognormal distribution with scatter $\sigma_{SMHM}$:
\begin{align}
 P(M_{star}|M_h)& \equiv SMHM =\\ \nonumber
 &=\frac{1}{\sqrt{2\pi \sigma_{SMHM}^{2}}} exp \Bigl [ -  \frac{(logM_{star}-\langle (logM_{star} \rangle)^2} {2\sigma_{SMHM}^2} \Bigr].
\label{eq:SMHM}
\end{align}
The overall scatter $\sigma_{SMHM}$ stems from a convolution of observational errors $\sigma_{*}$ and the intrinsic scatter $\sigma_{int}$ that may be related to the stochasticity of the formation histories of galaxies within dark matter haloes (\citealt{Tinker+17_SMHM_massivegal}). We use the parametrization of the SMHM from \citet{Behroozi+13}, which reads:
\begin{equation}
\langle\log M_{star}\rangle = \log (\epsilon M_{10}) + g(x) -g(0),
\end{equation}
where
\begin{equation}
g(x)= \delta \frac{\log(1+e^{x})^{\gamma}}{1+e^{10^{-x}}} 
- \log(10^{\alpha x} +1)
\end{equation}
and $x=\log (M_{h}/M_{10})$. 
 We also assume that $\sigma_{SMHM}=0.16$ dex\footnote{We note that our choice for a relatively small scatter in the SMHM relation is a conservative one. Larger values of the scatter $\sigma_{SMHM}$ would strengthen our main result for the need of a tight $R_e$-$R_h$ relation.}, as suggested by other studies at low redshift (e.g. \citealt{Tinker+17_SMHM_massivegal}), with no dependence on halo mass, which is a very good approximation especially at the high mass end of the SMF (\citealt{Shankar+14_SMHM}, \citealt{Rodriguez-Puebla+15}).
The parameters of the SMHM are $\vec{p} = (\epsilon,M_{10},\delta, \alpha, \gamma)$.\\

According to AM, the SMF is retrieved from the SMHM by computing the integral
\begin{equation}
\phi(M_{star}) = \int SMHM(M_{h};\vec{p}) \phi(M_{h})dM_{h}
\end{equation}
where $\phi(M_{h})$ is the halo mass function. Hence, abundance matching reproduces the observed galaxy SMF by design, and it can therefore be used to produce realistic mock catalogues. \\
We run a Markov Chain Monte Carlo\footnote{We use the publicly available Python package \textsc{emcee}, \citep{Foreman-Mackey+13}} to fit the parameters of the SMHM to the SMF of central galaxies in SDSS adopting the \citet{Tinker+08} halo mass function for central haloes only by maximizing the likelihood $\mathcal{L} \propto exp(-\chi^2)$. 
The parameters of our $z \sim0.1$ SMHM are the following:
\begin{align}
\centering
M_{10} &= 11.632^{+0.008}_{-0.009} \\
\epsilon_0 &= -1.785^{+0.010}_{-0.008} \\
\alpha_0 &= -2.352^{+0.026}_{-0.021} \\
\delta_0 &= 3.797^{+0.052}_{-0.052}\\
\gamma_0 &= 0.600^{+0.10}_{-0.013}\\
\sigma_{SMHM} &=0.16 \quad (\text{fixed}).
\end{align}
It is perhaps not surprising that the uncertainty on the inferred parameters is so low compared to other works, given the very small error bars on the SMF. Moreover, here the fit is performed at one redshift only, as opposed to, e.g., \citet{Rodriguez-Puebla+17}. 

While many studies include satellite galaxies in their models (\citealt{Behroozi+13}, \citealt{Behroozi+18}; \citealt{Rodriguez-Puebla+15},\citealt{Rodriguez-Puebla+17}, \citealt{Grylls+19}), we choose to restrict our analysis to central galaxies only. \citet{Hearin+17} have shown that the sizes of satellite galaxies are linked to their halo mass at infall time, which is not straightforwardly available in analytic halo catalogues. 

We note that in principle LTGs and ETGs may occupy different loci in the $M_{star}-M_{halo}$ plane, as suggested by some studies (\citealt{Rodriguez-Puebla+15},\citealt{Dutton+10}, \citealt{Moster+18}, \citealt{More+11}). However, as pointed out in \citet{Wechsler&Tinker2018}, there is no agreement between different studies, which sometimes even reach opposite results (\citealt{Berhoozi+18}, \citealt{Moster+18}, \citealt{Rodriguez-Puebla+15}). In what follows, we will therefore adopt the same SMHM for both LTGs and ETGs. We will show in Appendix  \ref{sect:Caveats_SMHM} that our core results are largely independent of the choice of SMHM. 
\subsection{Galaxy sizes}
\label{Sect:model_sizes}
 We now proceed to assign to our mock galaxies a size. This is done according to theoretically or empirically justified models that  link galaxy sizes to the size of their host dark matter halo, $R_h$. 
 
 \medskip 
 
We adopt three models of galaxy sizes:
\begin{enumerate}
\item The \emph{MMW model} (or $\lambda$ model). This model is inspired by the classical picture in which galaxies are born as disks out of cooling from the hot gas in the halo (MMW, see Sect. \ref{Sect:introduction}). We recast equation \ref{eq:MMW_tot} as
\begin{equation}
\label{eq:lambdamodel}
R_e =  \frac{1.68}{\sqrt{2}} f_c f_j f_R \lambda R_h = 1.68 A_\lambda \lambda R_h, 
\end{equation}
where we define $A_\lambda=f_c f_j f_R /\sqrt{2}$ and the factor $1.68$ comes from $R_e \approx 1.68  R_d$, appropriate for an exponential profile. 
 We will take $A_\lambda=1$ and rescale our results as needed to match the data. We discuss the implications of this assumption in Section \ref{Sect:discussion:size_distrib}. The spin parameter
$\lambda$ is defined either by \citet{Peebles1969} (eq. \ref{eq:spinparamP}) or \citet{Bullock+01} (eq. \ref{eq:spinparamB}), see Sect. \ref{Sect:introduction}.
 Note that the MMW model was devised to explain the formation of the \emph{baryonic} size of galactic disks, while we will compare it to the \emph{stellar} sizes. In particular, the factor $f_j$ addresses the angular momentum retention of \emph{baryons} and not stars. We will discuss the implications of this difference in Section \ref{Sect:discussion_MMW}.
\item The \emph{K13 model}. This model is based on the empirical findings by \citet{Kravtsov2013}. The author adopted abundance matching techniques similar to the ones presented here\footnote{We note that \citet{Kravtsov2013} \emph{backwards} models, that is in his work dark matter haloes are assigned to galaxies via the inverse of the SMHM relation, without accounting for its scatter. \citet{Somerville+18} have shown that doing so would result in severe biases in the estimate of the host halo masses. These authors stress that the \emph{forward} modelling approach that we adopt here is instead more accurate (see also discussion in \citealt{Shankar+17_bulgehalo} and \citealt{Rodriguez-Puebla+17})} and found evidence that:
\begin{equation}
R_e = A_k R_h.
\label{eq:K13model}
\end{equation}
Here $A_k$ is the normalization which may vary with halo mass or galaxy stellar mass. We add to eq. \ref{eq:K13model} an intrinsic log-normal scatter $\sigma_K$, which, as $A_k$, is a free parameter that can be tuned to match observations.  The K13 model is hence purely empirical and will be applied to both LTGs and ETGs. Note that the physical meaning of both $A_k$ and $\sigma_K$ is not known a priori. However the K13 model reduces precisely to the MMW model when applied to LTGs in  the case $\sigma_K=\sigma_{log \lambda}\approx 0.25$ dex. The K13 model, in this respect, is more flexible. In fact, being empirically-based, it can allow for any input scatter.  As suggested by \citet{Kravtsov2013}, constraining the value of $\sigma_K$ can be crucial to probe models of galaxy formation. 
\item The \emph{concentration model}. Recently, based both on observational and numerical studies, some groups have suggested that galaxy sizes should scale in a way that is inversionally proportional to halo concentration  (\citealt{Desmond+18_SPARC}, \citealt{Jiang+17}). Following \citet{Jiang+17}, mathematically this model can be expressed as
\begin{equation}
R_e = A_c \bigl (\frac{c}{10} \bigr) ^{\gamma} R_h, 
\label{eq:concmodel}
\end{equation}
with $\gamma<0$.
Similarly to what assumed in the other two models, we initially adopt $A_c=0.012$ \citep{Jiang+17} and then rescale our results to match data.
We also adopt the concentration-mass relation by \citet{Dutton&Maccio2014},
\begin{equation}
\label{eq:concD14}
log c = a+ b\log{M_h [M\odot]/10^{12}/h}
\end{equation}
with $a(z)=0.537 + (1.025-0.537)exp(-0.718z^{1.08})$ and $b(z) =-0.097 +0.024  z$. \citet{Dutton&Maccio2014} report a log-normal scatter of about $\sim 0.11$ dex at $z\sim 0$, which is independent on halo mass. Hydrodynamical simulations suggest that the intrinsic scatter $\sigma_{CM}$ in this model is lower than in the K13 model \citep{Jiang+17}. Indeed, we will not include any other source of scatter in the concentration model other than the scatter in concentration (i.e., $\sigma_{CM}=0$, see discussion in Sect. \ref{Sect:scatters}).\\
 We find that the concentration model may in fact be interpreted as a further generalization of the K13 model. Indeed, combining eq. \ref{eq:concD14} and eq. \ref{eq:concmodel}, and bearing in mind that $M_h \propto R_h^{1/3}$, yields
\begin{equation}
    R_e \propto R_h^{b\gamma/3+1},
    \label{eq:concmodel_expanded}
\end{equation}
which reduces to the K13 model when $\gamma=0$.
The scatter implied by this version of the concentration model and the difference with that produced by eq. \ref{eq:concmodel} is discussed at the end of the next Section.

\end{enumerate}
Although , following the seminal approach by K13, we model the link between galaxies and their haloes in terms of the projected effective radius $R_e$, such relation would be more physically motivated when expressed in terms of the 3D physical half mass radii of galaxies $R_{e,3D}$. However, the deprojection of galaxy shapes is a very hard task (Jiang et al. in prep). In any event, as discussed in Appendix \ref{app:deproj}, projection effects tend to increase the variance in the measured effective radii, implying even tighter distributions in intrinsic sizes $R_{e,3D}$. Accounting for deprojection effects would then further tighten the measured distribution of 3D galaxy sizes, which would constitute an even harder challenge for models. In Appendix \ref{app:deproj} we give an estimate of the (small) biases induced by assuming that $R_{e,2D}=R_{e,3D}$ based on mock observations of galaxies from the Illustris TNG simulation (\citealt{Nelson+18_datarelease}, \citealt{Rodriguez-Gomez+19}, \citealt{Huertas-Company+19}).

\subsection{Sources of scatter in our models}
\label{Sect:scatters}
\begin{figure}
\centering
\includegraphics[width=0.4\textwidth]{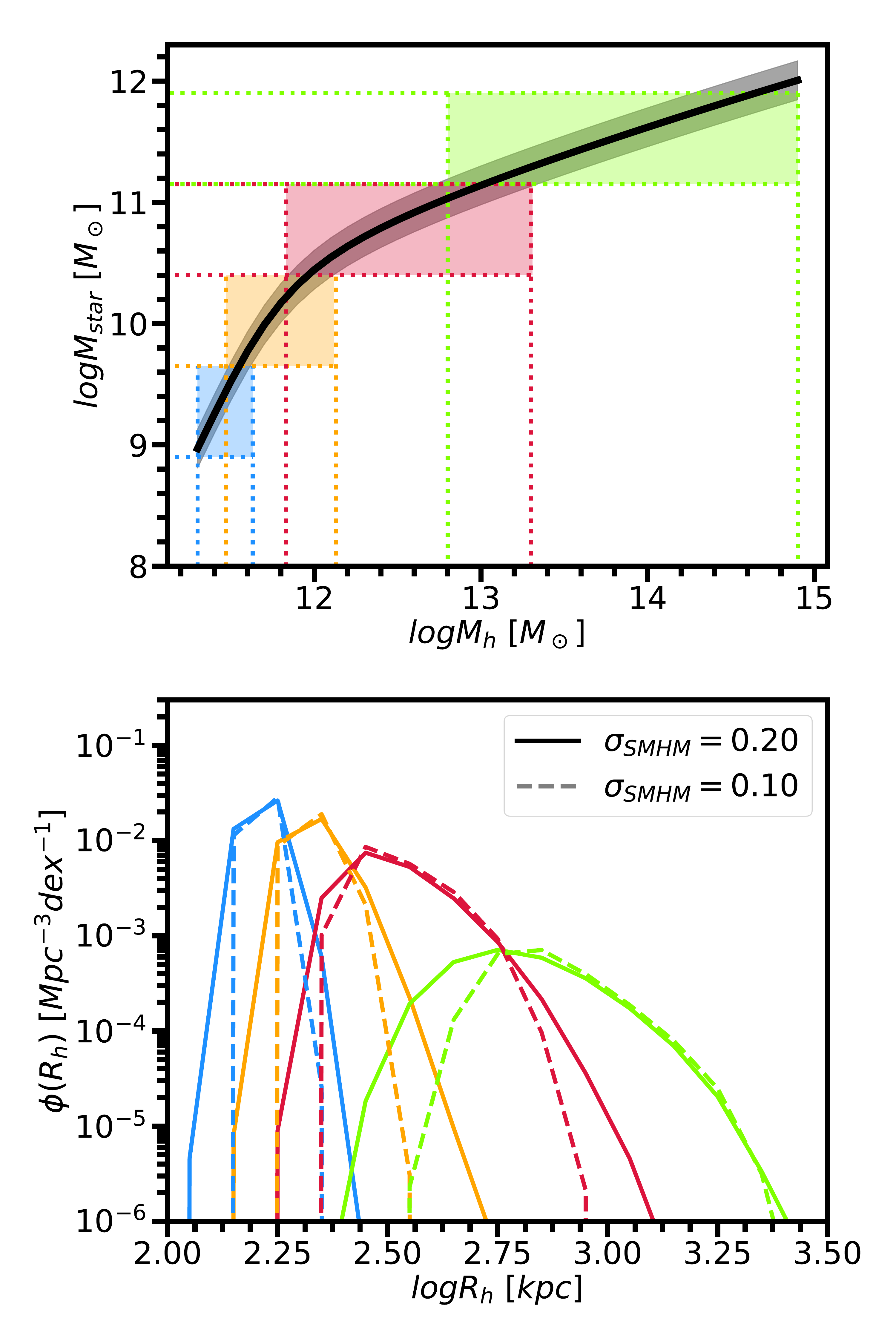}
\caption{Role of the shape of the SMHM and its $\sigma_{SMHM}$ in setting the scatter in halo size (and hence in galaxy size according to our models). \emph{Upper panel.} The black line is the SMHM retrieved from MCMC fitting of the total SMF in SDSS (Sect. \ref{Sect:model}). Different cuts in stellar mass highlight different regions of the SMHM with different colours. Each coloured band corresponds to a stellar mass cut of the same width (0.75 dex). Their projections onto the x axis select the halo mass range in which galaxies of a given stellar mass are expected to reside. 
\emph{Lower panel.} The halo size functions resulting from the stellar mass cuts applied in the upper panel, with the same color code. Dashed and solid lines indicate predictions for $\sigma_{SMHM}=0.10$ dex and $\sigma_{SMHM}=0.20$ dex. No additional scatter in size is added. Higher stellar mass cuts are naturally mapped in broader distributions. Larger values of $\sigma_{SMHM}$ correspond to broader distributions with an effect that is larger the higher the stellar mass cut.
}
\label{fig:cartoon_scatter}
\end{figure}

 At fixed bin in stellar mass, the width of the implied size distribution resulting from our three adopted models depends on a combination of different effects. In all models, there is always a contribution from the intrinsic scatter in the SMHM, as shown in figure \ref{fig:cartoon_scatter}. In fact, at fixed stellar mass there is a distribution of possible host haloes, a feature that is usually described in terms of the \emph{halo occupation distribution function}  $P(M_{h}|M_{star})$  \footnote{Which is \emph{different} from the inverse of $P(M_{star}|M_{h})$ (\citealt{Shankar+14_SMHM}, \citealt{Somerville+18})}, which translates into a distribution in halo sizes $P(R_h|M_{star})$ (see eq. \ref{eq:Rhalo}), the main ingredient in all our models. The distributions get progressively broader for higher stellar mass cuts, given the shallow slope of the SMHM at high halo masses in combination with its intrinsic scatter. As this feature is mainly driven by the double power-law shape of the SMHM, it would be present even in the case of $\sigma_{SMHM}=0$.

 The origin of the double power-law shape of the SMHM is thought to be linked to the efficiency of star formation, which is suppressed below and above a certain halo mass, where Supernova and  a combination of AGN feedback and virial shocks, respectively, are believed to be most efficient (e.g., \citealt{Shankar+06}, \citealt{Pillepich+18}, \citealt{Faqucher-giguere+11}). If zero intrinsic scatter $\sigma_K$ in the K13 model was required to match observations, it could be argued that the same physics that shapes the SMHM is responsible for the width of the observed size distributions. On the other hand, wherever $\sigma_K>0$ is needed to match the data, there must be some physical processes unrelated to the build-up of the shape of the SMHM at play in determining the broadness of the observed size functions.

 \vspace{1em}
 
\begin{figure}
\includegraphics[width=0.4 \textwidth]{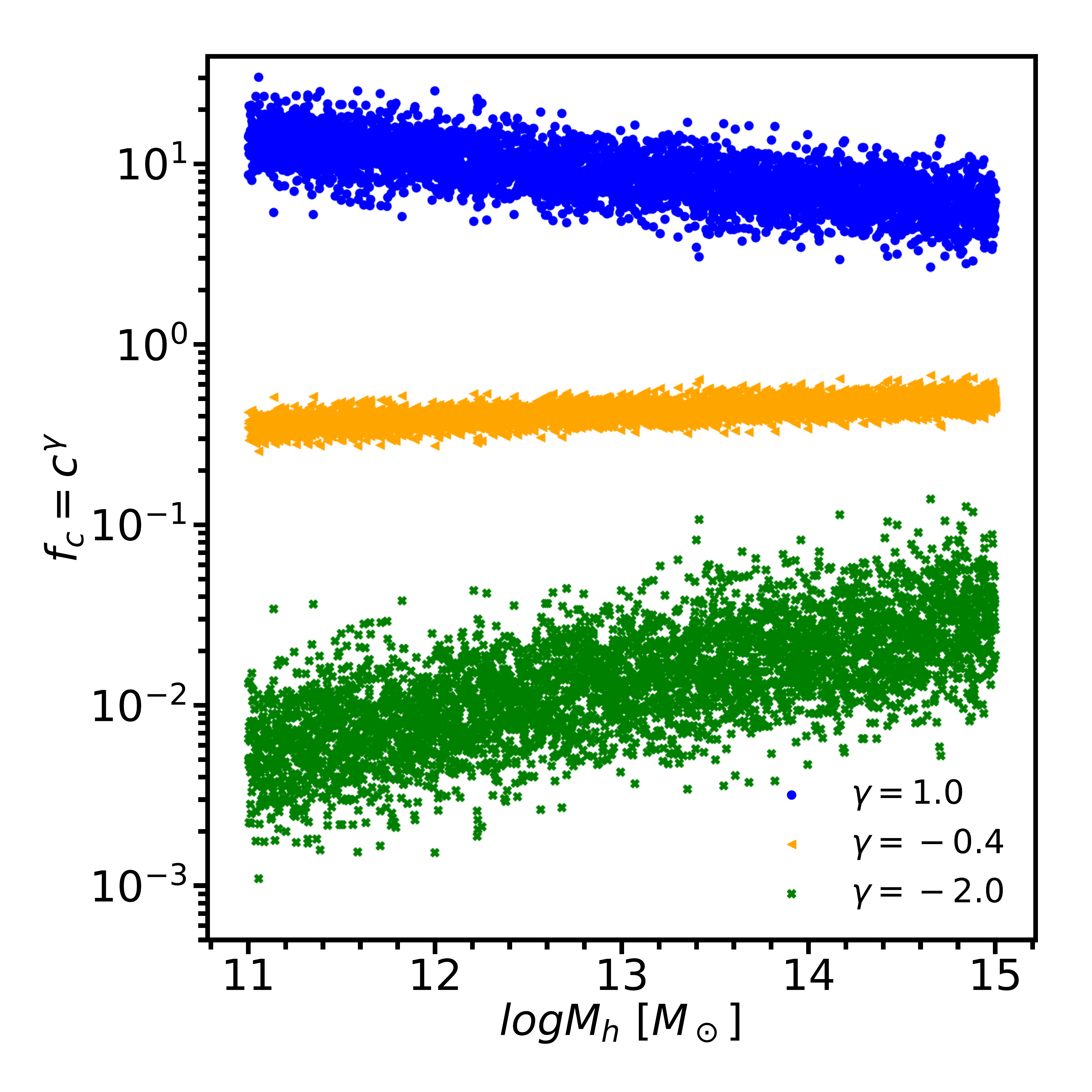}
\caption{Scatter induced by different choices of $\gamma$ in the factor $f(c)=c^\gamma$ as a function of halo mass. Blue dots, orange triangles and green crosses are for $\gamma=1,-0.4$ and $-2.0$ respectively. Concentrations are from \citet{Dutton&Maccio2014}. }
\label{fig:conc_scatters}
\end{figure}

 In the MMW and concentration models, the scatter is due to both the halo occupation distribution (described above) and the internal properties of the dark matter hosts. 
In fact, most of the scatter of the \emph{MMW model} derives from the distribution of the spin parameter $\lambda$, with a typical dispersion  of $\sigma_{log\lambda}\approx0.25$ dex and very weak dependence on halo mass.  
Interestingly, we find that for the concentration model one additional source of scatter derives from the factor $c^{\gamma}$ in eq. \ref{eq:concmodel}. As shown in figure \ref{fig:conc_scatters}, the (quite tight) distribution in concentration at fixed halo mass (blue dots) is modified for different values of $\gamma$. As we will see in Section \ref{Sect:results:K13&conc}, adopting  lower values of $\gamma$ will result in broader distributions. Such effect is degenerate with the intrinsic scatter in the concentration model $\sigma_{CM}$. We set $\sigma_{CM}$=0 in this work, noting that having $\sigma_{CM}>0$ would require higher values of $\gamma$ to match the observed size functions.
Therefore our constraints are \emph{lower limits} to $\gamma$.
 
 \vspace{1em}
 
As a final note, we recall that the concentration model may be seen as a further generalization of the K13 model (see Section \ref{Sect:model_sizes}). It must however be noted that at fixed $R_h$ the concentrations follow a lognormal distribution, while the expression in eq. \ref{eq:concmodel_expanded} has been derived assuming only the \emph{mean} of eq. \ref{eq:concmodel} and \ref{eq:concD14}. Hence, the arguments about the scatter in the concentration model presented in this Section would not apply straightforwardly to eq. \ref{eq:concmodel_expanded}. However, studying this issue in more detail is outside of the scope of our paper. 

\section{Results}

 \label{Sect:results}
 We now proceed to a careful comparison of our three models to the size functions extracted from the SDSS  photo-morphological catalogues (see Sect. \ref{Sect:data}). \\
In our models we do not differentiate between ETGs and LTGs, and so the size function from the model should be compared to the \emph{total} observed distribution. In each bin of stellar mass we retrieve the size function from our model and rescale it to match the observed ones for different morphological types,
\begin{align}
\phi(R_e|M_{star})_{obs}^{LTGs} &= f_L(M_{star})\phi(R_e|M_{star})_{model}^{tot}\\
\phi(R_e|M_{star})_{obs}^{ETGs} &= (1-f_L(M_{star}))\phi(R_e|M_{star})_{model}^{tot}
\end{align}
where $f_L(M_{star})$ is the fraction of late type galaxies as defined is Section \ref{Sect:data}. Note that here we are implicitly assuming that ETGs and LTGs at fixed stellar mass live on average in the same dark matter haloes. 

\vspace{1 em}

\begin{table}

\begin{tabular}{c|c|c|c|c|c|c}
   $M_{star}$      & 9.25  & 9.75  & 10.25 & 10.75 & 11.25 & 11.75 \\
\hline
$A_k$        & 0.018 & 0.019 & 0.019 & 0.019 & 0.024 & 0.024 \\
\hline
$A_c$        & 0.034 & 0.030 & 0.027 & 0.026 & 0.021 & 0.015  \\
\hline
$A_\lambda$  & 0.60 & 0.60 & 0.60 & 0.60 & 0.60 & 0.47
\end{tabular}
\caption{Values of $A_k$, $A_c$ and $A_\lambda$ in different bins of $M_{star}$, for LTGs. Compare to table \ref{table:ETGs}.}
\label{table:LTGs}
\end{table}

Figures \ref{fig:disksLambda}, \ref{fig:disksK13} and \ref{fig:disksConc} show a comparison between  the observed size functions $\phi(R_e)$ of  LTGs and our models (the MMW, K13 and concentration models respectively). We bin both our model galaxies and data in bins of 0.5 dex in stellar mass.
In all the figures the model size functions are shifted to match the location of the peaks of the observed distributions. The normalizations of the different models ($A_\lambda, A_K, A_c$) in each stellar mass bin are reported in Table \ref{table:LTGs}. Results for ETGs are given in Appendix \ref{app:ETGs}.\\

\subsection{The MMW model}
\label{Sect:MMW_results}
In figure \ref{fig:disksLambda}  it can be seen that the classical $\lambda$-disk model by MMW does not provide a good fit to data, irrespective of the definition of spin parameter adopted (log-normal or Schechter-like, see Section \ref{Sect:introduction}). This effect becomes gradually more severe as more massive populations of LTGs are considered. As for the normalization $A_\lambda$  we note that the values listed in Table \ref{table:LTGs}, recalling eq. \ref{eq:lambdamodel} and that $R_e \approx 1.68 R_d$ for LTGs, are consistent with  $R_d\approx 0.3\lambda R_h$, in agreement with the study by \citet{Lapi+18_disks}. Notably, given that $A_\lambda=f_j f_R f_c /\sqrt{2}$, this is fully consistent with the MMW model with an angular momentum retaining factor $f_j$ of about $0.5-0.7$.

\subsubsection{The case of bulgeless galaxies}
\label{Sect:MMW_bulgeless}
To select LTGs from the catalogues by DS18 we applied the cut $TType>0$. We note that this cut might still include galaxies with prominent bulges, which may have a non negligible contribution in determining the half light radius of the whole galaxy, especially at high masses \citep{Kormendy2016}. On the contrary, the MMW model is expected to work for pure disk galaxies only and therefore comparing the MMW model with LTGs selected as above may not be entirely accurate. \\

\begin{figure*}
    \centering
    \subfloat[\label{fig:disksLambda}]{{\includegraphics[width=0.45\textwidth]{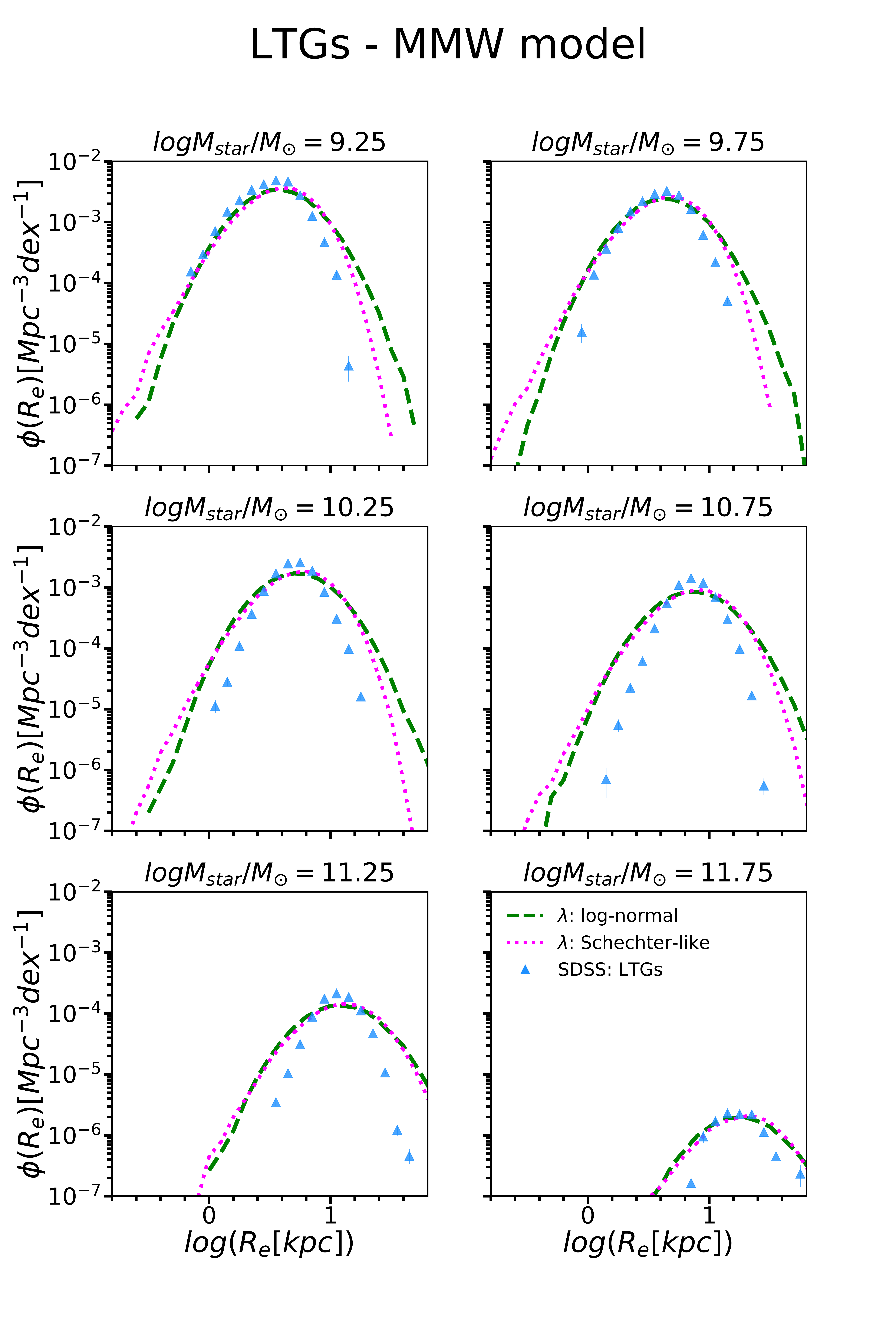} }}%
    \subfloat[\label{fig:TTfunct}]{{\includegraphics[width=0.45\textwidth]{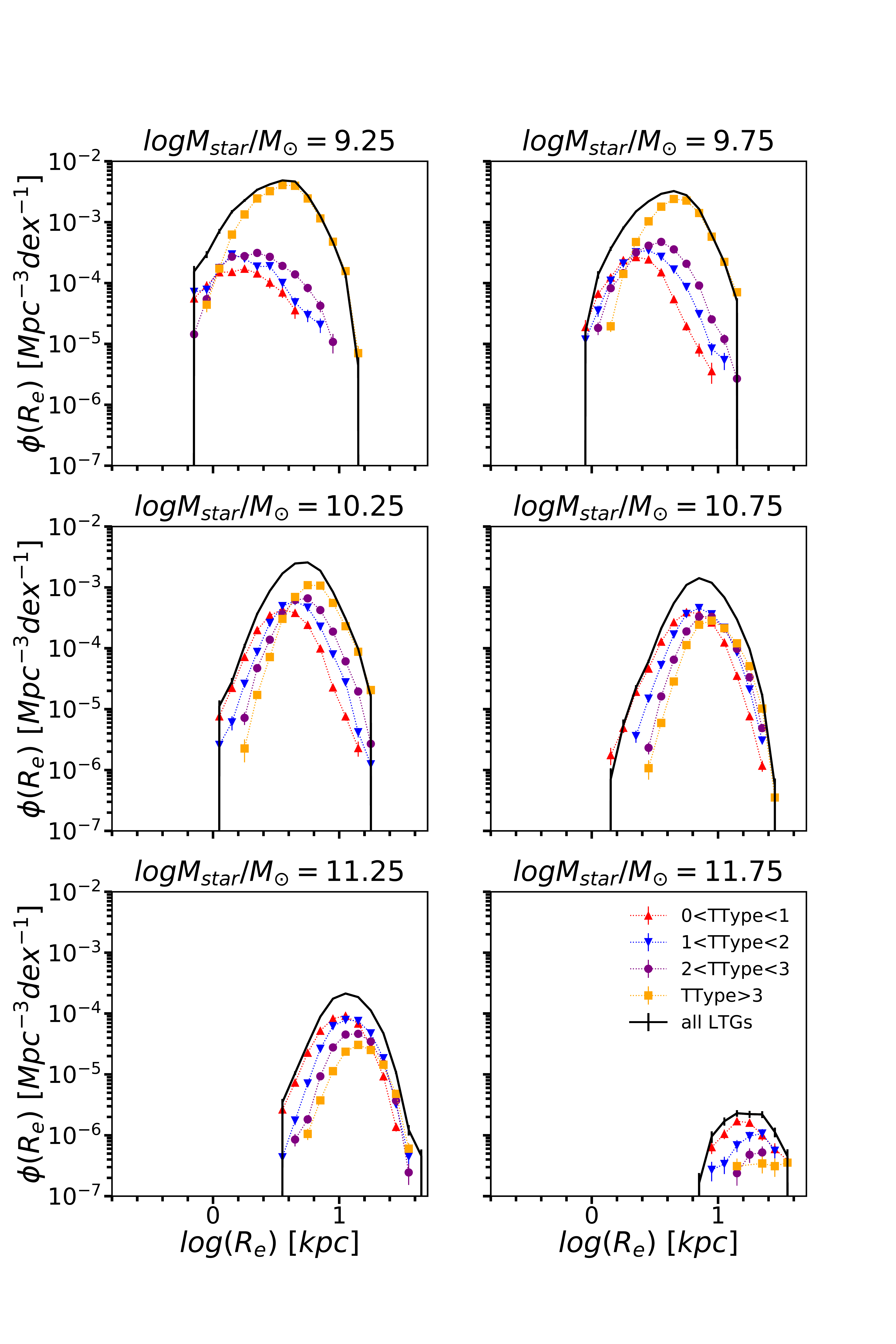} }}
    \caption{\emph{Left}: Size functions from the MMW model ($\lambda$ model, eq. \ref{eq:lambdamodel}). The spin parameter $\lambda$ is retrieved either from the log-normal (pink dotted lines) or Schechter-like (purple dashed lines) fits from  \citet{Rodriguez-Puebla+16}. Data points are LTGs from the photo+morphological SDSS catalogues described in Section \ref{Sect:data}.  \emph{Right}: Size functions for LTGs divided in bins of TType. The total distribution is shown with solid black lines, the distributions for $0<TType<1$, $1<TType<2$, $2<TType<3$ and $TType>3$  are instead shown with red upward triangles, blue downward triangles, purple circles and yellow squares plus dotted lines respectively.}
    \label{fig:pixelcnn_figure}
\end{figure*}

In figure  \ref{fig:TTfunct} we show the size functions of LTGs in our SDSS photo+morphological catalogues divided by TType. 
 We find that for $M_{star}\lesssim10^{10.5} M_\odot$ the population is entirely dominated by galaxies with $TType>3$, which represent the disk dominated Sb-Sc-Sd galaxies according to the \citet{Nair+10} classification against which the CNN in DS18 was trained. At higher masses earlier types start dominating, with the peaks of their size functions being located at lower $R_e$ due to the progressively important contribution of the bulge. Interestingly, LTGs with $TType>3$ display an even tighter size distribution than that of the overall population, which dismisses our concerns\footnote{Note that the skeweness of the size function is partially explained by the morphological mix of LTGs, but that for the later types the skeweness still persists.}.  It might however be argued that the comparison between model and data may not be ideally set up since not all Sb-Sc-Sd can be fitted by a pure exponential disk. To check for the latter effect, we further restricted our analysis to LTGs with TType>3 and $B/T<0.2$ and still did not find significant changes in the width of the size distributions.
 
\subsection{The K13 and concentration models}
\label{Sect:results:K13&conc}
\begin{figure*}
    \centering
    \subfloat[\label{fig:disksK13}]{{\includegraphics[width=0.45\textwidth]{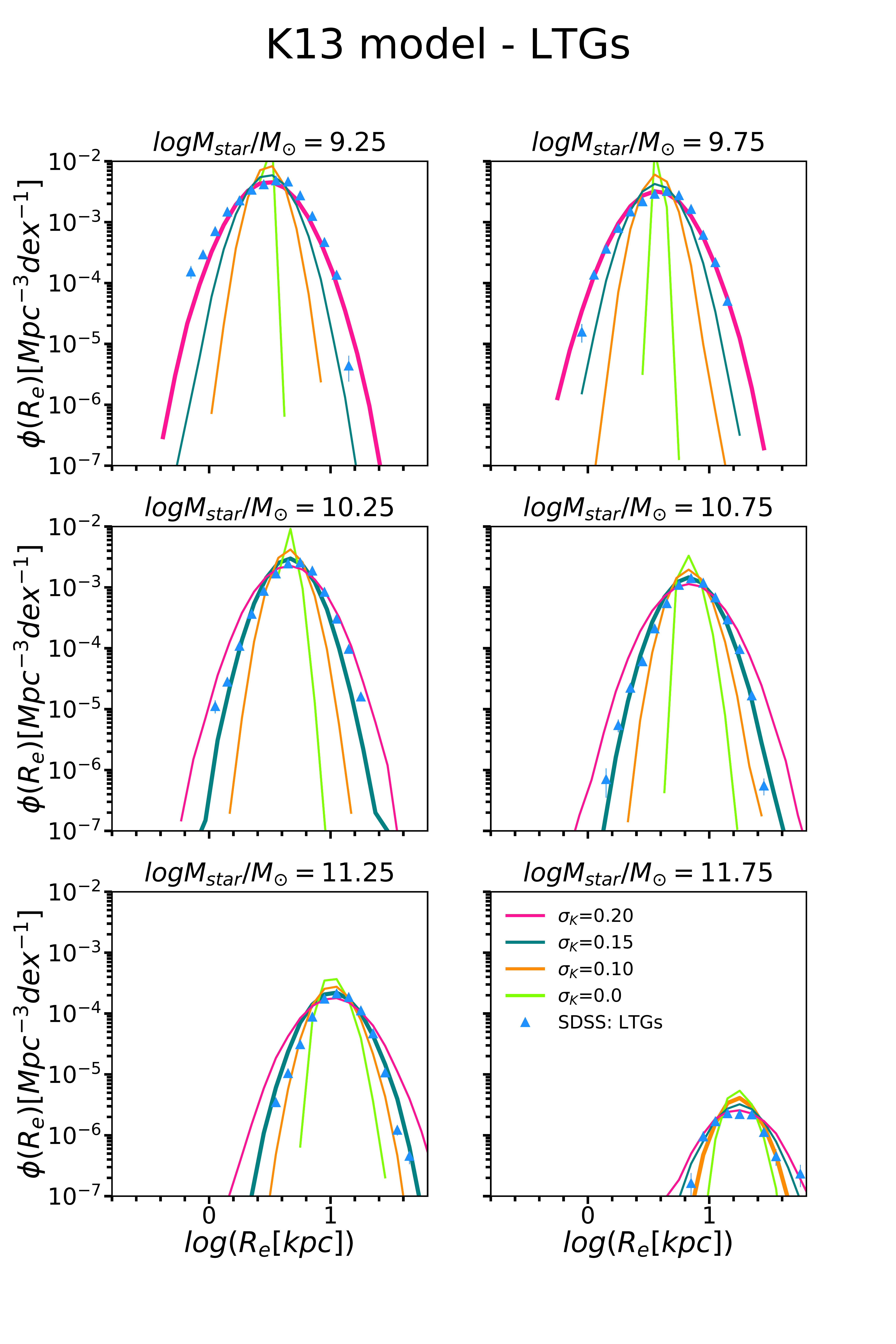} }}%
    \subfloat[\label{fig:disksConc}]{{\includegraphics[width=0.45\textwidth]{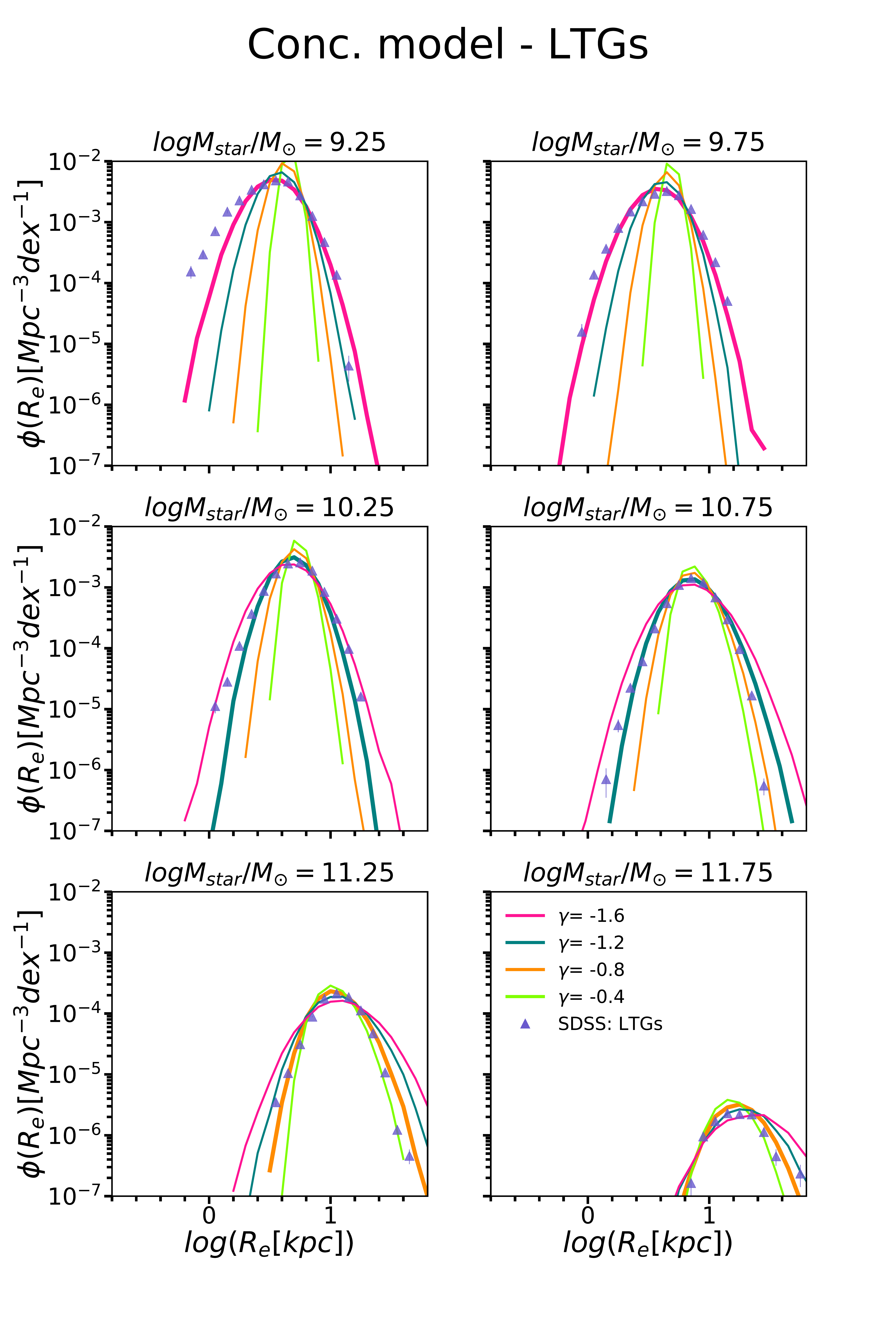} }}
    \caption{\emph{Left}:  Size functions from the K13 model (eq. \ref{eq:K13model}) for values of $\sigma_K=0.00,0.10,0.15,0.20$.  \emph{Right}: Size functions from the concentration model (eq. \ref{eq:concmodel}) for values of $\gamma=-1.6,-1.2,-0.8,-0.4$. Models that work best for a given stellar mass bin are highlighted in each panel by a thicker line.
 Data points are LTGs from the photo+morphological SDSS catalogues described in Section \ref{Sect:data}.}
    \label{fig:LTGs_K13conc}
\end{figure*}

The size distributions from the K13 and concentration models are reported in figures \ref{fig:disksK13} and \ref{fig:disksConc}. The free parameters in these models are ($A_k$, $\sigma_K$) and  ($A_c$, $\gamma$)  respectively. The values of $A_k$ and $A_c$ are reported in Table \ref{table:LTGs}. Although we do not aim to give exact fits for $\gamma$ and $\sigma_K$, we show how the models depend on these parameters by plotting results for models with different values of $\sigma_K$ and $\gamma$ as labelled. As it can be seen in figures \ref{fig:disksK13} and \ref{fig:disksConc}, varying $\sigma_K$ and $\gamma$ leads to quite drastic differences in the model distributions. In each panel of the figures we highlight with a thicker line the parameter that seems to best reproduce observations. For the K13 model an intrinsic scatter larger than $\sim0.20$ dex would be strongly disfavoured by current data. 
Likewise, K13 models with $\sigma_K \lesssim 0.1$ dex provide a poor agreement with data. However, our model suggests a trend in which $\sigma_K$ decreases as higher stellar mass bins are considered, with $\sigma_K \sim 0.20$ dex for the lowest masses and $\sigma_K \sim 0.10$ dex for the most massive galaxies.

Turning to the \emph{concentration model},  at lower stellar masses lower values of $\gamma$ are preferred, while for more massive galaxies $\gamma \sim -0.8$ gives a better match to data. Adopting $\gamma \gtrsim -0.4$ or $\gamma \lesssim -1.6$ would produce distributions that are too tight or too wide respectively, compared to the observed ones.\\ 
  It is worth pointing out here that the same identical considerations about $\gamma$ and $\sigma_K$ can be applied to ETGs, as shown in Appendix \ref{app:ETGs}. In Table \ref{table:ETGs} we report the values of $A_c$ and $A_K$, which instead are significantly lower than those of LTGs (compare to Table \ref{table:LTGs}). Thus, ETGs and LTGs define two separate relations in the $R_e-R_h$ plane, qualitatively in agreement with the findings of \citet{Huang+17} (see Sect. \ref{Sect:introduction}). However, we recall that in our framework $R_e\equiv R_{e,2D}$ is the 2D projection on the sky of the intrinsic galaxy shape. While in this work we do not model deprojection explicitly, we will show that cosmological models where the intrinsic galaxy shape is available still produce a  rather marked dichotomy in the $R_{e,3D}-R_h$ relation (see Sect. \ref{Sect:discussion_stateoftheart}), qualitatively in agreement with our empirical findings for $R_{e,2D}$.

%


\section{What drives the tightness of the observed size distributions?}
\label{Sect:discussion:size_distrib}

\subsection{ Implications for Ultra-Massive galaxies}
We recall that part of the scatter $\sigma_K$ originates from the shape of the $SMHM$ (i.e. the halo occupation distribution, see fig. \ref{fig:cartoon_scatter}). The latter contributes very little to the observed size functions at low masses (see figures \ref{fig:disksK13}, \ref{fig:ETGsK13}) and further scatter is needed to obtain a good match to data. On the other hand for UMGs (Ultra Massive Galaxies, for which $M_{\odot}>10^{11.5} M_\odot$ -  and which are mostly ETGs, see Fig. \ref{fig:sizefunctions_SDSS}) the contribution of the halo occupation distribution is the most relevant source of scatter. Therefore only a very small intrinsic additional scatter $\sigma_K \lesssim0.1$ is necessary to match observations. Thus, essentially, the K13 model predicts that the width of the galaxy size distribution at the high mass end can be entirely interpreted in terms of their halo occupation distribution, that is, \emph{the information about the broadness of the size distribution of UMGs is already contained in the SMHM}. We also note that adopting a flatter high mass end slope in the SMHM, as proposed by other groups (e.g., \citealt{Behroozi+13}, \citealt{Moster+13}), would result in an even larger source of scatter, perhaps in tension with the width of the observed size function of UMGs.
\subsection{On the validity of the ``MMW framework''}
\label{Sect:discussion_MMW}
Although the MMW model strictly predicts galaxy sizes only, it sets a framework in which also galaxy angular momenta can be predicted. Thus, in the following we will refer to the general notion of angular momentum conservation, which works for both galaxy angular momentum and sizes, as the \emph{MMW framework}.

The prediction for galaxy angular momenta in the MMW framework is straightforward. In the context of a biased collapse scenario (e.g. \citealt{Romanowsky&Fall2012}, \citealt{Kassin+12_RadiusBaryonicCollapse}, \citealt{Posti+18_angmom}, \citealt{Lapi+18_ETGs},) 
\begin{equation}
j_{inf}=f_{inf}j_{h}= f_{inf} \sqrt{2}\lambda V_h R_h
\end{equation}
where $f_{inf}$ is the fraction of gas that is able to cool efficiently \citep{Shi+17}. If angular momentum was strictly conserved during gas collapse, the distribution of $j_{inf}$ should be such that $\sigma_{j_{inf}} \approx \sigma_{log\lambda}\approx 0.25$ dex.  Studies have constrained $f_{inf}\approx 1$ for LTGs (e.g., \citealt{Shi+17}), so that the gas that cools has the same specific angular momentum of the host halo.  The factor $f_j$ that appears in eq. \ref{eq:lambdamodel} corresponds exactly to $f_{inf}$ in the biased collapse scenario.

\subsubsection{Is it safe to compare the MMW model to stellar sizes?}

The MMW model gives a clear prediction for the \emph{baryonic} sizes of LTGs, but here  we compare to the stellar effective radius instead. It could therefore be argued that from this comparison it is not possible to draw conclusions about the MMW model. We now show that instead $R_{e,star}$ is a good proxy of $R_{e,bar}$.\\

\citet{Kravtsov2013} have shown that in a sample of local LTGs from \citet{Leroy+08_THINGS}, the gas and stellar mass surface densities are well described by exponential profiles with $R_{d,gas} \approx 2.5 R_{d, star}$. Using this information, it can be shown that at fixed radius, $\Sigma_{bar}$ and $\Sigma_{star}$ differ by only $\lesssim$ 10\%, and hence $R_{d,bar} \approx R_{d,star}$. However, this is not sufficient to confirm that observations of $R_{d,star}$ can be compared to the predictions of the MMW model for $R_{d,bar}$. The reason for this is that the factor $f_j$ that appears in eq. \ref{eq:lambdamodel} strictly refers to the angular momentum retained by all \emph{baryons}, which might well be different from that retained by stars $f_{j,star}$, since gas is so spread out in the outskirts of LTGs with substantially high velocities traced by HI emission. 
On the other hand, using constraints from chemical abundances and star formation efficiency, \citet{Shi+17} have shown that $f_{inf} \approx f_{j,star}$ for LTGs, and we can therefore conclude that the MMW model can be also extended to the stellar component as well. More details on this subject can be found in Section \ref{sec:zoldan}, where we compare our Semi-Empirical Model to a state-of-the-art Semi-Analytic Model.

\subsubsection{Is the MMW model consistent with observed LTGs scaling relations?}
\label{Sect:discussion_MMW:scalingrelations}
 We now show that the observed proportionality between $R_e$ and $R_h$, as well as its scatter, is fully consistent with observations of galaxy angular momenta and a high angular momentum retention factor $f_j \sim0.5$ in the MMW framework (\citealt{Romanowsky&Fall2012}, \citealt{Posti+18_angmom}). \\
We recall the mathematical form of the MMW model,
\begin{equation}
R_d \simeq \frac{\lambda}{\sqrt{2}} f_c f_j f_R R_h.    
\end{equation}
We now note that $f_j = j_{star}/j_{h}$ with $j_{h}=\sqrt{2}\lambda V_h R_h \propto \lambda M_{h}^{2/3}$  (see, e.g., \citet{Romanowsky&Fall2012}). With this in mind, the equation above reads
\begin{equation}
R_d \simeq B \frac{j_{star}}{ M_{h}^{2/3}} R_h
\label{eq:jstar_Re_rh}
\end{equation}
where, critically, $\lambda$ disappears, which will be an important point in the discussion that follows. Here $B$ is a factor that encloses all the dependencies not relevant for our discussion.
\citet{Posti+18_angmom} have shown that in the mass range $9<logM_{star}/M_\odot<11$ the ratio $\Tilde{f_j} \equiv Bj_{star}/M_{h}^{2/3}$  depends very weakly on stellar mass\footnote{Actually \citet{Posti+18_angmom} constrain $f_j = j_{star}/j_{h} \approx0.5$, but since $\lambda$ is mass independent the same applies to $\Tilde{f_j}$}$^{,}$\footnote{A close look at their figure 5 for the \citet{Dutton+10} SMHM reveals that at most $f_j \sim M_{star}^{0.1}$. Moreover, the factor $f_c \propto c^{-0.2}$ (see \citet{MoMaoWhite98} and \citet{Jiang+17}) depends very weakly on halo mass ($c\propto M_h^{-0.1}$, \citealt{Dutton&Maccio2014}) and therefore on stellar mass. }. 
The only dependence left on stellar mass is in the factor $R_h\propto M_h^{1/3}\propto M_{star}^{1/6}$ (\citealt{Dutton+10}), for LTGs with $M_{star}<10^{11}M_\odot$.\\
With all this in mind, eq. \ref{eq:MMW_tot} reads
\begin{equation}
    R_d  \propto \Tilde{f_j} M_{star}^{1/6}.
\end{equation}
A slope of $1/6$ is consistent with measurements of the slope of the $R_e-M_{star}$ relation of LTGs (see \citealt{Shen+03}, \citealt{Bernardi+14}), plus minor corrections mainly due to the factor $\Tilde{f_j}$. We now note that the normalization $A_\lambda$ is remarkably constant over the whole mass range studied here (see Table \ref{table:LTGs}). Furthermore, the scatter in this relation is entirely governed by $j_{star}$, as for the mass range under consideration the halo occupation distribution is not critical  (fig. \ref{fig:cartoon_scatter})  and therefore for this purpose $\sigma_{\Tilde{f_j}} \approx \sigma_{jstar} \approx 0.20$ dex \citep{Posti+18_letter}. Notably, this is consistent with the scatter of the $R_e-R_h$ relation that we calibrate with the K13 model $\sigma_K\lesssim0.2$ dex. Moreover, the scatter that would come from $f_c$ is negligible (see fig. \ref{fig:conc_scatters}). \\ We note that to compute $j_{star}$ some authors adopt the simple scaling \begin{equation}
\label{eq:jstar_empirical}
    j_{star}\approx R_e V_c,
\end{equation} 
where $V_c$ is the circular velocity of a galaxy assuming a flat rotation curve \citep{Romanowsky&Fall2012}. In this case the observed scatter in $R_e$ would drive the one in $j_{star}$, making the argument above circular. However, the constraints on the scatter in $j_{star}$ by \citet{Posti+18_letter} quoted above, are found by direct integration of the observed rotation curves in the SPARC sample \citep{Lelli+16_SPARC}. It is also intriguing that, to first order, eq. \ref{eq:jstar_Re_rh} is consistent with the empirical finding of eq. \ref{eq:jstar_empirical}.
Indeed, assuming for simplicity an isothermal profile for dark matter haloes, for which the circular velocity $V_h$ is proportional to the halo radius $R_h$, $V_h\propto R_h$, eq. \ref{eq:jstar_empirical} can be easily inferred from eq. \ref{eq:jstar_Re_rh} assuming that $V_c\approx V_h$ \citep{Lapi+18_disks}.\\
Our conclusions above further corroborate the theoretical link between galaxy sizes and their angular momentum. It is interesting to investigate whether the origin of such a connection lies in the MMW framework. Indeed, \citet{Cervantes-Sodi+13} and  \citet{Burkert+16} have observationally constrained the quantity $\lambda f_j$ and have found that its dispersion is $\lesssim 0.2$ dex. 
  In the light of the discussion above, where we have shown that $\lambda f_j$ does not actually depend on $\lambda$, we argue that what these authors have found is in fact the scatter of the distribution of $\Tilde{f_j}$, $\sigma_{\Tilde{f_j}}$ , which essentially boils down to the distribution of galaxy stellar angular momenta, $\sigma_{jstar}$.   As a caveat, it should be noted that actually in both studies it is the gas kinematics that is probed, which may differ from the stellar kinematics. Nevertheless, in a recent study  \citet{Aquino-Ortiz+19} have shown that for a sample of local LTGs from the CALIFA survey \citep{Sanchez+12_CALIFA} gas and stellar kinematics show similar scaling relations. In the discussion above, we have tentatively assumed that this is also true for the sample of high redshift galaxies  used in \citet{Burkert+16}.  
  
To summarize, our work suggests that the MMW taken at face value is able to recover the median values of the observables, but it fails at reproducing the width of their distributions. Conversely, observations of galaxy angular momenta combined with the MMW model recover our semi-empirically determined constraints on the $R_e-R_h$ relation $\sigma_K \lesssim 0.20$ dex. Moreover, we have analytically shown that the MMW model naturally gives the slope of the $R_e-M_{star}$ relation. Our discussion confirms and extends to greater detail the results of \citet{Obreschkow&Glazebrook2014}, who have shown that if angular momentum and mass are known for LTGs, then the size-mass relation, as well as other observable LTG scaling relations, are automatically reproduced.  

 \subsection{Reconciling the MMW framework and observations}
 We have seen in the previous Section that both galaxy sizes and angular momenta predicted in the MMW framework suffer from the same shortcomings. In particular, although a mean $\langle \lambda \rangle\approx 0.035$ seems to work  well in predicting the mean of the observables, the width of the predicted distributions is $\approx 0.25$ dex, whereas independent observations sistematically find evidence for a width of $\lesssim 0.10-0.20$ dex. 
We thus conclude that either the MMW model is an oversimplification of an underlying more complex problem (Sect. \ref{Sect:danovich}), or we must introduce additional physical processes that limit the acceptable values of $\lambda$ (Sect. \ref{Sect:toymodel}).

\subsubsection{Insights from hydrodynamic cosmological zoom-in simulations}
\label{Sect:danovich}
Using hydrodynamical cosmological zoom-in simulations, \citet{Danovich+15} have traced the buildup of galaxy angular momentum in four phases that are linked to different spatial scales, from the cosmic web ($R \approx 2 R_h$) to the innermost part of the halo where $R \lesssim 0.1 R_h$. The region where $0.1 \lesssim R/R_h \lesssim 0.3$, termed as the ``messy region'' \citep{Ceverino+10}, is particularly interesting. This is the zone where the cold streams coming from 3-5 different independent directions start to interact. These streams have had their angular momentum set at $R \approx 2 R_h$, which does not significantly vary during its transport down to the ``messy region''. In this region substantial angular momentum exchange and torquing occurs, which eventually drive the baryons down to  $R\lesssim 0.1 R_h$. The resulting dynamics is such that the stellar spin parameter of the disc stars, defined as $\lambda_{star}= j_{star}/(\sqrt{2}R_h V_h)$,  is well described by a lognormal distribution with $\langle \lambda_{star} \rangle =0.019$ and dispersion of $0.2$ dex. \\

The results by \citet{Danovich+15} have several implications for our work. First of all, they agree with the estimate that $f_j = j_{star}/j_h \equiv \lambda_{star}/\lambda \approx 0.5$. Secondly, they provide a tighter distribution of angular momenta than that predicted from the MMW model, qualitatively in agreement with observations (\citealt{Cervantes-Sodi+13} and \citealt{Burkert+16}) and with our empirical constraints on the scatter of the size functions of LTGs (although possibly still too large for very massive LTGs and indeed to wide if only pure disks are considered, see Sect. \ref{Sect:MMW_bulgeless}). Third, recall that our results for the MMW model imply that $R_d \approx 0.3 \lambda R_h$. Here the factor $0.3 R_h$, is reminiscent of the outer boundary of the ``messy region'' seen in \citet{Ceverino+10} and \citet{Danovich+15}, where, effectively, the final angular momentum of baryons that will settle in a disk at $R \lesssim 0.1 R_h$ originates. On the other hand, in the MMW model the factor $0.3 R_h$ boils down to $f_j R_h / \sqrt{2}$ with $f_j\approx 0.5$ constrained in various ways, which is related to the angular momentum ``conserved during the gas collapse". Thus, the longstanding success of the MMW framework in predicting the normalization of LTG scaling relations (\citealt{Lapi+18_disks}, \citealt{Marasco+19}, \citealt{Somerville+08} \citealt{Straatman+17_tullyFisher} among many others) might be attributed only to the fact that the relevant physics is set at $0.3 R_h$, which also regulates the normalization of the MMW model. Note that this is not a matter of semantic, but of the underlying  physics. The scenario envisaged in the MMW model is that of a rather smooth formation history. The gas is assumed to be tight to the \emph{overall} spin parameter of dark matter, and slowly accretes onto the protogalaxy at the centre of the halo. Conversely, the simulations described in \citet{Danovich+15} reveal a quite more violent scenario where the gas is funneled towards the inner halo in only a few streams with an angular momentum higher than that of dark matter, which is then lowered by gravitational torques in the ``messy region". Indeed, the value of $f_j\approx 0.5$ can be understood in the light of these torques. Notably, in the MMW framework $f_j\approx 0.5$ provides a good fit to the mean observed size and angular momentum distributions, but it is not possible to predict it from first principles.  \\

\subsubsection{Shrinking the predicted size distributions: a toy model} 
\label{Sect:toymodel}
A narrower observed size distribution for LTGs could still be reconciled with the MMW model if we consider that not all the values of the spin parameter are physically acceptable.  \\

Low values of $\lambda$ are for example disfavoured by the standard disk instability scenario (e.g, \citet{Efstathiou+82}) according to which a disk becomes unstable to its own self-gravity if
\begin{equation}
    V_{disk}^{2} < \epsilon^2 \frac{GM_{d}}{R_d}.
\end{equation}
 Here we will consider disks that are dominated  by stars, for which $\epsilon \sim 0.9$ (e.g., \citealt{Christodoulou+95}), so that $M_d\approx M_{star}$. Using the definition $M_{star}/f_b M_h=f_*$ \footnote{$f_b\approx0.16$ is the cosmological baryon fraction, \citet{Planck_cosmopar}}  and eq. \ref{eq:MMW_tot}, the condition above reads
 \begin{equation}
 \label{eq:lambdaDI}
     \lambda< \lambda_{DI} \equiv \sqrt{2}\frac{\epsilon^2}{f_j} \frac{f_b}{f_v} f_* f_R f_c
 \end{equation}
where $f_v\approx1.07 V_h/V(3R_d)$ (\citet{Lapi+18_disks}) and $V_h$ is the circular velocity of the halo.\\

On the other hand, it could be envisaged that the high-spin tail of the $\lambda$ distribution provides a substantial centrifugal barrier that prevents the gas from collapsing and forming stars. In such a scenario, the gas would set in a rotationally-supported disk at $R_{rot}>R_{d,gas}$, where  $R_{rot}$ is given by solving the following expression,
\begin{equation}
    \frac{j_{gas}^2}{R_{rot}^3} = \frac{(G M_{gas}(<R_{rot}) + M_h(<R_{rot}))}{R_{rot}^2}
\end{equation}
 Here $R_{rot}$ is the size of the galaxy that cannot form stars, while $R_{d,gas}$  the size that would be predicted by the MMW model, if the gas could collapse beyond the centrifugal barrier.
 In the biased collapse scenario, $M_{gas}\equiv M_{inf}=f_{inf}f_b M_h$ and $M_h(<R_{rot}) \approx 10 M_h \bigl (R_{rot}/R_h)^2$ (\citealt{Lapi+18_ETGs}) and $j_{gas}=f_{inf}j_h$. The solution to the equation above is 
\begin{equation}
    R_{rot} = 2 \lambda^2 \frac{f_{inf}}{f_b} R_h,
\end{equation}
and therefore the condition $R_{rot}>R_{d,gas}$ reads
\begin{equation}
\label{eq:lambdaCB}
    \lambda > \lambda_{CB} = \frac{f_b}{2\sqrt{2}} \frac{1}{f_R f_c f_v}.
\end{equation}

\begin{figure*}
    \centering
    \includegraphics[width=0.6\textwidth]{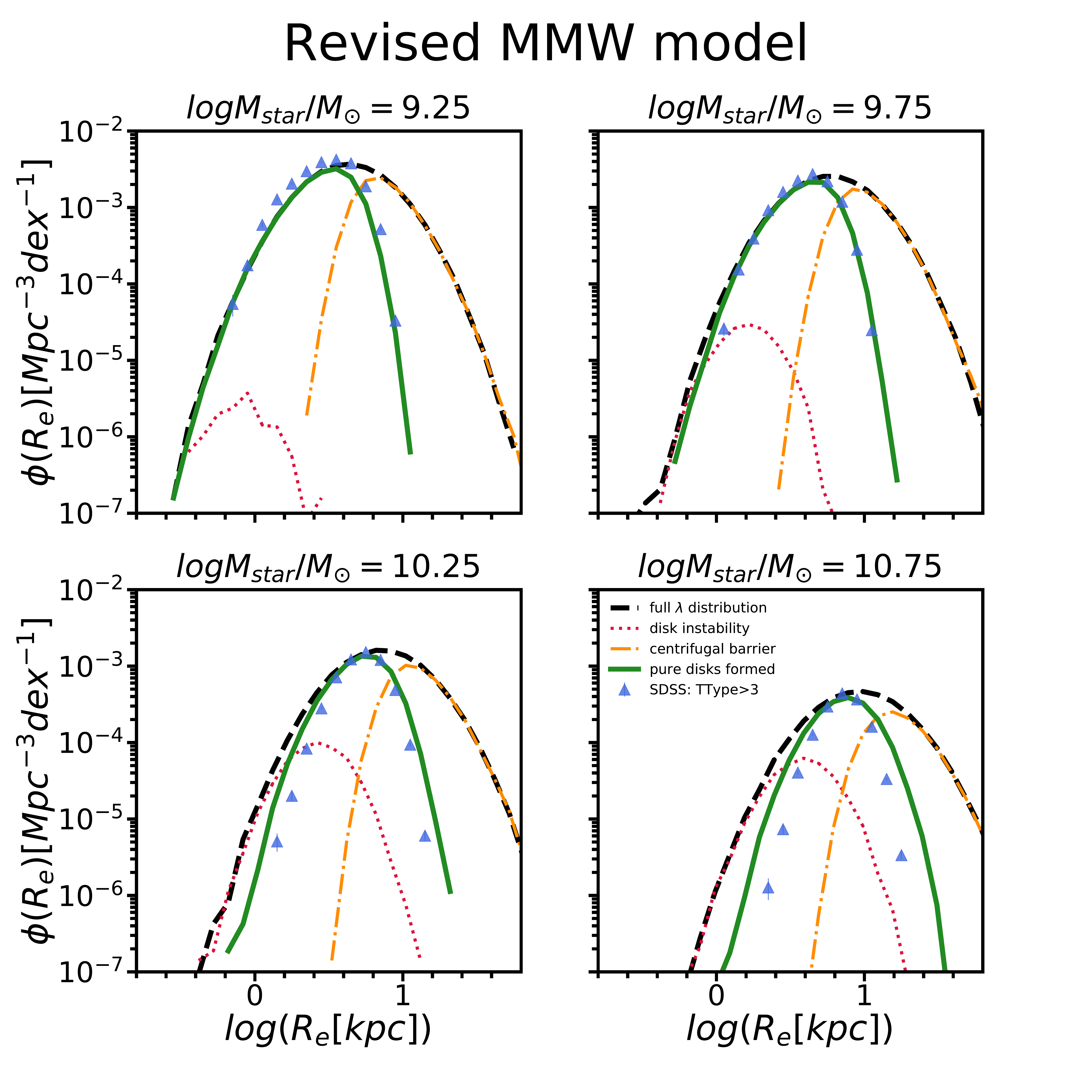}
    \caption{Result of the toy model outlined in Section \ref{Sect:toymodel}. The full distribution of $\lambda$ is indicate with dashed black lines.  The contributions to the total size function that would come from haloes with $\lambda<\lambda_{DI}$ and $\lambda>\lambda{CB}$ are shown in red dotted and orange dash-dotted lines. Only galaxies within the range $\lambda_{DI}<\lambda<\lambda{CB}$ are able to form (green solid lines).}
    \label{fig:MMW_stable}
\end{figure*}

Figure \ref{fig:MMW_stable} shows the results of this toy model, compared to the observed size function of bulgeless galaxies with $M_{star}<10^{11}M_\odot$. In this case, the values of $A_\lambda$ in table \ref{table:LTGs} would need to be increased by 0.1 dex, which could be interpreted as a higher $f_j$ for pure disks, which agrees with \citet{Romanowsky&Fall2012} and \citet{Obreschkow&Glazebrook2014}. Overall, it can be seen that our simple framework improves tremendously on the shape of the size function that would be predicted from the MMW model. Indeed, the cuts that we apply are able to tighten the predicted size function in a mass dependent fashion, as suggested by the mass dependence of $\sigma_K$, with disk instabilities being important only for more massive galaxies \citep{DeLucia+11_bulgeFormation}. Moreover, these cuts behave very differently at the low- and high-spin ends respectively, in a way that almost fully recovers the skeweness of the observed size functions (see also Sect. \ref{Sect:data}).

\vspace{1 em}

The power of this very simple toy model is that it is able to shrink not only the predicted size functions, but also the angular momentum distributions, since $j_{gas} \propto \lambda$.
The resulting angular momentum distribution would retain the same skeweness seen in the size distributions, and future data would be a powerful test for this model.

\subsection{Comments on the concentration model}
\label{Sect:discussion_concentration}

As regards to the concentration model, we have seen (cfr. Sect. \ref{Sect:results}) that lower values of $\gamma$ produce wider distributions, and that $\gamma$ may be tuned to match the size functions without adding any intrinsic scatter $\sigma_{CM}$ in the concentration model. There are two factors at play here. The first is that halo concentration has already an intrinsic scatter that amounts to $\sigma_{logc}\sim 0.11$ dex. The second is that, as it turns out, adopting different values of $\gamma$ also ends up contributing to the total scatter at different levels (see Sect. \ref{Sect:scatters}). 
We note that in principle some degeneracy may be expected in the contribution to the total scatter from $\gamma$ and $\sigma_{CM}$.
Inspired by the results of \citet{Jiang+17}, we discuss what the consequences of having  a mass independent $\gamma\sim-0.7$ would be. It is clear from figures \ref{fig:disksConc} and \ref{fig:ETGsConc} that such a value of $\gamma$ would account for some of the observed width of the size functions. In fact, it can be seen that the scatters produced by $\gamma=-1.6,-1.2,-0.8$ and $-0.4$ are roughly equivalent to those given by $\sigma_K=0.20,0.15,0.10$ and $0.0$. A constant value of $\gamma=-0.7$ from \citet{Jiang+17} would be able to account for $\approx 13\%$, $\approx 25\%$ and all of the scatter observed for galaxies with $M_{star}<10^{10} M_\odot$, $10^{10}<M_{star}/M_\odot<10^{11}$ and $10^{11}<M_{star}/M_\odot<10^{12}$ respectively. Note that if $\gamma$ is independent of $M_{star}$, then a mass-dependent $\sigma_{CM}$ is expected due to the mass dependence of the width of the size functions. Overall, the concentration model could be favoured due to its lower intrinsic scatter, however its explanation from a theoretical standpoint remains a challenge.  

\section{The K13 model in state-of-the-art simulations and the formation of ETGs }
\label{Sect:discussion_stateoftheart}

We now proceed to test whether current cosmological models of galaxy formation are consistent with  the semi-empirical constraints outlined in the previous Sections, i.e. the existence of a tight relationship $R_e-R_h$, between galaxy size and host halo radius, and a lower normalization in the relationship $R_e-R_h$ relation for ETGs compared to LTG. To this purpose, we will use the Rome semi-analytic model (the Rome SAM herafter) and the Illustris TNG simulation.

\subsection{The Rome SAM}
We updated the Rome SAM (described in detail in \citealt{Menci+05}, \citealt{Menci+08}, \citealt{Menci+14}) with a standard prescription for galaxy sizes \citep{Cole+00} as adopted in many other semi-analytic models (e.g. \citealt{Guo+11}, \citealt{Shankar+13}). We refer the reader to the original papers for the details of the SAM. Here we just recall that in the Rome SAM galaxies are initialized as disks following the MMW model, with with $\langle \lambda \rangle = 0.035$ and $\sigma_{log\lambda}=0.25$ dex from dark matter only simulations (e.g. \citealt{Rodriguez-Puebla+16} Although in principle both internal torques (i.e., disk instabilities) and galaxy mergers may contribute to the evolution of galaxy sizes, in our SAM we deliberately choose to ignore the former to isolate the sole role of mergers in setting galaxy sizes. In our SAM after a merger the size of the remnant is computed from energy conservation and the virial theorem,
\begin{equation}
    \frac{M_{fin}^2}{R_{fin}} = \frac{M_{1}^2}{R_1} + \frac{M_2^2}{R_2}
\end{equation}
where $M_{fin}, M_1,M_2$ and $R_{fin},R_1,R_2$ are the masses and half-mass radii of the remnant and the merging partners. Here we neglect the term of gravitational interaction between the merging galaxies which corresponds to assuming that all mergers occur on parabolic orbits. We have checked that at this level of the modelling, including the gravitational interaction term mostly impacts the relative normalization of galaxy sizes, but not their distribution, which is the main aim of this work.  Major mergers ($M_1/M_2>0.3$) are assumed to completely destroy the stellar disk, while minor mergers leave the disk intact and grow the galactic bulge only. The total size of a galaxy is computed as the mass-weighted mean of the disk and bulge radii. During a major merger  substantial energy dissipation may occur \citep{Covington+11}, which will modify the size of the remnant as \citep{Shankar+13}
\begin{equation}
    R(dissipation) = \frac{R(dissipationless)}{1+f_{gas}/0.2}
\end{equation}
where $f_{gas}$ is the gas fraction of the merging pair. We run the Rome SAM with and without the implementation of such process, and we will show that it is not crucial to our conclusions.\\
In the Rome SAM we classify galaxies according to their B/T ratio, with ``pure disks'' being the galaxies with $B/T<0.3$, and ``pure bulges'' those with $B/T>0.7$. To enable a closer comparison to observations, the sizes of our semi-analytic galaxies are convolved with a measurement error of $\sim0.1$ dex.

\subsection{Illustris TNG}
The Illustris TNG project (described in detail in \citealt{Pillepich+18}, \citealt{Pillepich+18_TNGdescription}, \citealt{Nelson+18_datarelease}), is a suite of cosmological simulations run with the same parameters in three boxes with side $50/h, 100/h $ and $300/h$ Mpc. The simulation builds and improves on the previous Illustris project (\citealt{Vogelsberger+14}, \citealt{Genel+14}, for a review of the differences between Illustris and Illustris TNG see \citealt{Pillepich+18_TNGdescription}), and has proven to achieve a good agreement with observed galaxy sizes (\citealt{Genel+18}), SMF  (\citealt{Pillepich+18}) and morphologies (\citealt{Huertas-Company+19}). Here we use the box of $100 Mpc/h$ a side, which is publicly available\footnote{ \url{www.tng-project.org/data/}} and described in \citet{Nelson+18_datarelease}. For a complete description of the numerical implementation of Illustris TNG we refer the reader to \citet{Springel+10}, \citet{Marinacci+18}, \citet{Naiman+18}, \citet{Springel+18}, \citet{Pillepich+18_TNGdescription}, \citet{Weinberger+17}, \citet{Nelson+18_datarelease}.\\
 For IllustrisTNG100, we make use of the catalog of optical morphologies presented in \citet{Huertas-Company+19} based on \texttt{statmorph} \citep{Rodriguez-Gomez+19}, a Python package for calculating non-parametric morphologies of galaxy images  as well as fitting 2D S\'ersic profiles.\footnote{Available at \url{https://statmorph.readthedocs.io}.}. 
 Briefly, \citet{Huertas-Company+19} have selected galaxies with $M_{star}>10^{9.5} M_\odot$ in the snapshot 95 at $z\sim0.045$ and processed their images with the radiative transfer code \texttt{SKIRT}  (\citealt{Baes+11_SKIRT}, \citealt{Camps&Baes2015}). The mock images are then observed in the SDSS r-band filter and further realism is added using \texttt{RealSim}\footnote{Available at \url{https://github.com/cbottrell/RealSim}.} (\citealt{Bottrell+17a},\citealt{Bottrell+17b}, \citealt{Bottrell+19}). The full procedure is outlined in \citet{Rodriguez-Gomez+19} and \citet{Huertas-Company+19}. Morphological information is also available for the mock galaxies. Using the \citet{Nair+10} catalog as training set, \citet{Huertas-Company+19} have trained an ensamble of Convolutional Neural Networks in binary classification mode to distinguish LTGs from ETGs; a finer within-class classification is also available. The mock images from Illustris TNG galaxies are then classified as LTGs or ETGs using the same neural networks. \citet{Huertas-Company+19} have found that the morphologically classified Illustris TNG galaxies follow the same scaling relations of SDSS galaxies almost everywhere. The issue of how exactly the morphologies of simulated galaxies resemble observations is the subject of a forthcoming paper (L. Zanisi et al. in prep). We match the catalog with the SubFind catalog of Illustris TNG and select central galaxies only. This leaves us with a total of 7222 galaxies.\\ 
For each of our IllustrisTNG100 galaxies, we compute the specific angular momentum of the stellar particles:
\begin{equation}
j_{star} = \frac{1}{\sum_{n} m^{(n)}} \sum_{n} m^{(n)}\boldsymbol{x}^{(n)} \times \boldsymbol{v}^{(n)}
\end{equation}
where $\boldsymbol{v}^{(n)}$ is the velocity of the $n^{th}$ particle relative to the centre of mass for the galaxy. $\boldsymbol{x}^{(n)}$ is the position of a given particle with respect to the position of the most gravitationally bound particle in the galaxy. We choose this definition since the centre of mass velocity can be biased by structure at large radii and hence may not represent the true rotational centre. We compute the angular momentum relative to the centre of mass since the rest frame as defined by the most bound particle is often noise dominated. See \citet{Duckworth+19_misalignment} for more details.\\
As for the the sizes of IllustrisTNG100 galaxies, we adopt four of the available \texttt{statmorph} estimates. First, we use the semi-major axis size of the best S\'ersic fit, $R_{e,maj}$, ensuring that the flag \texttt{flag\textunderscore sersic} is equal to zero to include only good photometric fits. We will also show results for $R_{50}$ and $R_{80}$, the radii of a circular area that encloses 50\% and 80\% of the light contained in 1.5 times the Petrosian radius, where no prior assumption on the light profile is made. Finally, from the SubFind \citep{Springel+01_subfind} catalogue we extract the physical size $R_{e,3D}$. The correlation between $R_{e,maj}$ and $R_{e,3D}$ is shown in Appendix \ref{app:deproj}.

\subsection{Results}
\begin{figure*}
    \centering
    \includegraphics[width=\textwidth]{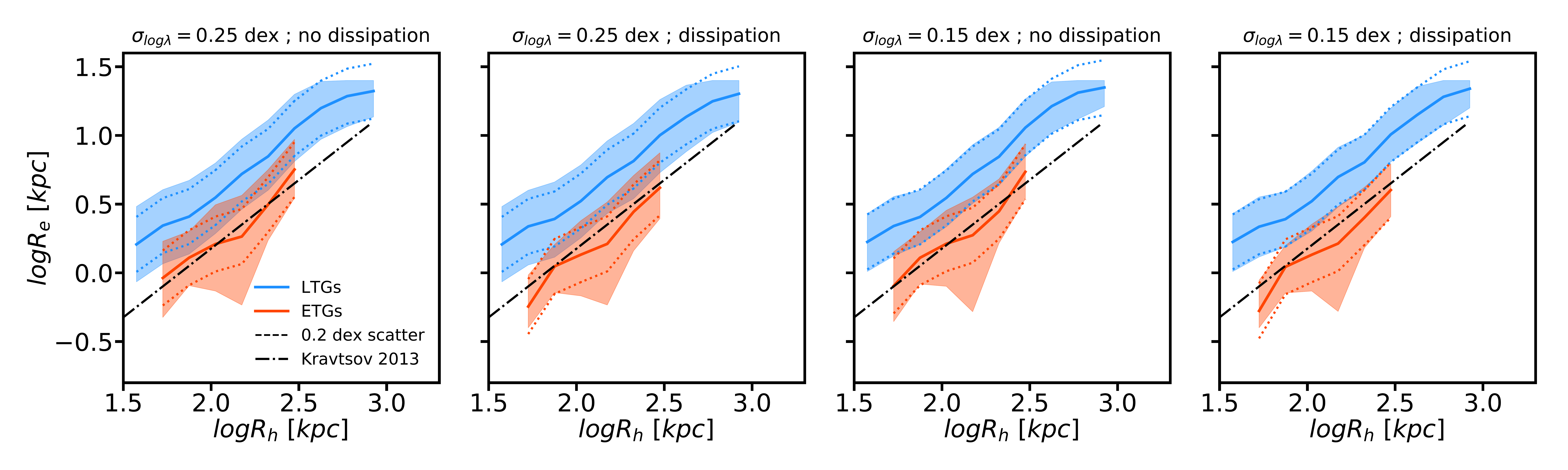}
    \caption{$R_e-R_h$ relation in the Rome SAM. Each panel represent a run of the model where $\sigma_{log\lambda}$ is varied or dissipation is included, as labeled.
    The red and blue lines are for LTGs and ETGs respectively, while the cyan and salmon shaded areas indicate the 16th and 84th percentiles of the distributions at fixed $R_{h}$. Dashed lines show a scatter of 0.2 dex from the mean, consistent with the upper limit provided our semi-empirical model. The relation by \citet{Kravtsov2013} is shown as dot-dashed lines for comparison. The predicted $R_e$ are convolved with an observational scatter of 0.1 dex.}
    \label{fig:K13_SAM}
\end{figure*}

\begin{figure*}
    \centering
    \includegraphics[width=\textwidth]{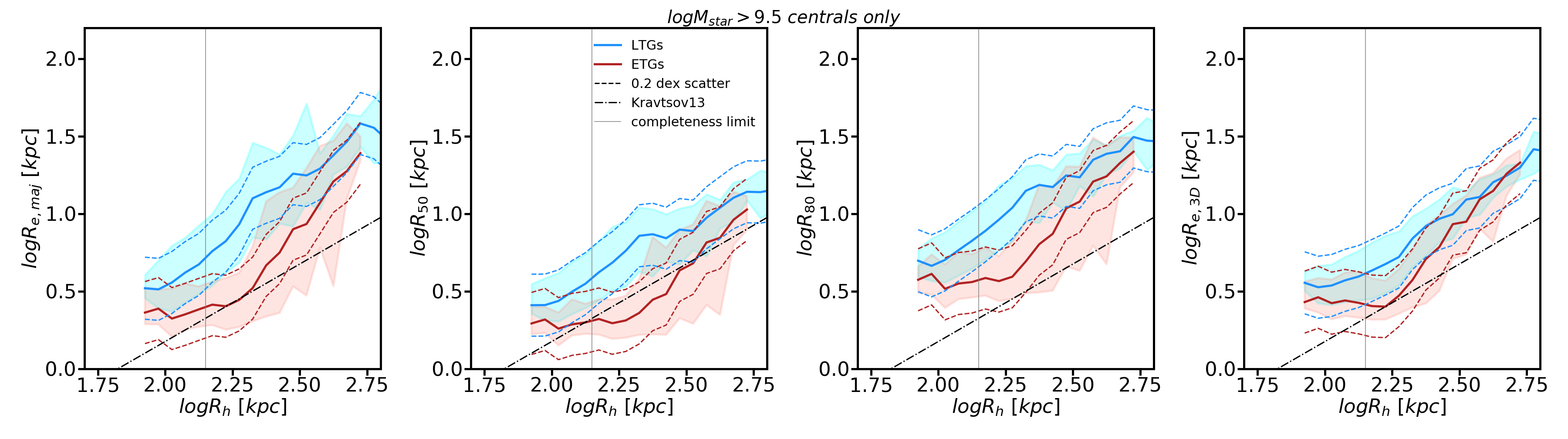}
    \caption{\emph{First panel:} \texttt{statmorph} S\'ersic semi-major axis sizes of the mock observed Illustris TNG galaxies as a function of $R_{h}$. \emph{Second panel:} \texttt{statmorph} estimates for $R_{50}$ of the mock observed Illustris TNG galaxies as a function of $R_{h}$. \emph{Third panel:} \texttt{statmorph} estimates for $R_{80}$ of the mock observed Illustris TNG galaxies as a function of $R_{h}$.  \emph{Fourth panel:} Physical 3D radius $R_{e,3D}$ of the same Illutris TNG galaxies as a function of $R_{h}$. Red and blue lines are for LTGs and ETGs respectively, while the cyan and salmon shaded areas indicate the 16th and 84th percentiles of the distributions at fixed $R_{h}$. Dashed lines show a scatter of 0.2 dex from the mean, consistent with the upper limit provided our semi-empirical model. The relation by \citet{Kravtsov2013} is shown as dot-dashed lines for comparison. The completeness limit on $R_{h}$ induced by the stellar mass cut is shown as a vertical gray line. The difference between the left and right panels may be understood in the light of Figure \ref{fig:projection_effects_TNG}.}
    \label{fig:TNGsizes}
\end{figure*}

We produce $R_e-R_h$ relations and their scatter at $z=0$ for central galaxies only from our SAM (Figure \ref{fig:K13_SAM}) and the mock observations of IllustrisTNG100 (Figure \ref{fig:TNGsizes}). At a first glance, we do not see much difference between the $R_e-R_h$ relation found in TNG100 and in the Rome SAM. It is indeed pleasing that both models predict that ETGs and LTGs lie on two separate relations, in agreement with our semi-empirical constraints. We now discuss the outcomes of the two models in more detail.

In the SAM, the relation of ETGs is offset by $\sim 0.3 $ dex and $\sim 0.4$ dex with respect to that of LTGs in the dissipationless and dissipative scenarios, respectively.  The two left panels of Figure \ref{fig:K13_SAM} show that using the distribution of spin parameters taken from dark matter only simulations result in a scatter $\sigma_K >0.2$ for both ETGs and LTGs, which is larger than that found by our semi-empirical approach. In the two right panels, instead, we have assumed that the distribution of spin parameters from which LTGs can form is $\sigma_{log\lambda}=0.15$ which, once convolved with measurement uncertainty, is consistent with the upper limits to $\sigma_K$ that we give in Section \ref{Sect:results} (Figure \ref{fig:disksK13}). In this case, the scatter in the $R_e-R_h$ relation of ETGs is somewhat reduced, and it becomes consistent with our semi-empirical findings, especially at high values of $R_h$. We also note that dissipation does not affect the scatter in either case.

In IllustrisTNG100 we can see that using the semi-major axis size $R_{e,maj}$ gives a scatter that is somewhat larger than the one we find with our semi-empirical model, while the size $R_{50}$ of mock-observed LTGs follows more closely our constraints on the scatter of the $R_e-R_h$ relation. Indeed, it can be seen that for intermediate values of $R_h$ the scatter is just about $0.2$ dex, declining with increasing $R_h$. However, it seems that for ETGs the scatter is larger than $0.2$ dex in both cases. The right panel of Figure \ref{fig:TNGsizes} shows that the distribution of physical sizes at fixed halo radius is indeed already of the order of $0.2$ dex for ETGs even before the mock observations are performed. On the positive side, we observe that such scatter decreases as $R_h$ increases also for ETGs in all cases. In passing, we also note that the relationship between the physical sizes of very large central galaxies and that of their dark matter haloes is extremely tight, of the order of $0.05$ dex, in agreement with our constraints (see Figure \ref{fig:ETGsK13})
 
\subsection{Discussion} 
  The difficulty of mantaining a tight scatter in the observed structural scaling relations of ETGs implied by a pure merger scenario has been discussed in, e.g., \citet{Nipoti+12}, \citet{Nipoti+09} (see also discussion in \citealt{Shankar+14_sizes}).  Using our SAM we find that a pure merger scenario, where ETGs are only formed as a consequence of merging of LTGS, requires a very tight distribution for the sizes of LTGs, of the order of $\sim 0.15 $ dex, which is supported by the estimate of the intrinsic scatter of the size distributions of star forming galaxies at $z>0$ provided in \citet{vanderwel+2014}. We discuss a comparison with another SAM in Section \ref{sec:zoldan}.
  
  \vspace{1 em}
  
\begin{figure*}
    \centering
    \includegraphics[width=0.7\textwidth]{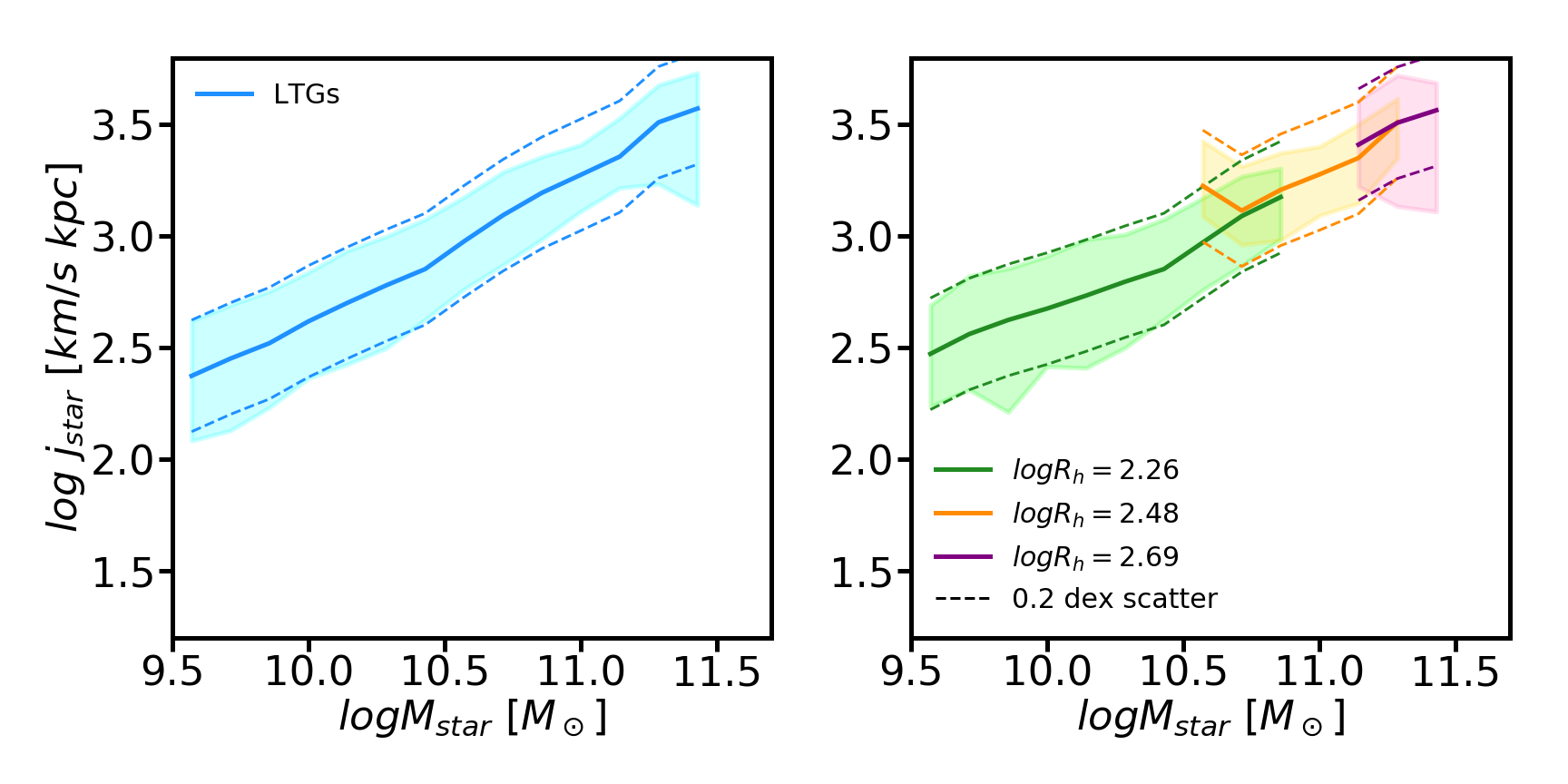}
    \caption{The relationship between stellar angular momentum and stellar mass for Illustris TNG LTGs (left), binned in three ranges of halo radius (right).  }
    \label{fig:jstar_Mstar_all}
\end{figure*}

\begin{figure*}
    \centering
    \includegraphics[width=0.8\textwidth]{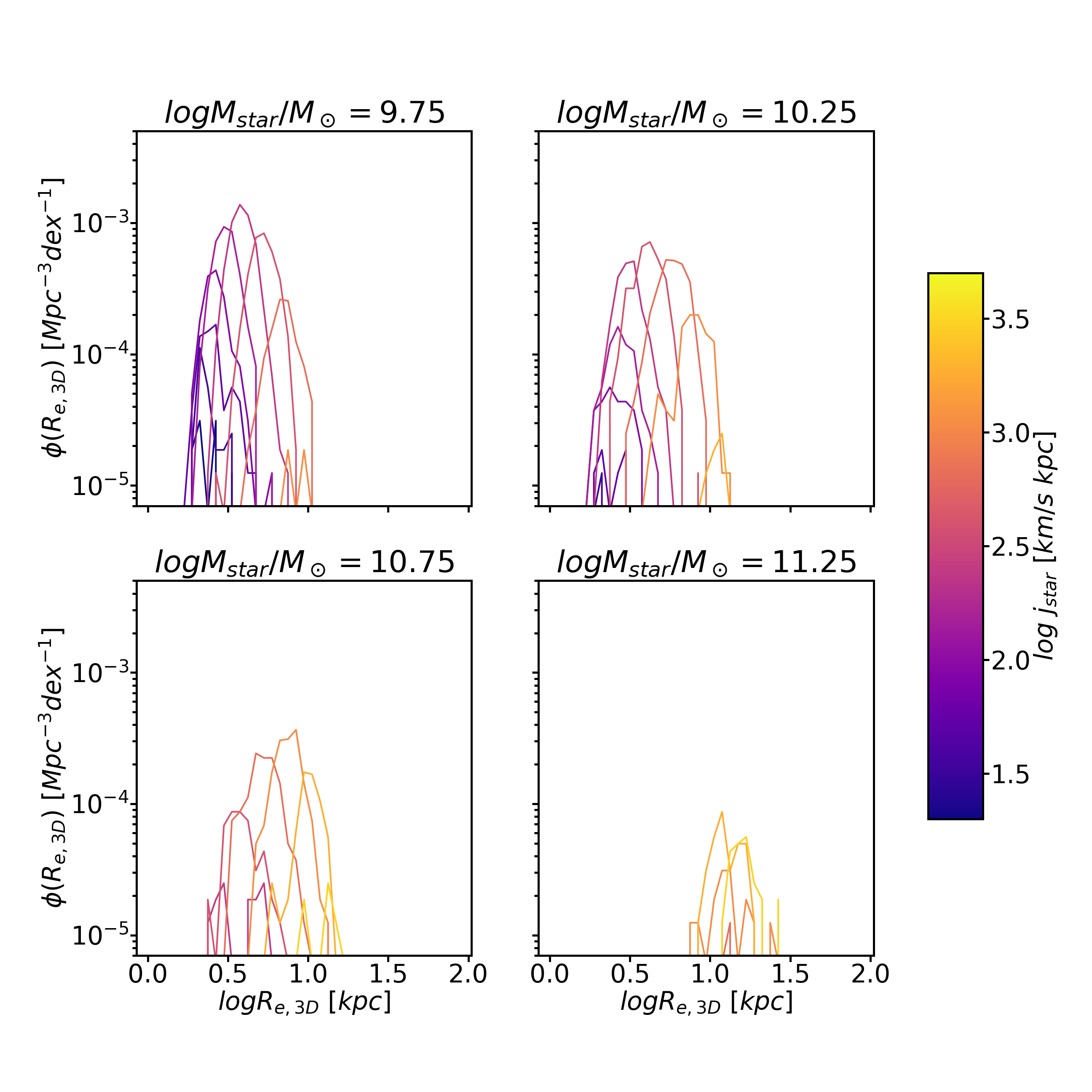}
    \caption{Size functions of IllustrisTNG100 galaxies in bins of stellar mass and color coded by the value of specific stellar angular momentum. The tightness of the size functions at fixed $j_{star}$ is remarkable.}
    \label{fig:sizefunct_jstarall}
\end{figure*}

   On the other hand, in hydrodynamical simulations internal torques and mergers arise naturally from the local and global gravitational fields respectively. The implementation of IllustrisTNG100 achieves naturally a tight relation between $R_e$ and $R_h$ for LTGs but not so for ETGs. We now speculate on the possible reasons behind the success of the simulation in reproducing the semi-empirical trends for LTGs. In Section \ref{Sect:discussion_MMW:scalingrelations} we have shown that the MMW model may be consistent with the observed scaling relations of LTGs if the stellar angular momentum, rather than the halo spin parameter, is used. Being consistent with our determination on the $R_e-R_h$ relation for LTGs, Illustris TNG offers an ideal testbed for this hypothesis. We recall that empirically, and using the MMW model, our argument would predict that at fixed stellar mass and halo radius the scatter in $R_e$ should be completely driven by that in $j_{star}$. Therefore,  a variance of about 0.2 dex in the distribution of $j_{star}$ at fixed $M_{star}$ and $R_h$ would support our argument.\\
  In Figure \ref{fig:jstar_Mstar_all} we show the relationship between the stellar angular momentum $j_{star}$ and stellar mass in IllustrisTNG100 in bins of $R_h$. We also show the relation for all LTGs since two highest bins in $R_h$ suffer from low number statistics.  It can be seen that the predicted scatter is about 0.2 dex and decreasing with increasing stellar mass and bin of $R_h$. This is consistent with our argument, and also with the decrease in scatter in the $R_e-R_h$ relation at high halo radii. We test more directly the connection between galaxy size and stellar angular momentum in Figure \ref{fig:sizefunct_jstarall}, where we show the size functions of IllustrisTNG100 LTGs in narrow bins of $j_{star}$. The first striking feature of Figure \ref{fig:sizefunct_jstarall}
 is that in a given bin of $M_{star}$ larger galaxies have a larger specific stellar angular momentum. Even more remarkable is the fact that the tightness of the size functions\footnote{Here we use $R_{e,3D}$ since we want to investigate the intrinsic relationship between size and angular momentum.} is extraordinarily narrow at fixed $j_{star}$, with a scatter of the order of $\lesssim0.1$ dex. These findings suggest that the link between galaxy sizes and their stellar angular momentum is extremely tight. We therefore advocate that an empirically motivated model where the relationship between $R_e$ and $R_h$ is mediated by stellar angular momentum seems to be supported by our analysis of IllustrisTNG100.
  
  \vspace{1em}
  
  To conclude, we briefly note that mechanical feedback from the AGN may also ``puff-up" the galaxy structure (\citealt{Fan+08}, \citealt{Fan+10}, \citealt{Ragone-Figueroa&Granato2011}),
  which may be critical to decrease the scatter in the ETGs scaling relations (\citealt{Lapi+18_ETGs}). This is not included in the Rome SAM, but it is modeled in Illustris TNG (\citealt{Weinberger+17}, \citealt{Pillepich+18_TNGdescription}). It seems however that both the purely hierarchical scenario adopted in the version of the Rome SAM where $\sigma_{log\lambda}=0.15$ and IllustrisTNG100 follow our semi-empirical constraints, which makes it difficult to disentangle the effects of mergers and AGN puffing-up here.

\section{Discussion and comparison to previous work}
\label{sec:comparisons}
\subsection{Comparison to other Semi-Analytic Models}
\label{sec:zoldan}
The constraints that we give on the $R_e-R_h$ connection with our semi-empirical model stem from single-epoch abundance matching. As such, we are unable to follow galaxies during their formation history in an evolutionary context. In particular, the MMW model is directly applied to dark matter haloes at $z\sim 0$, while disk galaxies are likely to have grown their stellar mass steadily in time during the last 10 Gyr or so (e.g. \citealt{Patel+13_milkyWayLikeEvolution}). The implementation of the Rome SAM  is such that whenever gas cooling occurs, the galaxy is assigned a size according the MMW model. Admittedly, this model is not able to capture the inside-out growth of galaxy disks (e.g.  \citealt{Sanchez+18_RMX}). 
  Recently, \citet{Zoldan+18} and \citet{Zoldan+19} have presented the result of a semi-analytic model (based on \citealt{Hirschmann+16}, \citealt{DeLucia+14}, \citealt{Xie+17}) in which stellar angular momentum and galaxy sizes are evolved self-consistently. In their model it is assumed that the stellar angular momentum is built up gradually as star formation proceeds and depletes the gas disks. The size of the stellar disk is then computed at each step assuming that a close analogue of eq. \ref{eq:jstar_empirical} holds. The Zoldan et al. model is built on the assumption that angular momentum is conserved during both star formation and disk instabilities, which are also included, contrary to the Rome SAM. 
  Their studies point out that mergers are the primary drivers of the observed LTGs-ETGs dichotomy in the $j_{star}-M_{star}$ relation. This qualitatively agrees with our result that bulge growth via mergers may lead to the same bimodality but in the $R_e-R_h$ plane. These authors also obtain a tight scatter in the $R_e-R_h$ relation for ETGs,  as shown in their Figure 7. Whether such scatter is consistent with our constraints remains to be seen. \\

\subsection{Comparison to other Semi-Empirical Models}
 Using a semi-empirical technique similar to ours, \citet{Somerville+18} found that the \emph{total} size distributions observed in GAMA and CANDELS are in agreement with the MMW model. These findings are suggestive that both the population of ETGs and LTGs may be described in the MMW framework. For example, ETGs could be formed in dark matter haloes with preferentially lower $\lambda $, which would account for the fact that the distribution of ETGs is peaked at lower $R_e$ than that of LTGs. However, this model would not be able to explain the observed angular momenta of ETGs, as shown by \citet{Romanowsky&Fall2012} and \citet{Posti+18_angmom}. Alternatively, one could note that the normalization  of the MMW model bears the dependence on the fraction of the halo angular momentum $f_j$ that was retained by the collapsing gas, since $A_\lambda\propto f_j$ (see Sect. \ref{Sect:model}). In principle ETGs and LTGs could then be two populations that retained preferentially lower and higher $f_j$ respectively but that did not form in haloes with different value of $\lambda$. Such a scenario may also be able explain why ETGs always have smaller sizes than LTGs. However, although ETGs are sometimes envisioned to form mostly in-situ (e.g., \citealt{Shi+17}, \citealt{Lapi+18_ETGs}), it is often suggested that they have likely undergone merger events,  which may have led to lower $f_j$ on average \citep{Romanowsky&Fall2012}. We would thus be cautious in interpreting $f_j$ for ETGs in the context of the MMW model, at least at low redshift. Indeed, we have shown that a purely hierarchical model is able to produce smaller sizes for ETGs, while preserving the linearity of the MMW model. As a side note, we recall that the total size function shown in figure \ref{fig:sizefunctions_SDSS} is wider than those of ETGs and LTGs taken individually and therefore it might well be that the agreement between the MMW model and the total size function found by \citet{Somerville+18} occurs only by chance.\\
 Another possible explanation for the difference in the normalization of the $R_e-R_h$ relation is that the size of a galaxy is more tightly bound to that of its halo at the redshift of formation than to the size of the halo at the time of observation. In particular, given the older ages of ETGs (e.g. \citealt{Bernardi+10}), they must have formed at high redshift where haloes were smaller (see eq. \ref{eq:Rhalo}). The late evolution of ETGs, which seems to be dominated by minor dry mergers (e.g. \citealt{Shankar+13}, \citealt{Oser+10}), will however modify the $R_e-R_h$ relation onto which ETGs formed. Unfortunately, with the  Semi-Empirical model used in this work we are able to constrain only the present day relation, and therefore we cannot directly infer any information about the formation of ETGs.
\subsection{Using $R_{80}$ instead of $R_e$}
\label{sec:R80}
In recent work (\citealt{Mowla+19_SMHM_R80}, \citealt{Miller+19_R80}) it has been proposed to use as proxy for galaxy size $R_{80}$, the size that encloses 80\% of the light, rather than the half-light radius $R_e$. This claim was made on the grounds that: i) the sizes of passive and star-forming galaxies tend to collapse on the same size-stellar mass relation in the case where $R_{80}$ is used \citep{Miller+19_R80}; ii) $R_{80}$ is more closely linked to the size of the host dark matter halo $R_e$ \citep{Mowla+19_SMHM_R80}. \\
Figure \ref{fig:sizefunctions_SDSS} shows a comparison between the size functions computed for $R_e$ and $R_{80}$ for both ETGs and LTGs. We observe that the difference in the size functions of ETGs and LTGs computed using $R_e$ is only slightly reduced when using $R_{80}$. While such difference appears to be somewhat more pronounced at $M_{star}>10^{11}M_\odot$, the bimodality of the size functions $\phi(R_e)$ seems to be substantially conserved also for $\phi(R_{80})$ at lower masses. It is also noteworthy that the scatter of the individual size functions is not affected by the choice of the definition of galaxy size. Therefore, adopting $R_{80}$ rather than $R_e$ in our work would only require an overall higher normalization for the $R_e-R_h$ relations studied here, but the results for the implied scatters remain robust. In particular, such result would not undermine our empirical model where galaxy and halo sizes are mediated by stellar angular momenta.\\ The discussion above is in agreement with the fact that the \texttt{statmorph} estimate of $R_{80}$ for the mock-observed IllustrisTNG100 galaxies entails a similar scatter in the galaxy size-halo size relation of LTGs and ETGs compared to that of $R_{50}$ (see central panel of Figure \ref{fig:TNGsizes}). Moreover, the relations for the two morphological classes keep being separated also in the $R_{80}-R_h$ plane also in the case of IllustrisTNG100.\\
We stress that our analysis of SDSS makes use of a mix of \texttt{S\'ersic} and \texttt{S\'ersic+Exponential} fits, contrary to \citet{Mowla+19_SMHM_R80} and \citet{Miller+19_R80} where only \texttt{S\'ersic} profiles are assumed, while in IllustrisTNG100  $R_{80}$ is the size of a region that contains 80\% of the light inside an area of 1.5 the Petrosian radius. We refer the reader to \citet{Bernardi+13, Bernardi+14, Bernardi+17_ICL} for a detailed discussion of the implications of using different fits to photometric light profiles. \\

\section{Conclusions}
\label{Sect:conclusions}

 In this work we have used a semi-empirical approach to study three models of galaxy sizes, where the sizes of galaxies are linked to that of their haloes by means of the dynamical (the MMW model, eq. \ref{eq:MMW_tot}) or structural (the concentration model, eq. \ref{eq:concmodel}) properties of the dark matter halo in which they are hosted, or by a simple constant (the K13 model, eq. \ref{eq:K13model}) the origin of which is a priori unknown.\\
Our results can be summarized as follows:
\begin{enumerate}
\item The scatter in the K13 model must decrease for more massive galaxies, irrespective of galaxy morphology. This implies that most of the information on the size distributions of the most massive galaxies is fully dependent on the shape of the SMHM and hence on the physical processes that determine it.
\item In In the concentration model we find that $\gamma$ is  degenerate with the model intrinsic scatter $\sigma_{CM}$. This suggests that a lower $\sigma_{CM}$ may be needed to account for the width of the size functions, and that $\gamma$ must be low for massive galaxies. A lower $\sigma_{CM}$ might make the concentration model more fundamental than any other model studied here, however its physical origin  remains unclear. 
\item Similarly to other studies (\citealt{Huang+17}, \citealt{Lapi+18_disks}) we find that the normalization of both the K13 model and concentration model must be different for ETGs and LTGs. 
\item The classical disk model by MMW taken at face value overestimates the tails of the size and angular momentum distributions of disk galaxies, but is able to predict the correct normalizations of the scaling relations for LTGs. We discuss two scenarios that bring the model in better agreement with data:
\begin{itemize}
    \item We outline a model where only some values of $\lambda$ are physically acceptable. This model reproduces well the skeweness and tightness of the size functions of LTGs.
    \item Based on our constraints from the K13 model, we discuss a scenario where the link between the sizes of LTGs and their dark matter haloes is mediated by the stellar angular momentum, and where the halo spin parameter may not play any major role.
\end{itemize}
\end{enumerate}

 We also investigate whether our empirical constraints are reproduced in current cosmological models of galaxy formation and evolution. 
 \begin{enumerate}
     \item In the Rome SAM, which implements a purely hierarchical scenario where the MMW model is taken at face value, we find that mergers of LTGs alone are able to reproduce the dichotomy of the $R_e-R_h$ relation, but overestimate its scatter. We show that with a tighter scatter in the LTGs $R_e-R_h$ relation it could be possible to lower the inferred scatter in the sizes of ETGs at fixed halo radius to meet our semi-empirical constraints.
     \item In IllustrisTNG100, where both mergers and internal torques are at work, the morphological segregation in the $R_e-R_h$ plane is also present, with a scatter which is within the empirical constraints given in this work for LTGs, and somewhat higher for ETGs.
     \item We exploit the information about the dynamics available from IllustrisTNG100 to show that the scatter of the galaxy size-halo size connection of LTGs is consistent with being driven by the stellar specific angular momentum, which corroborates our empirical model based on the MMW model and the scatter of the K13 model.
\end{enumerate}

\section*{Acknowledgements}
We thank the referee for a constructive report that significantly improved this paper. L.Z. acknowledges support from the Science and Technologies Facilities Council and from the Royal Astronomical Society. F.S. acknowledges partial support from a Leverhulme Trust Research Fellowship and from the European Union's Horizon 2020 programme under the AHEAD project (grant agreement 654215).  AL acknowledges PRIN MIUR 2017 prot. 20173ML3WW002, Opening the ALMA window on the cosmic evolution ofgas, stars and supermassive black holes. L.Z. wishes to thank Avishai Dekel, Mauro Giavalisco, Helena Dom\'inguez-Sanchez,  Fangzhou Jiang, Gabriella De Lucia, Anna Zoldan, Vladimir Avila-Reese and Gigi Danese for helpful discussions and Dylan Nelson for guidance with the Illustris TNG simulation database.

\appendix

\section{Early Type galaxies}
\label{app:ETGs}
\begin{figure*}
    \centering
    \subfloat[\label{fig:ETGsK13}]{{\includegraphics[width=0.45\textwidth]{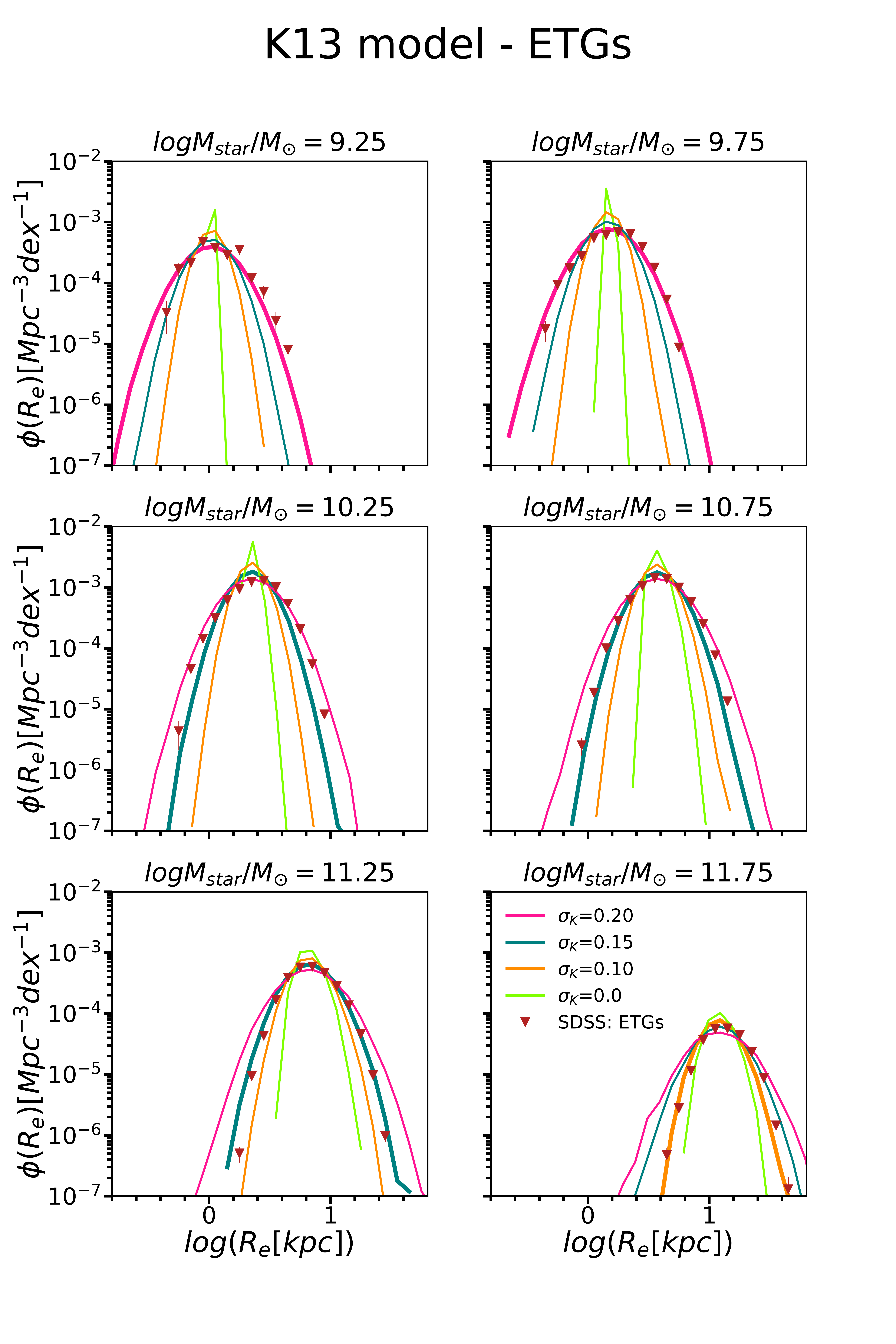} }}%
    \subfloat[\label{fig:ETGsConc}]{{\includegraphics[width=0.45\textwidth]{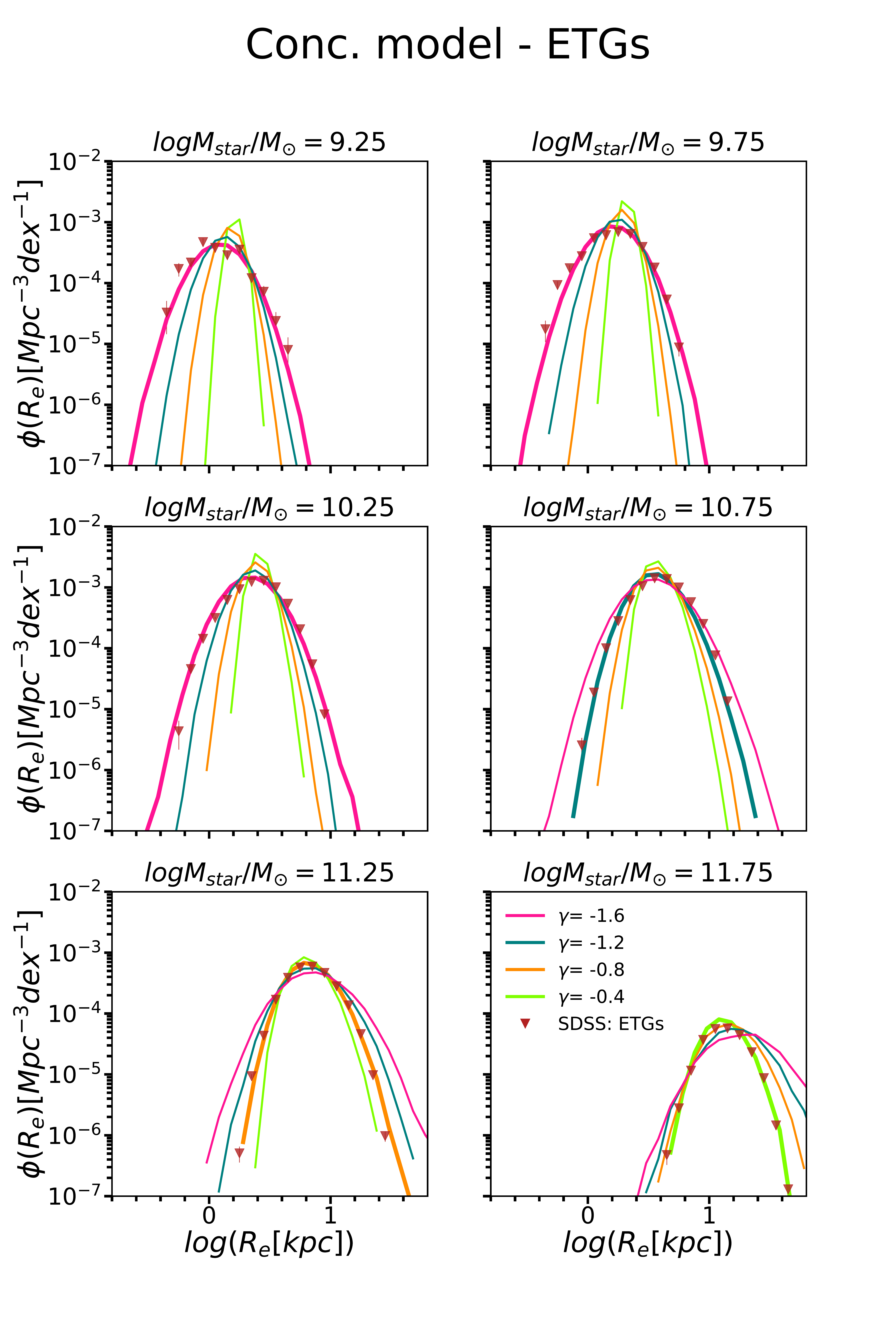} }}
    \caption{\emph{Left}:  Size functions from the K13 model (eq. \ref{eq:K13model}) for values of $\sigma_K=0.00,0.10,0.15,0.20$.  \emph{Right}: Size functions from the concentration model (eq. \ref{eq:concmodel}) for values of $\gamma=-1.6,-1.2,-0.8,-0.4$. Models that work best for a given stellar mass bin are highlighted in each panel by a thicker line.
 Data points are ETGs from the photo+morphological SDSS catalogues described in Section \ref{Sect:data}.}
    \label{fig:ETGs_K13conc}
\end{figure*}

Figures \ref{fig:ETGsK13} and \ref{fig:ETGsConc} show a comparison between data and the size functions $\phi(R_e)$ from our models (K13 and concentration models respectively). We bin both our model galaxies and data in bins of 0.5 dex in stellar mass.
In all the figures the model size functions are shifted to match the peaks of the observed distributions. The normalization of the different models in each stellar mass bin is reported in table \ref{table:LTGs}. The values in table \ref{table:ETGs} are not meant to be best fits, rather they only indicate that there is trend with stellar mass. \\
Similarly to what we did for LTGs, we report models for $\sigma_K=0.10,0.15,0.20$ and $\gamma=-1.6,-1.2,-0.8, -0.4$. In each panel of figures \ref{fig:ETGsConc} and \ref{fig:ETGsK13}  we highlight with a thicker line the parameter that seems to best reproduce observations, keeping in mind that this should not be considered as a fit to data. Qualitatively, we find that ETGs obey relations that are very similar to those of LTGs in terms of $\sigma_K$ and $\gamma$, with higher (lower) $\sigma_K$ ($\gamma$) for lower (higher) stellar masses.

\begin{table}

\begin{tabular}{c|c|c|c|c|c|c}
   $M_{star}$                            & 9.25  & 9.75  & 10.25 & 10.75 & 11.25 & 11.75 \\
   \hline
$A_k$  & 0.006 & 0.007 & 0.010 & 0.011 & 0.015 & 0.016 \\
\hline
$A_c$  & 0.012 & 0.013 & 0.013 & 0.013 & 0.013 & 0.013 
\end{tabular}
\caption{Values of $A_k$ and $A_c$ in different bins of $M_{star}$, for ETGs. Compare to table \ref{table:LTGs}.}
\label{table:ETGs}
\end{table}

 \section{Caveats}
\label{Sect:Caveats}
\begin{figure}
    \centering
    \includegraphics[width=0.5\textwidth]{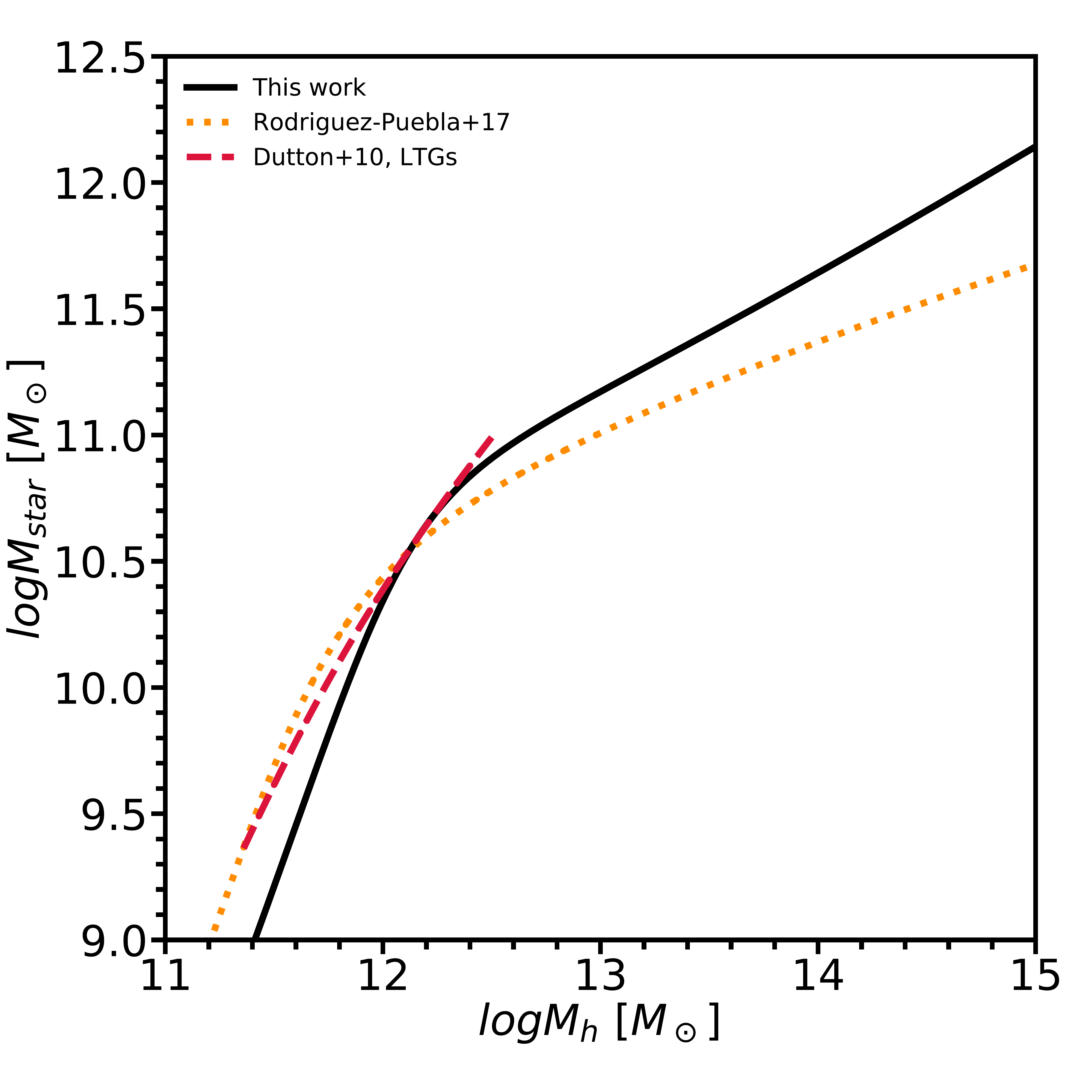}
    \caption{The SMHM fit to reproduce the SDSS SMF adopted in our work compared to that by \citet{Rodriguez-Puebla+17} and \citet{Dutton+10} for LTGs at $z\sim0.1$}
    \label{fig:SMHM}
\end{figure}
\subsection{Model assumptions}
\label{sect:Caveats_SMHM}
The backbone of our work is the SMHM and all our results depend on it. In figure \ref{fig:SMHM} we compare  our SMHM fitted to reproduce the SDSS M15/16 SMF to that of \citet{Rodriguez-Puebla+17} and of \citet{Dutton+10} for LTGs at $z\sim0.1$. Notably, the high mass slope of our SMHM is much steeper than that of \citet{Rodriguez-Puebla+17}.
Our estimate of the SMHM agrees with other studies where an improved photometry was used (\citealt{Shankar+17_bulgehalo}, \citealt{Grylls+19}). With a flatter high mass end slope in the SMHM the halo occupation distribution of massive galaxies would be wider (see Sect. \ref{Sect:scatters}).

 It can also be seen that the difference  between our SMHM and that of \citet{Dutton+10} for LTGs is not critical. In the \citet{Dutton+10} SMHM LTGs tend to live in haloes $\approx0.2$ dex less massive than in our determination of the SMHM, which would correspond to $\approx$5\% difference in halo size, which provides only a very minor corrections to our results.\\
 We also point out that some studies have reported the intrinsic scatter in the SMHM to be larger at lower halo masses (e.g. \citealt{Moster+18}, \citealt{Rodriguez-Puebla+15}), but we have adopted a constant $\sigma_{SMHM}=0.16$. Since larger $\sigma_{SMHM}$ would mean a larger scatter in the derived size distribution (fig. \ref{fig:cartoon_scatter}, this would exacerbate the tension between data and the MMW model, even after accounting for only the physically acceptable values of $\lambda$.\\
We note that the self gravity of baryons may lead to important modifications in the structure of the halo, as  suggested  by several works (\citealt{Desmond&Wechsler2017}, \citealt{Desmond+15}, \citealt{Gnedin+11}, \citealt{Blumenthal+86}, \citealt{Shankar+17_bulgehalo}). For example, \citet{Jiang+17} have shown that in a dark matter only simulation matched to a complementary hydrodynamical simulation the correlation of galaxy and halo spin is much less strong than in the hydrodynamical simulation in itself, proving that the effect of baryons in the inner regions of the halo may be crucial. The lack of this kind of information in our approach could potentially affect our conclusions.\\
It is also important to note that some authors suggest that LTGs live preferentially in haloes with lower concentration ( e.g. \citealt{Wojtak&Mamon2013}, \citealt{Desmond+17_RAR}). This would mean that the distribution of concentration at fixed halo mass for LTGs is tighter than the full $c-M_h$ relation, which would account for less scatter in the observed size functions. In such case, if $\gamma\sim -0.7$ is adopted, a larger intrinsic scatter in the concentration model would be needed. Whether this would still be lower than that in the K13 model remains to be seen, perhaps in the context of conditional abundance matching (e.g., \citealt{Hearin+17_halotools}).

\subsection{The role of projection effects}
\label{app:deproj}
\begin{figure}
    \centering
    \includegraphics[width=0.5\textwidth]{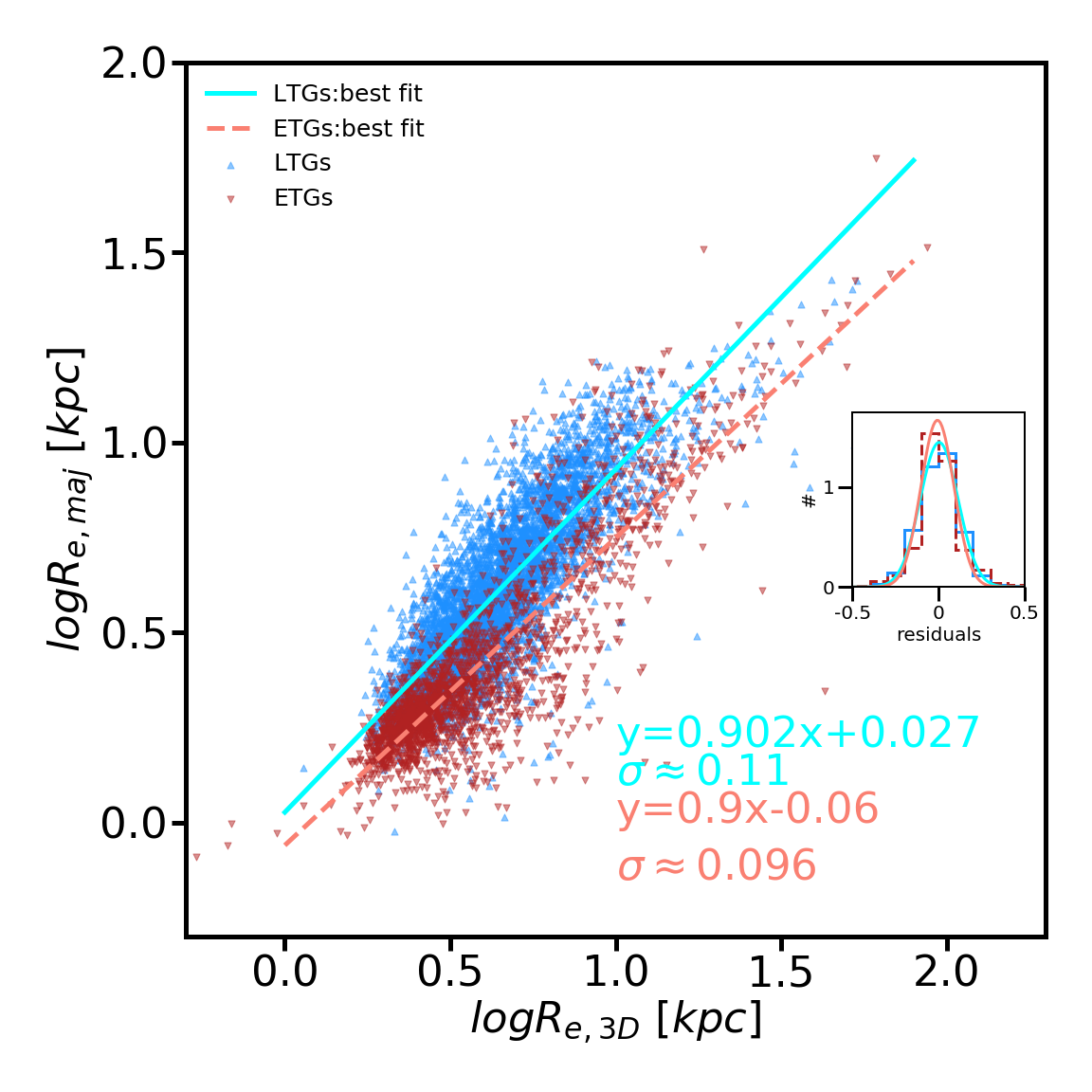}
    \caption{Correlation between 3D physical size and the semimajor axis sizes from \texttt{statmorph} (\citealt{Rodriguez-Gomez+19}, \citealt{Huertas-Company+19}) of galaxies in Illustris TNG morphologically classified as ETGs and LTGs using the threshold $P(Late)=0.5$. The flag \texttt{flag\textunderscore sersic} has been enforced to ensure that only good photometric S\'ersic fits are used. Red downward triangles and blue upward triangles indicate ETGs and LTGs respectively, while the solid cyan and salmon dashed lines are the best linear fit to the relations. The inset shows the distribution of residuals around the best fit for each relation, where the best fit gaussian to the residuals has been superimposed in both cases.}
    \label{fig:projection_effects_TNG}
\end{figure}

In our model we link the halo size directly to the observable 2D effective radius $R_e\equiv R_{e,2D}$, which is a projection of the true galaxy shape on the sky. However it is the physical half-light radius $R_{e,3D}$ the quantity that should be physically linked to $R_h$, which  may be different from $R_e$. Indeed, Jiang et al. (in prep) study the relation between $R_e$ and intrinsic 3D sizes, and find that a considerable scatter in the relation is present at fixed $R_{e,3D}$ and intrinsic shape, depending on the line of sight. Therefore the intrinsic scatter in the size distributions would be even tighter and would constitute a further challenge for models of galaxy formation and  evolution. \\
 To explore how projection effects affect our analysis of Illustris TNG, we have used the catalog of optical morphologies and photometric mock observations of IllustrisTNG100 presented in \citet{Huertas-Company+19} and briefly introduced in Section \ref{Sect:discussion_stateoftheart} and we plot $R_{e,maj}$ from the mock observations against the intrinsic 3D size $R_{e,3D}$ in Figure \ref{fig:projection_effects_TNG}. It can be seen that the measured $R_{e,maj}$ for  LTGs (ETGs) are only about $\sim 0.03$ ($0.06$) \ dex higher (lower) than their physical size, while the slope of the correlation is close to 1 in both cases. Interestingly, the dispersion of the residuals of both the $R_{e,maj}-R_{e,3D}$ relations are quite small, of the order of $\sim 0.1$ dex. This is even more striking in the light of the fact that the estimate of $R_{e,maj}$ is prone to both projection effects (as galaxies are mock observed along random lines of sight) and photometric errors. Based on the analysis above and on the fact that galaxy morphologies in IllustrisTNG100 are reasonably well reproduced \citep{Huertas-Company+19} we conclude that projection effects may not strongly bias the comparison between observations and models that predict the 3D sizes of galaxies.

\bibliographystyle{mnras}
\interlinepenalty=1000000
\bibliography{papersize.bib} 

\begin{thebibliography}{}
\makeatletter
\relax
\def\mn@urlcharsother{\let\do\@makeother \do\$\do\&\do\#\do\^\do\_\do\%\do\~}
\def\mn@doi{\begingroup\mn@urlcharsother \@ifnextchar [ {\mn@doi@}
  {\mn@doi@[]}}
\def\mn@doi@[#1]#2{\def\@tempa{#1}\ifx\@tempa\@empty \href
  {http://dx.doi.org/#2} {doi:#2}\else \href {http://dx.doi.org/#2} {#1}\fi
  \endgroup}
\def\mn@eprint#1#2{\mn@eprint@#1:#2::\@nil}
\def\mn@eprint@arXiv#1{\href {http://arxiv.org/abs/#1} {{\tt arXiv:#1}}}
\def\mn@eprint@dblp#1{\href {http://dblp.uni-trier.de/rec/bibtex/#1.xml}
  {dblp:#1}}
\def\mn@eprint@#1:#2:#3:#4\@nil{\def\@tempa {#1}\def\@tempb {#2}\def\@tempc
  {#3}\ifx \@tempc \@empty \let \@tempc \@tempb \let \@tempb \@tempa \fi \ifx
  \@tempb \@empty \def\@tempb {arXiv}\fi \@ifundefined
  {mn@eprint@\@tempb}{\@tempb:\@tempc}{\expandafter \expandafter \csname
  mn@eprint@\@tempb\endcsname \expandafter{\@tempc}}}

\bibitem[\protect\citeauthoryear{{Abazajian} et~al.,}{{Abazajian}
  et~al.}{2009}]{Abazajian+09}
{Abazajian} K.~N.,  et~al., 2009, \mn@doi [\apjs]
  {10.1088/0067-0049/182/2/543}, \href
  {http://adsabs.harvard.edu/abs/2009ApJS..182..543A} {182, 543}

\bibitem[\protect\citeauthoryear{{Aquino-Ort{\'\i}z}
  et~al.,}{{Aquino-Ort{\'\i}z} et~al.}{2018}]{Aquino-Ortiz+19}
{Aquino-Ort{\'\i}z} E.,  et~al., 2018, \mn@doi [\mnras]
  {10.1093/mnras/sty1522}, \href
  {https://ui.adsabs.harvard.edu/abs/2018MNRAS.479.2133A} {479, 2133}

\bibitem[\protect\citeauthoryear{{Baes}, {Verstappen}, {De Looze}, {Fritz},
  {Saftly}, {Vidal P{\'e}rez}, {Stalevski}  \& {Valcke}}{{Baes}
  et~al.}{2011}]{Baes+11_SKIRT}
{Baes} M.,  {Verstappen} J.,  {De Looze} I.,  {Fritz} J.,  {Saftly} W.,  {Vidal
  P{\'e}rez} E.,  {Stalevski} M.,   {Valcke} S.,  2011, \mn@doi [\apjs]
  {10.1088/0067-0049/196/2/22}, \href
  {https://ui.adsabs.harvard.edu/abs/2011ApJS..196...22B} {196, 22}

\bibitem[\protect\citeauthoryear{{Behroozi}, {Wechsler}  \&
  {Conroy}}{{Behroozi} et~al.}{2013}]{Behroozi+13}
{Behroozi} P.~S.,  {Wechsler} R.~H.,   {Conroy} C.,  2013, \mn@doi [\apj]
  {10.1088/0004-637X/770/1/57}, \href
  {http://adsabs.harvard.edu/abs/2013ApJ...770...57B} {770, 57}

\bibitem[\protect\citeauthoryear{{Behroozi}, {Wechsler}, {Hearin}  \&
  {Conroy}}{{Behroozi} et~al.}{2018a}]{Behroozi+18}
{Behroozi} P.,  {Wechsler} R.,  {Hearin} A.,   {Conroy} C.,  2018a, preprint,
  \href {http://adsabs.harvard.edu/abs/2018arXiv180607893B} {} (\mn@eprint
  {arXiv} {1806.07893})

\bibitem[\protect\citeauthoryear{{Behroozi}, {Wechsler}, {Hearin}  \&
  {Conroy}}{{Behroozi} et~al.}{2018b}]{Berhoozi+18}
{Behroozi} P.,  {Wechsler} R.,  {Hearin} A.,   {Conroy} C.,  2018b, preprint,
  \href {http://adsabs.harvard.edu/abs/2018arXiv180607893B} {} (\mn@eprint
  {arXiv} {1806.07893})

\bibitem[\protect\citeauthoryear{{Bernardi}, {Shankar}, {Hyde}, {Mei},
  {Marulli}  \& {Sheth}}{{Bernardi} et~al.}{2010}]{Bernardi+10}
{Bernardi} M.,  {Shankar} F.,  {Hyde} J.~B.,  {Mei} S.,  {Marulli} F.,
  {Sheth} R.~K.,  2010, \mn@doi [\mnras] {10.1111/j.1365-2966.2010.16425.x},
  \href {https://ui.adsabs.harvard.edu/abs/2010MNRAS.404.2087B} {404, 2087}

\bibitem[\protect\citeauthoryear{{Bernardi}, {Meert}, {Sheth}, {Vikram},
  {Huertas-Company}, {Mei}  \& {Shankar}}{{Bernardi}
  et~al.}{2013}]{Bernardi+13}
{Bernardi} M.,  {Meert} A.,  {Sheth} R.~K.,  {Vikram} V.,  {Huertas-Company}
  M.,  {Mei} S.,   {Shankar} F.,  2013, \mn@doi [\mnras]
  {10.1093/mnras/stt1607}, \href
  {http://adsabs.harvard.edu/abs/2013MNRAS.436..697B} {436, 697}

\bibitem[\protect\citeauthoryear{{Bernardi}, {Meert}, {Vikram},
  {Huertas-Company}, {Mei}, {Shankar}  \& {Sheth}}{{Bernardi}
  et~al.}{2014}]{Bernardi+14}
{Bernardi} M.,  {Meert} A.,  {Vikram} V.,  {Huertas-Company} M.,  {Mei} S.,
  {Shankar} F.,   {Sheth} R.~K.,  2014, \mn@doi [\mnras]
  {10.1093/mnras/stu1106}, \href
  {http://adsabs.harvard.edu/abs/2014MNRAS.443..874B} {443, 874}

\bibitem[\protect\citeauthoryear{{Bernardi}, {Meert}, {Sheth}, {Fischer},
  {Huertas-Company}, {Maraston}, {Shankar}  \& {Vikram}}{{Bernardi}
  et~al.}{2017a}]{Bernardi+17_highmassend}
{Bernardi} M.,  {Meert} A.,  {Sheth} R.~K.,  {Fischer} J.~L.,
  {Huertas-Company} M.,  {Maraston} C.,  {Shankar} F.,   {Vikram} V.,  2017a,
  \mn@doi [\mnras] {10.1093/mnras/stx176}, \href
  {https://ui.adsabs.harvard.edu/abs/2017MNRAS.467.2217B} {467, 2217}

\bibitem[\protect\citeauthoryear{{Bernardi}, {Fischer}, {Sheth}, {Meert},
  {Huertas-Company}, {Shankar}  \& {Vikram}}{{Bernardi}
  et~al.}{2017b}]{Bernardi+17_ICL}
{Bernardi} M.,  {Fischer} J.~L.,  {Sheth} R.~K.,  {Meert} A.,
  {Huertas-Company} M.,  {Shankar} F.,   {Vikram} V.,  2017b, \mn@doi [\mnras]
  {10.1093/mnras/stx677}, \href
  {https://ui.adsabs.harvard.edu/abs/2017MNRAS.468.2569B} {468, 2569}

\bibitem[\protect\citeauthoryear{{Blumenthal}, {Faber}, {Flores}  \&
  {Primack}}{{Blumenthal} et~al.}{1986}]{Blumenthal+86}
{Blumenthal} G.~R.,  {Faber} S.~M.,  {Flores} R.,   {Primack} J.~R.,  1986,
  \mn@doi [\apj] {10.1086/163867}, \href
  {http://adsabs.harvard.edu/abs/1986ApJ...301...27B} {301, 27}

\bibitem[\protect\citeauthoryear{Bottrell, Torrey, Simard  \& Ellison}{Bottrell
  et~al.}{2017a}]{Bottrell+17a}
Bottrell C.,  Torrey P.,  Simard L.,   Ellison S.~L.,  2017a, \mn@doi [MNRAS]
  {10.1093/mnras/stx017}, 467, 1033

\bibitem[\protect\citeauthoryear{{Bottrell}, {Torrey}, {Simard}  \&
  {Ellison}}{{Bottrell} et~al.}{2017b}]{Bottrell+17b}
{Bottrell} C.,  {Torrey} P.,  {Simard} L.,   {Ellison} S.~L.,  2017b, \mn@doi
  [\mnras] {10.1093/mnras/stx276}, \href
  {https://ui.adsabs.harvard.edu/abs/2017MNRAS.467.2879B} {467, 2879}

\bibitem[\protect\citeauthoryear{{Bottrell} et~al.,}{{Bottrell}
  et~al.}{2019}]{Bottrell+19}
{Bottrell} C.,  et~al., 2019, \mn@doi [\mnras] {10.1093/mnras/stz2934}, \href
  {https://ui.adsabs.harvard.edu/abs/2019MNRAS.tmp.2541B} {p.~2541}

\bibitem[\protect\citeauthoryear{{Bryan} \& {Norman}}{{Bryan} \&
  {Norman}}{1998}]{Bryan&Norman98}
{Bryan} G.~L.,  {Norman} M.~L.,  1998, \mn@doi [\apj] {10.1086/305262}, \href
  {http://adsabs.harvard.edu/abs/1998ApJ...495...80B} {495, 80}

\bibitem[\protect\citeauthoryear{{Bullock}, {Dekel}, {Kolatt}, {Kravtsov},
  {Klypin}, {Porciani}  \& {Primack}}{{Bullock} et~al.}{2001}]{Bullock+01}
{Bullock} J.~S.,  {Dekel} A.,  {Kolatt} T.~S.,  {Kravtsov} A.~V.,  {Klypin}
  A.~A.,  {Porciani} C.,   {Primack} J.~R.,  2001, \mn@doi [\apj]
  {10.1086/321477}, \href {http://adsabs.harvard.edu/abs/2001ApJ...555..240B}
  {555, 240}

\bibitem[\protect\citeauthoryear{{Burkert} et~al.,}{{Burkert}
  et~al.}{2016}]{Burkert+16}
{Burkert} A.,  et~al., 2016, \mn@doi [\apj] {10.3847/0004-637X/826/2/214},
  \href {http://adsabs.harvard.edu/abs/2016ApJ...826..214B} {826, 214}

\bibitem[\protect\citeauthoryear{{Camps} \& {Baes}}{{Camps} \&
  {Baes}}{2015}]{Camps&Baes2015}
{Camps} P.,  {Baes} M.,  2015, \mn@doi [Astronomy and Computing]
  {10.1016/j.ascom.2014.10.004}, \href
  {https://ui.adsabs.harvard.edu/abs/2015A&C.....9...20C} {9, 20}

\bibitem[\protect\citeauthoryear{{Cappellari}}{{Cappellari}}{2016}]{Cappellari2016_review}
{Cappellari} M.,  2016, \mn@doi [\araa] {10.1146/annurev-astro-082214-122432},
  \href {http://adsabs.harvard.edu/abs/2016ARA%26A..54..597C} {54, 597}

\bibitem[\protect\citeauthoryear{{Cervantes-Sodi}, {Li}, {Park}  \&
  {Wang}}{{Cervantes-Sodi} et~al.}{2013}]{Cervantes-Sodi+13}
{Cervantes-Sodi} B.,  {Li} C.,  {Park} C.,   {Wang} L.,  2013, \mn@doi [\apj]
  {10.1088/0004-637X/775/1/19}, \href
  {http://adsabs.harvard.edu/abs/2013ApJ...775...19C} {775, 19}

\bibitem[\protect\citeauthoryear{{Ceverino}, {Dekel}  \& {Bournaud}}{{Ceverino}
  et~al.}{2010}]{Ceverino+10}
{Ceverino} D.,  {Dekel} A.,   {Bournaud} F.,  2010, \mn@doi [\mnras]
  {10.1111/j.1365-2966.2010.16433.x}, \href
  {http://adsabs.harvard.edu/abs/2010MNRAS.404.2151C} {404, 2151}

\bibitem[\protect\citeauthoryear{{Ceverino}, {Klypin}, {Klimek},
  {Trujillo-Gomez}, {Churchill}, {Primack}  \& {Dekel}}{{Ceverino}
  et~al.}{2014}]{Ceverino+14}
{Ceverino} D.,  {Klypin} A.,  {Klimek} E.~S.,  {Trujillo-Gomez} S.,
  {Churchill} C.~W.,  {Primack} J.,   {Dekel} A.,  2014, \mn@doi [\mnras]
  {10.1093/mnras/stu956}, \href
  {http://adsabs.harvard.edu/abs/2014MNRAS.442.1545C} {442, 1545}

\bibitem[\protect\citeauthoryear{{Christodoulou}, {Shlosman}  \&
  {Tohline}}{{Christodoulou} et~al.}{1995}]{Christodoulou+95}
{Christodoulou} D.~M.,  {Shlosman} I.,   {Tohline} J.~E.,  1995, \mn@doi [\apj]
  {10.1086/175547}, \href {http://adsabs.harvard.edu/abs/1995ApJ...443..551C}
  {443, 551}

\bibitem[\protect\citeauthoryear{{Cole}, {Lacey}, {Baugh}  \& {Frenk}}{{Cole}
  et~al.}{2000}]{Cole+00}
{Cole} S.,  {Lacey} C.~G.,  {Baugh} C.~M.,   {Frenk} C.~S.,  2000, \mn@doi
  [\mnras] {10.1046/j.1365-8711.2000.03879.x}, \href
  {http://adsabs.harvard.edu/abs/2000MNRAS.319..168C} {319, 168}

\bibitem[\protect\citeauthoryear{{Covington}, {Primack}, {Porter}, {Croton},
  {Somerville}  \& {Dekel}}{{Covington} et~al.}{2011}]{Covington+11}
{Covington} M.~D.,  {Primack} J.~R.,  {Porter} L.~A.,  {Croton} D.~J.,
  {Somerville} R.~S.,   {Dekel} A.,  2011, \mn@doi [\mnras]
  {10.1111/j.1365-2966.2011.18926.x}, \href
  {https://ui.adsabs.harvard.edu/abs/2011MNRAS.415.3135C} {415, 3135}

\bibitem[\protect\citeauthoryear{{Danovich}, {Dekel}, {Hahn}, {Ceverino}  \&
  {Primack}}{{Danovich} et~al.}{2015}]{Danovich+15}
{Danovich} M.,  {Dekel} A.,  {Hahn} O.,  {Ceverino} D.,   {Primack} J.,  2015,
  \mn@doi [\mnras] {10.1093/mnras/stv270}, \href
  {http://adsabs.harvard.edu/abs/2015MNRAS.449.2087D} {449, 2087}

\bibitem[\protect\citeauthoryear{{De Lucia}, {Fontanot}, {Wilman}  \&
  {Monaco}}{{De Lucia} et~al.}{2011}]{DeLucia+11_bulgeFormation}
{De Lucia} G.,  {Fontanot} F.,  {Wilman} D.,   {Monaco} P.,  2011, \mn@doi
  [\mnras] {10.1111/j.1365-2966.2011.18475.x}, \href
  {http://adsabs.harvard.edu/abs/2011MNRAS.414.1439D} {414, 1439}

\bibitem[\protect\citeauthoryear{{De Lucia}, {Tornatore}, {Frenk}, {Helmi},
  {Navarro}  \& {White}}{{De Lucia} et~al.}{2014}]{DeLucia+14}
{De Lucia} G.,  {Tornatore} L.,  {Frenk} C.~S.,  {Helmi} A.,  {Navarro} J.~F.,
   {White} S. D.~M.,  2014, \mn@doi [\mnras] {10.1093/mnras/stu1752}, \href
  {https://ui.adsabs.harvard.edu/abs/2014MNRAS.445..970D} {445, 970}

\bibitem[\protect\citeauthoryear{{Desmond}}{{Desmond}}{2017}]{Desmond+17_RAR}
{Desmond} H.,  2017, \mn@doi [\mnras] {10.1093/mnras/stw2571}, \href
  {http://adsabs.harvard.edu/abs/2017MNRAS.464.4160D} {464, 4160}

\bibitem[\protect\citeauthoryear{{Desmond} \& {Wechsler}}{{Desmond} \&
  {Wechsler}}{2015}]{Desmond+15}
{Desmond} H.,  {Wechsler} R.~H.,  2015, \mn@doi [\mnras]
  {10.1093/mnras/stv1978}, \href
  {http://adsabs.harvard.edu/abs/2015MNRAS.454..322D} {454, 322}

\bibitem[\protect\citeauthoryear{{Desmond} \& {Wechsler}}{{Desmond} \&
  {Wechsler}}{2017}]{Desmond&Wechsler2017}
{Desmond} H.,  {Wechsler} R.~H.,  2017, \mn@doi [\mnras]
  {10.1093/mnras/stw2804}, \href
  {http://adsabs.harvard.edu/abs/2017MNRAS.465..820D} {465, 820}

\bibitem[\protect\citeauthoryear{{Desmond}, {Katz}, {Lelli}  \&
  {McGaugh}}{{Desmond} et~al.}{2018}]{Desmond+18_SPARC}
{Desmond} H.,  {Katz} H.,  {Lelli} F.,   {McGaugh} S.,  2018, preprint, \href
  {http://adsabs.harvard.edu/abs/2018arXiv180800271D} {} (\mn@eprint {arXiv}
  {1808.00271})

\bibitem[\protect\citeauthoryear{{Diemer}}{{Diemer}}{2017}]{DiemerColossus}
{Diemer} B.,  2017, preprint, \href
  {http://adsabs.harvard.edu/abs/2017arXiv171204512D} {} (\mn@eprint {arXiv}
  {1712.04512})

\bibitem[\protect\citeauthoryear{{Diemer}, {More}  \& {Kravtsov}}{{Diemer}
  et~al.}{2013}]{Diemer+13_pseudo}
{Diemer} B.,  {More} S.,   {Kravtsov} A.~V.,  2013, \mn@doi [\apj]
  {10.1088/0004-637X/766/1/25}, \href
  {http://adsabs.harvard.edu/abs/2013ApJ...766...25D} {766, 25}

\bibitem[\protect\citeauthoryear{{Dom{\'{\i}}nguez S{\'a}nchez},
  {Huertas-Company}, {Bernardi}, {Tuccillo}  \& {Fischer}}{{Dom{\'{\i}}nguez
  S{\'a}nchez} et~al.}{2018}]{Dominguez-Sanchez+18}
{Dom{\'{\i}}nguez S{\'a}nchez} H.,  {Huertas-Company} M.,  {Bernardi} M.,
  {Tuccillo} D.,   {Fischer} J.~L.,  2018, \mn@doi [\mnras]
  {10.1093/mnras/sty338}, \href
  {http://adsabs.harvard.edu/abs/2018MNRAS.476.3661D} {476, 3661}

\bibitem[\protect\citeauthoryear{{Driver} et~al.,}{{Driver}
  et~al.}{2011}]{Driver+11_GAMA}
{Driver} S.~P.,  et~al., 2011, \mn@doi [\mnras]
  {10.1111/j.1365-2966.2010.18188.x}, \href
  {http://adsabs.harvard.edu/abs/2011MNRAS.413..971D} {413, 971}

\bibitem[\protect\citeauthoryear{{Duckworth}, {Tojeiro}  \&
  {Kraljic}}{{Duckworth} et~al.}{2019}]{Duckworth+19_misalignment}
{Duckworth} C.,  {Tojeiro} R.,   {Kraljic} K.,  2019, arXiv e-prints, \href
  {https://ui.adsabs.harvard.edu/abs/2019arXiv191010744D} {p. arXiv:1910.10744}

\bibitem[\protect\citeauthoryear{{Dutton} \& {Macci{\`o}}}{{Dutton} \&
  {Macci{\`o}}}{2014}]{Dutton&Maccio2014}
{Dutton} A.~A.,  {Macci{\`o}} A.~V.,  2014, \mn@doi [\mnras]
  {10.1093/mnras/stu742}, \href
  {http://adsabs.harvard.edu/abs/2014MNRAS.441.3359D} {441, 3359}

\bibitem[\protect\citeauthoryear{{Dutton}, {Conroy}, {van den Bosch}, {Prada}
  \& {More}}{{Dutton} et~al.}{2010}]{Dutton+10}
{Dutton} A.~A.,  {Conroy} C.,  {van den Bosch} F.~C.,  {Prada} F.,   {More} S.,
   2010, \mn@doi [\mnras] {10.1111/j.1365-2966.2010.16911.x}, \href
  {http://adsabs.harvard.edu/abs/2010MNRAS.407....2D} {407, 2}

\bibitem[\protect\citeauthoryear{{Efstathiou}, {Lake}  \&
  {Negroponte}}{{Efstathiou} et~al.}{1982}]{Efstathiou+82}
{Efstathiou} G.,  {Lake} G.,   {Negroponte} J.,  1982, \mn@doi [\mnras]
  {10.1093/mnras/199.4.1069}, \href
  {http://adsabs.harvard.edu/abs/1982MNRAS.199.1069E} {199, 1069}

\bibitem[\protect\citeauthoryear{{Erfanianfar} et~al.,}{{Erfanianfar}
  et~al.}{2019}]{Erfanianfar+19}
{Erfanianfar} G.,  et~al., 2019, arXiv e-prints, \href
  {https://ui.adsabs.harvard.edu/abs/2019arXiv190801559E} {p. arXiv:1908.01559}

\bibitem[\protect\citeauthoryear{{Fan}, {Lapi}, {De Zotti}  \& {Danese}}{{Fan}
  et~al.}{2008}]{Fan+08}
{Fan} L.,  {Lapi} A.,  {De Zotti} G.,   {Danese} L.,  2008, \mn@doi [\apjl]
  {10.1086/595784}, \href {http://adsabs.harvard.edu/abs/2008ApJ...689L.101F}
  {689, L101}

\bibitem[\protect\citeauthoryear{{Fan}, {Lapi}, {Bressan}, {Bernardi}, {De
  Zotti}  \& {Danese}}{{Fan} et~al.}{2010}]{Fan+10}
{Fan} L.,  {Lapi} A.,  {Bressan} A.,  {Bernardi} M.,  {De Zotti} G.,   {Danese}
  L.,  2010, \mn@doi [\apj] {10.1088/0004-637X/718/2/1460}, \href
  {https://ui.adsabs.harvard.edu/abs/2010ApJ...718.1460F} {718, 1460}

\bibitem[\protect\citeauthoryear{{Faucher-Gigu{\`e}re}, {Kere{\v{s}}}  \&
  {Ma}}{{Faucher-Gigu{\`e}re} et~al.}{2011}]{Faqucher-giguere+11}
{Faucher-Gigu{\`e}re} C.-A.,  {Kere{\v{s}}} D.,   {Ma} C.-P.,  2011, \mn@doi
  [\mnras] {10.1111/j.1365-2966.2011.19457.x}, \href
  {https://ui.adsabs.harvard.edu/abs/2011MNRAS.417.2982F} {417, 2982}

\bibitem[\protect\citeauthoryear{{Fischer}, {Bernardi}  \& {Meert}}{{Fischer}
  et~al.}{2017}]{Fischer+17_truncation}
{Fischer} J.~L.,  {Bernardi} M.,   {Meert} A.,  2017, \mn@doi [\mnras]
  {10.1093/mnras/stx136}, \href
  {https://ui.adsabs.harvard.edu/abs/2017MNRAS.467..490F} {467, 490}

\bibitem[\protect\citeauthoryear{{Foreman-Mackey}, {Hogg}, {Lang}  \&
  {Goodman}}{{Foreman-Mackey} et~al.}{2013}]{Foreman-Mackey+13}
{Foreman-Mackey} D.,  {Hogg} D.~W.,  {Lang} D.,   {Goodman} J.,  2013, \mn@doi
  [\pasp] {10.1086/670067}, \href
  {http://adsabs.harvard.edu/abs/2013PASP..125..306F} {125, 306}

\bibitem[\protect\citeauthoryear{{Genel} et~al.,}{{Genel}
  et~al.}{2014}]{Genel+14}
{Genel} S.,  et~al., 2014, \mn@doi [\mnras] {10.1093/mnras/stu1654}, \href
  {http://adsabs.harvard.edu/abs/2014MNRAS.445..175G} {445, 175}

\bibitem[\protect\citeauthoryear{{Genel} et~al.,}{{Genel}
  et~al.}{2018}]{Genel+18}
{Genel} S.,  et~al., 2018, \mn@doi [\mnras] {10.1093/mnras/stx3078}, \href
  {http://adsabs.harvard.edu/abs/2018MNRAS.474.3976G} {474, 3976}

\bibitem[\protect\citeauthoryear{{Gnedin}, {Ceverino}, {Gnedin}, {Klypin},
  {Kravtsov}, {Levine}, {Nagai}  \& {Yepes}}{{Gnedin} et~al.}{2011}]{Gnedin+11}
{Gnedin} O.~Y.,  {Ceverino} D.,  {Gnedin} N.~Y.,  {Klypin} A.~A.,  {Kravtsov}
  A.~V.,  {Levine} R.,  {Nagai} D.,   {Yepes} G.,  2011, preprint, \href
  {http://adsabs.harvard.edu/abs/2011arXiv1108.5736G} {} (\mn@eprint {arXiv}
  {1108.5736})

\bibitem[\protect\citeauthoryear{{Grylls}, {Shankar}, {Zanisi}  \&
  {Bernardi}}{{Grylls} et~al.}{2019}]{Grylls+19}
{Grylls} P.~J.,  {Shankar} F.,  {Zanisi} L.,   {Bernardi} M.,  2019, \mn@doi
  [\mnras] {10.1093/mnras/sty3281}, \href
  {http://adsabs.harvard.edu/abs/2019MNRAS.483.2506G} {483, 2506}

\bibitem[\protect\citeauthoryear{{Guo} et~al.,}{{Guo} et~al.}{2011}]{Guo+11}
{Guo} Q.,  et~al., 2011, \mn@doi [\mnras] {10.1111/j.1365-2966.2010.18114.x},
  \href {http://adsabs.harvard.edu/abs/2011MNRAS.413..101G} {413, 101}

\bibitem[\protect\citeauthoryear{{Hearin}, {Behroozi}, {Kravtsov}  \&
  {Moster}}{{Hearin} et~al.}{2017a}]{Hearin+17}
{Hearin} A.,  {Behroozi} P.,  {Kravtsov} A.,   {Moster} B.,  2017a, preprint,
  \href {http://adsabs.harvard.edu/abs/2017arXiv171110500H} {} (\mn@eprint
  {arXiv} {1711.10500})

\bibitem[\protect\citeauthoryear{{Hearin} et~al.,}{{Hearin}
  et~al.}{2017b}]{Hearin+17_halotools}
{Hearin} A.~P.,  et~al., 2017b, \mn@doi [\aj] {10.3847/1538-3881/aa859f}, \href
  {http://adsabs.harvard.edu/abs/2017AJ....154..190H} {154, 190}

\bibitem[\protect\citeauthoryear{{Hirschmann}, {De Lucia}  \&
  {Fontanot}}{{Hirschmann} et~al.}{2016}]{Hirschmann+16}
{Hirschmann} M.,  {De Lucia} G.,   {Fontanot} F.,  2016, \mn@doi [\mnras]
  {10.1093/mnras/stw1318}, \href
  {https://ui.adsabs.harvard.edu/abs/2016MNRAS.461.1760H} {461, 1760}

\bibitem[\protect\citeauthoryear{{Huang} et~al.,}{{Huang}
  et~al.}{2017}]{Huang+17}
{Huang} K.-H.,  et~al., 2017, \mn@doi [\apj] {10.3847/1538-4357/aa62a6}, \href
  {http://adsabs.harvard.edu/abs/2017ApJ...838....6H} {838, 6}

\bibitem[\protect\citeauthoryear{{Hubble}}{{Hubble}}{1926}]{Hubble1926}
{Hubble} E.~P.,  1926, \mn@doi [\apj] {10.1086/143018}, \href
  {http://adsabs.harvard.edu/abs/1926ApJ....64..321H} {64}

\bibitem[\protect\citeauthoryear{{Huertas-Company} et~al.,}{{Huertas-Company}
  et~al.}{2019}]{Huertas-Company+19}
{Huertas-Company} M.,  et~al., 2019, arXiv e-prints, \href
  {http://adsabs.harvard.edu/abs/2019arXiv190307625H} {}

\bibitem[\protect\citeauthoryear{{Jiang} et~al.,}{{Jiang}
  et~al.}{2018}]{Jiang+17}
{Jiang} F.,  et~al., 2018, preprint, \href
  {http://adsabs.harvard.edu/abs/2018arXiv180407306J} {} (\mn@eprint {arXiv}
  {1804.07306})

\bibitem[\protect\citeauthoryear{{Kassin}, {Devriendt}, {Fall}, {de Jong},
  {Allgood}  \& {Primack}}{{Kassin}
  et~al.}{2012}]{Kassin+12_RadiusBaryonicCollapse}
{Kassin} S.~A.,  {Devriendt} J.,  {Fall} S.~M.,  {de Jong} R.~S.,  {Allgood}
  B.,   {Primack} J.~R.,  2012, \mn@doi [\mnras]
  {10.1111/j.1365-2966.2012.21219.x}, \href
  {http://adsabs.harvard.edu/abs/2012MNRAS.424..502K} {424, 502}

\bibitem[\protect\citeauthoryear{{Klypin}, {Trujillo-Gomez}  \&
  {Primack}}{{Klypin} et~al.}{2011}]{Kyplin+11_bolshoi}
{Klypin} A.~A.,  {Trujillo-Gomez} S.,   {Primack} J.,  2011, \mn@doi [\apj]
  {10.1088/0004-637X/740/2/102}, \href
  {http://adsabs.harvard.edu/abs/2011ApJ...740..102K} {740, 102}

\bibitem[\protect\citeauthoryear{{Klypin}, {Yepes}, {Gottl{\"o}ber}, {Prada}
  \& {He{\ss}}}{{Klypin} et~al.}{2016}]{Kyplin+16_multidark}
{Klypin} A.,  {Yepes} G.,  {Gottl{\"o}ber} S.,  {Prada} F.,   {He{\ss}} S.,
  2016, \mn@doi [\mnras] {10.1093/mnras/stw248}, \href
  {http://adsabs.harvard.edu/abs/2016MNRAS.457.4340K} {457, 4340}

\bibitem[\protect\citeauthoryear{{Koekemoer} et~al.,}{{Koekemoer}
  et~al.}{2011}]{Koekemoer+11}
{Koekemoer} A.~M.,  et~al., 2011, \mn@doi [\apjs] {10.1088/0067-0049/197/2/36},
  \href {http://adsabs.harvard.edu/abs/2011ApJS..197...36K} {197, 36}

\bibitem[\protect\citeauthoryear{{Komatsu} et~al.,}{{Komatsu}
  et~al.}{2011}]{Komatsu+11_cosmoparWmap}
{Komatsu} E.,  et~al., 2011, \mn@doi [\apjs] {10.1088/0067-0049/192/2/18},
  \href {http://adsabs.harvard.edu/abs/2011ApJS..192...18K} {192, 18}

\bibitem[\protect\citeauthoryear{{Kormendy}}{{Kormendy}}{2016}]{Kormendy2016}
{Kormendy} J.,  2016, in {Laurikainen} E.,  {Peletier} R.,   {Gadotti} D.,
  eds,  Astrophysics and Space Science Library Vol. 418, Galactic Bulges.
  p.~431 (\mn@eprint {arXiv} {1504.03330}),
  \mn@doi{10.1007/978-3-319-19378-6_16}

\bibitem[\protect\citeauthoryear{{Kravtsov}}{{Kravtsov}}{2013}]{Kravtsov2013}
{Kravtsov} A.~V.,  2013, \mn@doi [\apjl] {10.1088/2041-8205/764/2/L31}, \href
  {http://adsabs.harvard.edu/abs/2013ApJ...764L..31K} {764, L31}

\bibitem[\protect\citeauthoryear{{Kravtsov}, {Vikhlinin}  \&
  {Meshcheryakov}}{{Kravtsov} et~al.}{2018}]{Kravtsov+18}
{Kravtsov} A.~V.,  {Vikhlinin} A.~A.,   {Meshcheryakov} A.~V.,  2018, \mn@doi
  [Astronomy Letters] {10.1134/S1063773717120015}, \href
  {http://adsabs.harvard.edu/abs/2018AstL...44....8K} {44, 8}

\bibitem[\protect\citeauthoryear{{Lange} et~al.,}{{Lange}
  et~al.}{2015}]{Lange+15}
{Lange} R.,  et~al., 2015, \mn@doi [\mnras] {10.1093/mnras/stu2467}, \href
  {http://adsabs.harvard.edu/abs/2015MNRAS.447.2603L} {447, 2603}

\bibitem[\protect\citeauthoryear{{Lapi} et~al.,}{{Lapi}
  et~al.}{2018a}]{Lapi+18_ETGs}
{Lapi} A.,  et~al., 2018a, \mn@doi [\apj] {10.3847/1538-4357/aab6af}, \href
  {http://adsabs.harvard.edu/abs/2018ApJ...857...22L} {857, 22}

\bibitem[\protect\citeauthoryear{{Lapi}, {Salucci}  \& {Danese}}{{Lapi}
  et~al.}{2018b}]{Lapi+18_disks}
{Lapi} A.,  {Salucci} P.,   {Danese} L.,  2018b, \mn@doi [\apj]
  {10.3847/1538-4357/aabf35}, \href
  {http://adsabs.harvard.edu/abs/2018ApJ...859....2L} {859, 2}

\bibitem[\protect\citeauthoryear{{Lelli}, {McGaugh}  \& {Schombert}}{{Lelli}
  et~al.}{2016}]{Lelli+16_SPARC}
{Lelli} F.,  {McGaugh} S.~S.,   {Schombert} J.~M.,  2016, \mn@doi [\aj]
  {10.3847/0004-6256/152/6/157}, \href
  {http://adsabs.harvard.edu/abs/2016AJ....152..157L} {152, 157}

\bibitem[\protect\citeauthoryear{{Leroy}, {Walter}, {Brinks}, {Bigiel}, {de
  Blok}, {Madore}  \& {Thornley}}{{Leroy} et~al.}{2008}]{Leroy+08_THINGS}
{Leroy} A.~K.,  {Walter} F.,  {Brinks} E.,  {Bigiel} F.,  {de Blok} W.~J.~G.,
  {Madore} B.,   {Thornley} M.~D.,  2008, \mn@doi [\aj]
  {10.1088/0004-6256/136/6/2782}, \href
  {http://adsabs.harvard.edu/abs/2008AJ....136.2782L} {136, 2782}

\bibitem[\protect\citeauthoryear{{Liske} et~al.,}{{Liske}
  et~al.}{2015}]{Liske+15_GAMA}
{Liske} J.,  et~al., 2015, \mn@doi [\mnras] {10.1093/mnras/stv1436}, \href
  {http://adsabs.harvard.edu/abs/2015MNRAS.452.2087L} {452, 2087}

\bibitem[\protect\citeauthoryear{{Marasco}, {Fraternali}, {Posti}, {Ijtsma},
  {Di Teodoro}  \& {Oosterloo}}{{Marasco} et~al.}{2019}]{Marasco+19}
{Marasco} A.,  {Fraternali} F.,  {Posti} L.,  {Ijtsma} M.,  {Di Teodoro} E.~M.,
    {Oosterloo} T.,  2019, \mn@doi [\aap] {10.1051/0004-6361/201834456}, \href
  {http://adsabs.harvard.edu/abs/2019A%26A...621L...6M} {621, L6}

\bibitem[\protect\citeauthoryear{{Marinacci} et~al.,}{{Marinacci}
  et~al.}{2018}]{Marinacci+18}
{Marinacci} F.,  et~al., 2018, \mn@doi [\mnras] {10.1093/mnras/sty2206}, \href
  {https://ui.adsabs.harvard.edu/abs/2018MNRAS.480.5113M} {480, 5113}

\bibitem[\protect\citeauthoryear{{Matthee}, {Schaye}, {Crain}, {Schaller},
  {Bower}  \& {Theuns}}{{Matthee} et~al.}{2017}]{Matthee+17}
{Matthee} J.,  {Schaye} J.,  {Crain} R.~A.,  {Schaller} M.,  {Bower} R.,
  {Theuns} T.,  2017, \mn@doi [\mnras] {10.1093/mnras/stw2884}, \href
  {http://adsabs.harvard.edu/abs/2017MNRAS.465.2381M} {465, 2381}

\bibitem[\protect\citeauthoryear{{Meert}, {Vikram}  \& {Bernardi}}{{Meert}
  et~al.}{2015}]{Meert+15}
{Meert} A.,  {Vikram} V.,   {Bernardi} M.,  2015, \mn@doi [\mnras]
  {10.1093/mnras/stu2333}, \href
  {http://adsabs.harvard.edu/abs/2015MNRAS.446.3943M} {446, 3943}

\bibitem[\protect\citeauthoryear{{Meert}, {Vikram}  \& {Bernardi}}{{Meert}
  et~al.}{2016}]{Meert+16}
{Meert} A.,  {Vikram} V.,   {Bernardi} M.,  2016, \mn@doi [\mnras]
  {10.1093/mnras/stv2475}, \href
  {http://adsabs.harvard.edu/abs/2016MNRAS.455.2440M} {455, 2440}

\bibitem[\protect\citeauthoryear{{Menci}, {Fontana}, {Giallongo}  \&
  {Salimbeni}}{{Menci} et~al.}{2005}]{Menci+05}
{Menci} N.,  {Fontana} A.,  {Giallongo} E.,   {Salimbeni} S.,  2005, \mn@doi
  [\apj] {10.1086/432788}, \href
  {http://adsabs.harvard.edu/abs/2005ApJ...632...49M} {632, 49}

\bibitem[\protect\citeauthoryear{{Menci}, {Fiore}, {Puccetti}  \&
  {Cavaliere}}{{Menci} et~al.}{2008}]{Menci+08}
{Menci} N.,  {Fiore} F.,  {Puccetti} S.,   {Cavaliere} A.,  2008, \mn@doi
  [\apj] {10.1086/591438}, \href
  {http://adsabs.harvard.edu/abs/2008ApJ...686..219M} {686, 219}

\bibitem[\protect\citeauthoryear{{Menci}, {Gatti}, {Fiore}  \&
  {Lamastra}}{{Menci} et~al.}{2014}]{Menci+14}
{Menci} N.,  {Gatti} M.,  {Fiore} F.,   {Lamastra} A.,  2014, \mn@doi [\aap]
  {10.1051/0004-6361/201424217}, \href
  {http://adsabs.harvard.edu/abs/2014A%26A...569A..37M} {569, A37}

\bibitem[\protect\citeauthoryear{{Mendel}, {Simard}, {Palmer}, {Ellison}  \&
  {Patton}}{{Mendel} et~al.}{2014}]{Mendel+14}
{Mendel} J.~T.,  {Simard} L.,  {Palmer} M.,  {Ellison} S.~L.,   {Patton} D.~R.,
   2014, \mn@doi [\apjs] {10.1088/0067-0049/210/1/3}, \href
  {http://adsabs.harvard.edu/abs/2014ApJS..210....3M} {210, 3}

\bibitem[\protect\citeauthoryear{{Miller}, {van Dokkum}, {Mowla}  \& {van der
  Wel}}{{Miller} et~al.}{2019}]{Miller+19_R80}
{Miller} T.~B.,  {van Dokkum} P.,  {Mowla} L.,   {van der Wel} A.,  2019,
  \mn@doi [\apjl] {10.3847/2041-8213/ab0380}, \href
  {https://ui.adsabs.harvard.edu/abs/2019ApJ...872L..14M} {872, L14}

\bibitem[\protect\citeauthoryear{{Mo}, {Mao}  \& {White}}{{Mo}
  et~al.}{1998}]{MoMaoWhite98}
{Mo} H.~J.,  {Mao} S.,   {White} S.~D.~M.,  1998, \mn@doi [\mnras]
  {10.1046/j.1365-8711.1998.01227.x}, \href
  {http://adsabs.harvard.edu/abs/1998MNRAS.295..319M} {295, 319}

\bibitem[\protect\citeauthoryear{{More}, {van den Bosch}, {Cacciato}, {Skibba},
  {Mo}  \& {Yang}}{{More} et~al.}{2011}]{More+11}
{More} S.,  {van den Bosch} F.~C.,  {Cacciato} M.,  {Skibba} R.,  {Mo} H.~J.,
  {Yang} X.,  2011, \mn@doi [\mnras] {10.1111/j.1365-2966.2010.17436.x}, \href
  {http://adsabs.harvard.edu/abs/2011MNRAS.410..210M} {410, 210}

\bibitem[\protect\citeauthoryear{{Moster}, {Naab}  \& {White}}{{Moster}
  et~al.}{2013}]{Moster+13}
{Moster} B.~P.,  {Naab} T.,   {White} S.~D.~M.,  2013, \mn@doi [\mnras]
  {10.1093/mnras/sts261}, \href
  {http://adsabs.harvard.edu/abs/2013MNRAS.428.3121M} {428, 3121}

\bibitem[\protect\citeauthoryear{{Moster}, {Naab}  \& {White}}{{Moster}
  et~al.}{2018}]{Moster+18}
{Moster} B.~P.,  {Naab} T.,   {White} S.~D.~M.,  2018, \mn@doi [\mnras]
  {10.1093/mnras/sty655}, \href
  {http://adsabs.harvard.edu/abs/2018MNRAS.477.1822M} {477, 1822}

\bibitem[\protect\citeauthoryear{{Mowla}, {van der Wel}, {van Dokkum}  \&
  {Miller}}{{Mowla} et~al.}{2019}]{Mowla+19_SMHM_R80}
{Mowla} L.,  {van der Wel} A.,  {van Dokkum} P.,   {Miller} T.~B.,  2019,
  \mn@doi [\apjl] {10.3847/2041-8213/ab0379}, \href
  {https://ui.adsabs.harvard.edu/abs/2019ApJ...872L..13M} {872, L13}

\bibitem[\protect\citeauthoryear{{Naiman} et~al.,}{{Naiman}
  et~al.}{2018}]{Naiman+18}
{Naiman} J.~P.,  et~al., 2018, \mn@doi [\mnras] {10.1093/mnras/sty618}, \href
  {https://ui.adsabs.harvard.edu/abs/2018MNRAS.477.1206N} {477, 1206}

\bibitem[\protect\citeauthoryear{{Nair} \& {Abraham}}{{Nair} \&
  {Abraham}}{2010}]{Nair+10}
{Nair} P.~B.,  {Abraham} R.~G.,  2010, \mn@doi [\apjs]
  {10.1088/0067-0049/186/2/427}, \href
  {http://adsabs.harvard.edu/abs/2010ApJS..186..427N} {186, 427}

\bibitem[\protect\citeauthoryear{{Navarro}, {Frenk}  \& {White}}{{Navarro}
  et~al.}{1996}]{NavarroFrenkWhite95}
{Navarro} J.~F.,  {Frenk} C.~S.,   {White} S.~D.~M.,  1996, \mn@doi [\apj]
  {10.1086/177173}, \href {http://adsabs.harvard.edu/abs/1996ApJ...462..563N}
  {462, 563}

\bibitem[\protect\citeauthoryear{{Nelson} et~al.,}{{Nelson}
  et~al.}{2018}]{Nelson+18_datarelease}
{Nelson} D.,  et~al., 2018, arXiv e-prints, \href
  {http://adsabs.harvard.edu/abs/2018arXiv181205609N} {}

\bibitem[\protect\citeauthoryear{{Nipoti}, {Treu}, {Auger}  \&
  {Bolton}}{{Nipoti} et~al.}{2009}]{Nipoti+09}
{Nipoti} C.,  {Treu} T.,  {Auger} M.~W.,   {Bolton} A.~S.,  2009, \mn@doi
  [\apjl] {10.1088/0004-637X/706/1/L86}, \href
  {http://adsabs.harvard.edu/abs/2009ApJ...706L..86N} {706, L86}

\bibitem[\protect\citeauthoryear{{Nipoti}, {Treu}, {Leauthaud}, {Bundy},
  {Newman}  \& {Auger}}{{Nipoti} et~al.}{2012}]{Nipoti+12}
{Nipoti} C.,  {Treu} T.,  {Leauthaud} A.,  {Bundy} K.,  {Newman} A.~B.,
  {Auger} M.~W.,  2012, \mn@doi [\mnras] {10.1111/j.1365-2966.2012.20749.x},
  \href {http://adsabs.harvard.edu/abs/2012MNRAS.422.1714N} {422, 1714}

\bibitem[\protect\citeauthoryear{{Obreschkow} \& {Glazebrook}}{{Obreschkow} \&
  {Glazebrook}}{2014}]{Obreschkow&Glazebrook2014}
{Obreschkow} D.,  {Glazebrook} K.,  2014, \mn@doi [\apj]
  {10.1088/0004-637X/784/1/26}, \href
  {http://adsabs.harvard.edu/abs/2014ApJ...784...26O} {784, 26}

\bibitem[\protect\citeauthoryear{{Oser}, {Ostriker}, {Naab}, {Johansson}  \&
  {Burkert}}{{Oser} et~al.}{2010}]{Oser+10}
{Oser} L.,  {Ostriker} J.~P.,  {Naab} T.,  {Johansson} P.~H.,   {Burkert} A.,
  2010, \mn@doi [\apj] {10.1088/0004-637X/725/2/2312}, \href
  {https://ui.adsabs.harvard.edu/abs/2010ApJ...725.2312O} {725, 2312}

\bibitem[\protect\citeauthoryear{{Patel} et~al.,}{{Patel}
  et~al.}{2013}]{Patel+13_milkyWayLikeEvolution}
{Patel} S.~G.,  et~al., 2013, \mn@doi [\apj] {10.1088/0004-637X/778/2/115},
  \href {http://adsabs.harvard.edu/abs/2013ApJ...778..115P} {778, 115}

\bibitem[\protect\citeauthoryear{{Peebles}}{{Peebles}}{1969}]{Peebles1969}
{Peebles} P.~J.~E.,  1969, \mn@doi [\apj] {10.1086/149876}, \href
  {http://adsabs.harvard.edu/abs/1969ApJ...155..393P} {155, 393}

\bibitem[\protect\citeauthoryear{{Pillepich} et~al.,}{{Pillepich}
  et~al.}{2018a}]{Pillepich+18}
{Pillepich} A.,  et~al., 2018a, \mn@doi [\mnras] {10.1093/mnras/stx2656}, \href
  {http://adsabs.harvard.edu/abs/2018MNRAS.473.4077P} {473, 4077}

\bibitem[\protect\citeauthoryear{{Pillepich} et~al.,}{{Pillepich}
  et~al.}{2018b}]{Pillepich+18_TNGdescription}
{Pillepich} A.,  et~al., 2018b, \mn@doi [\mnras] {10.1093/mnras/stx2656}, \href
  {http://adsabs.harvard.edu/abs/2018MNRAS.473.4077P} {473, 4077}

\bibitem[\protect\citeauthoryear{{Planck Collaboration} et~al.,}{{Planck
  Collaboration} et~al.}{2016}]{Planck_cosmopar}
{Planck Collaboration} et~al., 2016, \mn@doi [\aap]
  {10.1051/0004-6361/201525830}, \href
  {http://adsabs.harvard.edu/abs/2016A%26A...594A..13P} {594, A13}

\bibitem[\protect\citeauthoryear{{Posti}, {Pezzulli}, {Fraternali}  \& {Di
  Teodoro}}{{Posti} et~al.}{2018a}]{Posti+18_angmom}
{Posti} L.,  {Pezzulli} G.,  {Fraternali} F.,   {Di Teodoro} E.~M.,  2018a,
  \mn@doi [\mnras] {10.1093/mnras/stx3168}, \href
  {http://adsabs.harvard.edu/abs/2018MNRAS.475..232P} {475, 232}

\bibitem[\protect\citeauthoryear{{Posti}, {Fraternali}, {Di Teodoro}  \&
  {Pezzulli}}{{Posti} et~al.}{2018b}]{Posti+18_letter}
{Posti} L.,  {Fraternali} F.,  {Di Teodoro} E.~M.,   {Pezzulli} G.,  2018b,
  \mn@doi [\aap] {10.1051/0004-6361/201833091}, \href
  {http://adsabs.harvard.edu/abs/2018A%26A...612L...6P} {612, L6}

\bibitem[\protect\citeauthoryear{{Prada}, {Klypin}, {Cuesta}, {Betancort-Rijo}
  \& {Primack}}{{Prada} et~al.}{2012}]{Prada+12}
{Prada} F.,  {Klypin} A.~A.,  {Cuesta} A.~J.,  {Betancort-Rijo} J.~E.,
  {Primack} J.,  2012, \mn@doi [\mnras] {10.1111/j.1365-2966.2012.21007.x},
  \href {http://adsabs.harvard.edu/abs/2012MNRAS.423.3018P} {423, 3018}

\bibitem[\protect\citeauthoryear{{Ragone-Figueroa} \&
  {Granato}}{{Ragone-Figueroa} \&
  {Granato}}{2011}]{Ragone-Figueroa&Granato2011}
{Ragone-Figueroa} C.,  {Granato} G.~L.,  2011, \mn@doi [\mnras]
  {10.1111/j.1365-2966.2011.18670.x}, \href
  {https://ui.adsabs.harvard.edu/abs/2011MNRAS.414.3690R} {414, 3690}

\bibitem[\protect\citeauthoryear{{Rodriguez-Gomez} et~al.,}{{Rodriguez-Gomez}
  et~al.}{2019}]{Rodriguez-Gomez+19}
{Rodriguez-Gomez} V.,  et~al., 2019, \mn@doi [\mnras] {10.1093/mnras/sty3345},
  \href {https://ui.adsabs.harvard.edu/abs/2019MNRAS.483.4140R} {483, 4140}

\bibitem[\protect\citeauthoryear{{Rodr{\'{\i}}guez-Puebla}, {Avila-Reese},
  {Yang}, {Foucaud}, {Drory}  \& {Jing}}{{Rodr{\'{\i}}guez-Puebla}
  et~al.}{2015}]{Rodriguez-Puebla+15}
{Rodr{\'{\i}}guez-Puebla} A.,  {Avila-Reese} V.,  {Yang} X.,  {Foucaud} S.,
  {Drory} N.,   {Jing} Y.~P.,  2015, \mn@doi [\apj]
  {10.1088/0004-637X/799/2/130}, \href
  {http://adsabs.harvard.edu/abs/2015ApJ...799..130R} {799, 130}

\bibitem[\protect\citeauthoryear{{Rodr{\'{\i}}guez-Puebla}, {Behroozi},
  {Primack}, {Klypin}, {Lee}  \& {Hellinger}}{{Rodr{\'{\i}}guez-Puebla}
  et~al.}{2016}]{Rodriguez-Puebla+16}
{Rodr{\'{\i}}guez-Puebla} A.,  {Behroozi} P.,  {Primack} J.,  {Klypin} A.,
  {Lee} C.,   {Hellinger} D.,  2016, \mn@doi [\mnras] {10.1093/mnras/stw1705},
  \href {http://adsabs.harvard.edu/abs/2016MNRAS.462..893R} {462, 893}

\bibitem[\protect\citeauthoryear{{Rodr{\'{\i}}guez-Puebla}, {Primack},
  {Avila-Reese}  \& {Faber}}{{Rodr{\'{\i}}guez-Puebla}
  et~al.}{2017}]{Rodriguez-Puebla+17}
{Rodr{\'{\i}}guez-Puebla} A.,  {Primack} J.~R.,  {Avila-Reese} V.,   {Faber}
  S.~M.,  2017, \mn@doi [\mnras] {10.1093/mnras/stx1172}, \href
  {http://adsabs.harvard.edu/abs/2017MNRAS.470..651R} {470, 651}

\bibitem[\protect\citeauthoryear{{Romanowsky} \& {Fall}}{{Romanowsky} \&
  {Fall}}{2012}]{Romanowsky&Fall2012}
{Romanowsky} A.~J.,  {Fall} S.~M.,  2012, \mn@doi [\apjs]
  {10.1088/0067-0049/203/2/17}, \href
  {http://adsabs.harvard.edu/abs/2012ApJS..203...17R} {203, 17}

\bibitem[\protect\citeauthoryear{{S{\'a}nchez} et~al.,}{{S{\'a}nchez}
  et~al.}{2012}]{Sanchez+12_CALIFA}
{S{\'a}nchez} S.~F.,  et~al., 2012, \mn@doi [\aap]
  {10.1051/0004-6361/201117353}, \href
  {https://ui.adsabs.harvard.edu/abs/2012A&A...538A...8S} {538, A8}

\bibitem[\protect\citeauthoryear{{S{\'a}nchez} et~al.,}{{S{\'a}nchez}
  et~al.}{2018}]{Sanchez+18_RMX}
{S{\'a}nchez} S.~F.,  et~al., 2018, \rmxaa, \href
  {https://ui.adsabs.harvard.edu/abs/2018RMxAA..54..217S} {54, 217}

\bibitem[\protect\citeauthoryear{{Shankar}, {Lapi}, {Salucci}, {De Zotti}  \&
  {Danese}}{{Shankar} et~al.}{2006}]{Shankar+06}
{Shankar} F.,  {Lapi} A.,  {Salucci} P.,  {De Zotti} G.,   {Danese} L.,  2006,
  \mn@doi [\apj] {10.1086/502794}, \href
  {http://adsabs.harvard.edu/abs/2006ApJ...643...14S} {643, 14}

\bibitem[\protect\citeauthoryear{{Shankar}, {Marulli}, {Bernardi},
  {Boylan-Kolchin}, {Dai}  \& {Khochfar}}{{Shankar}
  et~al.}{2010}]{Shankar+10_sizefunct}
{Shankar} F.,  {Marulli} F.,  {Bernardi} M.,  {Boylan-Kolchin} M.,  {Dai} X.,
  {Khochfar} S.,  2010, \mn@doi [\mnras] {10.1111/j.1365-2966.2010.16540.x},
  \href {http://adsabs.harvard.edu/abs/2010MNRAS.405..948S} {405, 948}

\bibitem[\protect\citeauthoryear{{Shankar}, {Marulli}, {Bernardi}, {Mei},
  {Meert}  \& {Vikram}}{{Shankar} et~al.}{2013}]{Shankar+13}
{Shankar} F.,  {Marulli} F.,  {Bernardi} M.,  {Mei} S.,  {Meert} A.,   {Vikram}
  V.,  2013, \mn@doi [\mnras] {10.1093/mnras/sts001}, \href
  {http://adsabs.harvard.edu/abs/2013MNRAS.428..109S} {428, 109}

\bibitem[\protect\citeauthoryear{{Shankar} et~al.,}{{Shankar}
  et~al.}{2014a}]{Shankar+14_sizes}
{Shankar} F.,  et~al., 2014a, \mn@doi [\mnras] {10.1093/mnras/stt2470}, \href
  {http://adsabs.harvard.edu/abs/2014MNRAS.439.3189S} {439, 3189}

\bibitem[\protect\citeauthoryear{{Shankar} et~al.,}{{Shankar}
  et~al.}{2014b}]{Shankar+14_SMHM}
{Shankar} F.,  et~al., 2014b, \mn@doi [\apjl] {10.1088/2041-8205/797/2/L27},
  \href {http://adsabs.harvard.edu/abs/2014ApJ...797L..27S} {797, L27}

\bibitem[\protect\citeauthoryear{{Shankar} et~al.,}{{Shankar}
  et~al.}{2017}]{Shankar+17_bulgehalo}
{Shankar} F.,  et~al., 2017, \mn@doi [\apj] {10.3847/1538-4357/aa66ce}, \href
  {http://adsabs.harvard.edu/abs/2017ApJ...840...34S} {840, 34}

\bibitem[\protect\citeauthoryear{{Shen}, {Mo}, {White}, {Blanton}, {Kauffmann},
  {Voges}, {Brinkmann}  \& {Csabai}}{{Shen} et~al.}{2003}]{Shen+03}
{Shen} S.,  {Mo} H.~J.,  {White} S.~D.~M.,  {Blanton} M.~R.,  {Kauffmann} G.,
  {Voges} W.,  {Brinkmann} J.,   {Csabai} I.,  2003, \mn@doi [\mnras]
  {10.1046/j.1365-8711.2003.06740.x}, \href
  {http://adsabs.harvard.edu/abs/2003MNRAS.343..978S} {343, 978}

\bibitem[\protect\citeauthoryear{{Shi}, {Lapi}, {Mancuso}, {Wang}  \&
  {Danese}}{{Shi} et~al.}{2017}]{Shi+17}
{Shi} J.,  {Lapi} A.,  {Mancuso} C.,  {Wang} H.,   {Danese} L.,  2017, \mn@doi
  [\apj] {10.3847/1538-4357/aa7893}, \href
  {http://adsabs.harvard.edu/abs/2017ApJ...843..105S} {843, 105}

\bibitem[\protect\citeauthoryear{{Somerville} \& {Dav{\'e}}}{{Somerville} \&
  {Dav{\'e}}}{2015}]{Somerville&Dave2015}
{Somerville} R.~S.,  {Dav{\'e}} R.,  2015, \mn@doi [\araa]
  {10.1146/annurev-astro-082812-140951}, \href
  {http://adsabs.harvard.edu/abs/2015ARA%26A..53...51S} {53, 51}

\bibitem[\protect\citeauthoryear{{Somerville} et~al.,}{{Somerville}
  et~al.}{2008}]{Somerville+08}
{Somerville} R.~S.,  et~al., 2008, \mn@doi [\apj] {10.1086/523661}, \href
  {http://adsabs.harvard.edu/abs/2008ApJ...672..776S} {672, 776}

\bibitem[\protect\citeauthoryear{{Somerville} et~al.,}{{Somerville}
  et~al.}{2018}]{Somerville+18}
{Somerville} R.~S.,  et~al., 2018, \mn@doi [\mnras] {10.1093/mnras/stx2040},
  \href {http://adsabs.harvard.edu/abs/2018MNRAS.473.2714S} {473, 2714}

\bibitem[\protect\citeauthoryear{{Springel}}{{Springel}}{2010}]{Springel+10}
{Springel} V.,  2010, \mn@doi [\mnras] {10.1111/j.1365-2966.2009.15715.x},
  \href {https://ui.adsabs.harvard.edu/abs/2010MNRAS.401..791S} {401, 791}

\bibitem[\protect\citeauthoryear{Springel, White, Tormen  \&
  Kauffmann}{Springel et~al.}{2001}]{Springel+01_subfind}
Springel V.,  White S. D.~M.,  Tormen G.,   Kauffmann G.,  2001, \mn@doi
  [Monthly Notices of the Royal Astronomical Society]
  {10.1046/j.1365-8711.2001.04912.x}, 328, 726

\bibitem[\protect\citeauthoryear{{Springel} et~al.,}{{Springel}
  et~al.}{2018}]{Springel+18}
{Springel} V.,  et~al., 2018, \mn@doi [\mnras] {10.1093/mnras/stx3304}, \href
  {https://ui.adsabs.harvard.edu/abs/2018MNRAS.475..676S} {475, 676}

\bibitem[\protect\citeauthoryear{{Straatman} et~al.,}{{Straatman}
  et~al.}{2017}]{Straatman+17_tullyFisher}
{Straatman} C.~M.~S.,  et~al., 2017, \mn@doi [\apj] {10.3847/1538-4357/aa643e},
  \href {http://adsabs.harvard.edu/abs/2017ApJ...839...57S} {839, 57}

\bibitem[\protect\citeauthoryear{{Tinker}, {Kravtsov}, {Klypin}, {Abazajian},
  {Warren}, {Yepes}, {Gottl{\"o}ber}  \& {Holz}}{{Tinker}
  et~al.}{2008}]{Tinker+08}
{Tinker} J.,  {Kravtsov} A.~V.,  {Klypin} A.,  {Abazajian} K.,  {Warren} M.,
  {Yepes} G.,  {Gottl{\"o}ber} S.,   {Holz} D.~E.,  2008, \mn@doi [\apj]
  {10.1086/591439}, \href {http://adsabs.harvard.edu/abs/2008ApJ...688..709T}
  {688, 709}

\bibitem[\protect\citeauthoryear{{Tinker} et~al.,}{{Tinker}
  et~al.}{2017}]{Tinker+17_SMHM_massivegal}
{Tinker} J.~L.,  et~al., 2017, \mn@doi [\apj] {10.3847/1538-4357/aa6845}, \href
  {http://adsabs.harvard.edu/abs/2017ApJ...839..121T} {839, 121}

\bibitem[\protect\citeauthoryear{{Vale} \& {Ostriker}}{{Vale} \&
  {Ostriker}}{2006}]{Vale&Ostriker2006}
{Vale} A.,  {Ostriker} J.~P.,  2006, \mn@doi [\mnras]
  {10.1111/j.1365-2966.2006.10605.x}, \href
  {http://adsabs.harvard.edu/abs/2006MNRAS.371.1173V} {371, 1173}

\bibitem[\protect\citeauthoryear{{Vogelsberger} et~al.,}{{Vogelsberger}
  et~al.}{2014}]{Vogelsberger+14}
{Vogelsberger} M.,  et~al., 2014, \mn@doi [\mnras] {10.1093/mnras/stu1536},
  \href {http://adsabs.harvard.edu/abs/2014MNRAS.444.1518V} {444, 1518}

\bibitem[\protect\citeauthoryear{{Wang}, {Dutton}, {Stinson}, {Macci{\`o}},
  {Penzo}, {Kang}, {Keller}  \& {Wadsley}}{{Wang} et~al.}{2015}]{Wang+15_NIHAO}
{Wang} L.,  {Dutton} A.~A.,  {Stinson} G.~S.,  {Macci{\`o}} A.~V.,  {Penzo} C.,
   {Kang} X.,  {Keller} B.~W.,   {Wadsley} J.,  2015, \mn@doi [\mnras]
  {10.1093/mnras/stv1937}, \href
  {http://adsabs.harvard.edu/abs/2015MNRAS.454...83W} {454, 83}

\bibitem[\protect\citeauthoryear{{Wechsler} \& {Tinker}}{{Wechsler} \&
  {Tinker}}{2018}]{Wechsler&Tinker2018}
{Wechsler} R.~H.,  {Tinker} J.~L.,  2018, preprint, \href
  {http://adsabs.harvard.edu/abs/2018arXiv180403097W} {} (\mn@eprint {arXiv}
  {1804.03097})

\bibitem[\protect\citeauthoryear{{Weinberger} et~al.,}{{Weinberger}
  et~al.}{2017}]{Weinberger+17}
{Weinberger} R.,  et~al., 2017, \mn@doi [\mnras] {10.1093/mnras/stw2944}, \href
  {http://adsabs.harvard.edu/abs/2017MNRAS.465.3291W} {465, 3291}

\bibitem[\protect\citeauthoryear{{Wojtak} \& {Mamon}}{{Wojtak} \&
  {Mamon}}{2013}]{Wojtak&Mamon2013}
{Wojtak} R.,  {Mamon} G.~A.,  2013, \mn@doi [\mnras] {10.1093/mnras/sts203},
  \href {http://adsabs.harvard.edu/abs/2013MNRAS.428.2407W} {428, 2407}

\bibitem[\protect\citeauthoryear{{Xie}, {De Lucia}, {Hirschmann}, {Fontanot}
  \& {Zoldan}}{{Xie} et~al.}{2017}]{Xie+17}
{Xie} L.,  {De Lucia} G.,  {Hirschmann} M.,  {Fontanot} F.,   {Zoldan} A.,
  2017, \mn@doi [\mnras] {10.1093/mnras/stx889}, \href
  {https://ui.adsabs.harvard.edu/abs/2017MNRAS.469..968X} {469, 968}

\bibitem[\protect\citeauthoryear{{Yang}, {Mo}, {van den Bosch}, {Pasquali},
  {Li}  \& {Barden}}{{Yang} et~al.}{2007}]{Yang+07}
{Yang} X.,  {Mo} H.~J.,  {van den Bosch} F.~C.,  {Pasquali} A.,  {Li} C.,
  {Barden} M.,  2007, \mn@doi [\apj] {10.1086/522027}, \href
  {http://adsabs.harvard.edu/abs/2007ApJ...671..153Y} {671, 153}

\bibitem[\protect\citeauthoryear{{Zoldan}, {De Lucia}, {Xie}, {Fontanot}  \&
  {Hirschmann}}{{Zoldan} et~al.}{2018}]{Zoldan+18}
{Zoldan} A.,  {De Lucia} G.,  {Xie} L.,  {Fontanot} F.,   {Hirschmann} M.,
  2018, \mn@doi [\mnras] {10.1093/mnras/sty2343}, \href
  {http://adsabs.harvard.edu/abs/2018MNRAS.481.1376Z} {481, 1376}

\bibitem[\protect\citeauthoryear{{Zoldan}, {De Lucia}, {Xie}, {Fontanot}  \&
  {Hirschmann}}{{Zoldan} et~al.}{2019}]{Zoldan+19}
{Zoldan} A.,  {De Lucia} G.,  {Xie} L.,  {Fontanot} F.,   {Hirschmann} M.,
  2019, arXiv e-prints, \href
  {http://adsabs.harvard.edu/abs/2019arXiv190210724Z} {}

\bibitem[\protect\citeauthoryear{{Zolotov} et~al.,}{{Zolotov}
  et~al.}{2015}]{Zolotov+15}
{Zolotov} A.,  et~al., 2015, \mn@doi [\mnras] {10.1093/mnras/stv740}, \href
  {http://adsabs.harvard.edu/abs/2015MNRAS.450.2327Z} {450, 2327}

\bibitem[\protect\citeauthoryear{{van der Wel} et~al.,}{{van der Wel}
  et~al.}{2014}]{vanderwel+2014}
{van der Wel} A.,  et~al., 2014, \mn@doi [\apj] {10.1088/0004-637X/788/1/28},
  \href {http://adsabs.harvard.edu/abs/2014ApJ...788...28V} {788, 28}

\makeatother
\end{thebibliography}




\bsp	
\label{lastpage}
\end{document}